\documentclass[aps,prb,amsmath,amssymb,superscriptaddress,twocolumn,floatfix]{revtex4-2}
\usepackage{graphicx}
\usepackage{dcolumn}
\usepackage{epsfig}
\usepackage{float}
\usepackage[bbgreekl]{mathbbol}
\usepackage{mathrsfs}
\usepackage{slashed}
\usepackage[cal=boondox,scr=boondoxo]{mathalfa}
\usepackage{trsym}

\newcommand\vex[1]{\mathbf{#1}}
\newcommand\gvex[1]{\boldsymbol{#1}}

\def\braket#1{\mathinner{\langle{#1}\rangle}}

\def\sgn{\mathrm{sgn}}

\def\tr{\mathrm{tr}}

\def\dd{\mathrm{d}}

\def\trans{{\raisebox{-1pt}{{\scriptsize\textsf{T}}}}}

\usepackage[T3,T1]{fontenc}
\DeclareSymbolFont{tipa}{T3}{cmr}{m}{n}
\DeclareMathAccent{\invbreve}{\mathalpha}{tipa}{16}

\usepackage{accents}
\newlength{\hhatheight}

\makeatother

\begin{document}
	
\title{
Lorentz violation in Dirac and Weyl semimetals 
}
	
\author{V.\ Alan Kosteleck\'y}
\affiliation{Department of Physics, Indiana University, Bloomington, Indiana 47405, USA}
\affiliation{Indiana University Center for Spacetime Symmetries, Bloomington, Indiana 47405, USA}
	
\author{Ralf Lehnert}
\affiliation{Department of Physics, Indiana University, Bloomington, Indiana 47405, USA}
\affiliation{Indiana University Center for Spacetime Symmetries, Bloomington, Indiana 47405, USA}
	
\author{Navin McGinnis}
\affiliation{TRIUMF, 4004 Westbrook Mall, Vancouver, BC, Canada V6T 2A3}
\affiliation{High Energy Physics Division, Argonne National Laboratory, Lemont, Illinois 60439, USA}
\affiliation{Department of Physics, Indiana University, Bloomington, Indiana 47405, USA}
	
\author{Marco Schreck}
\affiliation{Departamento de F\'isica, Universidade Federal do Maranh\~ao
Campus Universit\'ario do Bacanga, S\~ao Lu\'is (MA), 65085-580, Brazil}
	
\author{Babak Seradjeh}
\affiliation{Department of Physics, Indiana University, Bloomington, Indiana 47405, USA}
\affiliation{Indiana University Center for Spacetime Symmetries, Bloomington, Indiana 47405, USA}
\affiliation{Quantum Science and Engineering Center, Indiana University, Bloomington, Indiana 47405, USA}
	
\begin{abstract}
We propose a correspondence between the description of  
emergent Lorentz symmetry in condensed-matter systems
and the established general effective field theory 
for Lorentz violation in fundamental theories of spacetime and matter.
This correspondence has potential implications in both directions.
We illustrate the proposal by investigating its consequences 
for the spectral and transport properties of Dirac and Weyl semimetals. 
Particular realizations of this framework 
give rise to Dirac nodal spectra with nodal lines and rings. 
We demonstrate a bulk-boundary correspondence 
between bulk topological invariants and drumhead surface states 
of these Dirac nodal semimetals. 
We calculate their transport coefficients 
in leading-order perturbation theory,
thereby characterizing the unconventional electromagnetic response 
due to small deviations from emergent Lorentz invariance.
Some prospective future applications of the correspondence are outlined.
\end{abstract}
	
\date{\today}
	
{
\let\clearpage\relax
\maketitle
}

\section{Introduction}

The realization that space and time are closely intertwined
revolutionized physics over a century ago.
Since then,
the Lorentz symmetry of spacetime has become a cornerstone 
of our best theoretical description of fundamental particles and fields,
which is an amalgamation of General Relativity and the Standard Model.
In recent years,
another role for Lorentz symmetry has appeared,
as an emergent property of certain condensed-matter systems
such as semimetals and some unconventional superconductors. 
The existence of mathematically related symmetries
in these two different contexts
suggests intriguing prospects for interdisciplinary advances,
including the cross transferral and exploitation of concepts and methods.

The present work draws on a specific parallel between
prospective deviations from Lorentz symmetry in high-energy physics	
and departures from emergent Lorentz symmetry in condensed-matter systems.
In high-energy physics,
establishing a consistent unified theory of gravity and quantum physics
remains an open challenge.
Any such theory can be expected to generate 
small but observable deviations from known physics,
which could include tiny violations of Lorentz invariance.
In condensed-matter physics,
Lorentz symmetry is explicitly broken in low-energy phases of matter. 
However, 
the electronic energy bands in a crystalline solid can exhibit 
an emergent Lorentz symmetry at low energies 
that governs the dynamics of quasiparticle excitations above the ground state. 
The explicit or spontaneous breaking of this symmetry 
is then manifested as various electronic phases of the system.
Here,
we provide a general perspective for establishing the correspondence 
between these two types of Lorentz violation,
and we examine some specific consequences in the context of semimetals.

A powerful field-theoretic technique for describing low-energy signals
arising in an underlying high-energy theory
is effective field theory
\cite{Weinberg:2009bg}.
The prospective Lorentz violations emerging 
from a unified theory of gravity and quantum physics 
are described in a general and model-independent way 
by an effective field theory known as the Standard-Model Extension (SME)
\cite{Colladay:1996iz,Colladay_1998,Kostelecky:2003fs}.
This framework can be used to classify, enumerate, and interpret
the various possible physical effects of departures from Lorentz symmetry.
It also forms the basis for numerous precision experimental searches
for Lorentz violation
\cite{Kostelecky:2008edit},
with a reach in some cases exceeding sensitivity
to the Planck-scale effects expected to govern
the behavior of spacetime and matter in the underlying unified theory. 

A field-theoretic approach is also widely used
in studies of condensed-matter systems with emergent Lorentz symmetry.
For example,
Dirac materials with isolated band touching points 
host quasiparticles with dynamics governed by the Dirac equation,
thereby intrinsically implementing quantum electrodynamics 
in the presence of gauge potentials.
For descriptions of Dirac and Weyl semimetals
\cite{Lv:2015pya,Tamai:2016,Yan:2016euz,Armitage:2017cjs,Gao:2018xwm,Lee:2021},
it is thus natural to adopt field-theoretic methods
and their lattice implementations
\cite{Grushin:2012mt,Liu:2012hk,Zyuzin:2012vn,Zyuzin:2012tv,Goswami:2012db,%
	Hannukainen:2020sif,Baum:2015,Landsteiner:2015pdh, PhysRevX.6.041046,Elbistan:2016rla,%
	vanderWurff:2017fpv,Behrends_2019,Song:2019asj,Chernodub:2019blw,%
	Silva:2021fzh,Ji:2021aan,BitaghsirFadafan:2020lkh,Burkov:2011,%
	Pozo:2018yzs,Gomez:2021aez,Rodgers:2021azg}.
Examples with emergent Lorentz symmetry in 2+1 dimensions also exist,
including the high-temperature superconducting state of cuprates 
with two-dimensional $d$-wave pairing symmetry~\cite{Van-Harlingen_1995,Tsuei_Kirtley_2000,Franz_2002,Herbut_2002,Seradjeh_Herbut_2002}, 
surfaces of various topological insulators~\cite{Hasan_Kane_2010,Ryu_2010,Ando_Fu_2015,Xu_2016,Reja_2017}, 
and graphene~\cite{Novoselov:2004,Castro-Neto_2009}, 
a two-dimensional sheet of carbon atoms arranged on a honeycomb lattice. 

The thesis of the present work
is that the comprehensive SME framework for Lorentz violation
provides a basis for the classification and phenomenological exploration 
of general quasiparticle excitations 
in the band structures of Dirac and Weyl materials,
and that in turn the features of emergent Lorentz symmetry 
in these materials offer insights into aspects of Lorentz violation 
in the SME framework.
The implications of this thesis are substantial in both directions.
On one side,
it offers prospects for the description, realization, 
and perhaps even design of novel phases of matter 
based on the general SME framework.
On the other side,
the existence of phases of matter realizing an emergent Lorentz violation
implies the potential to shed light on open challenges in the SME context,
such as the theoretical issue of quantum stability
and the physical meaning of large coefficients for Lorentz violation
\cite{Kostelecky_2001},
and ambiguities in radiative corrections 
\cite{Colladay_1998,Chung:1998jv,Jackiw:1999yp,Perez-Victoria:1999erb,%
	Chung:1999gg,Chung:1999pt,Altschul:2003ce,Altschul:2004gs}. 
Potential mathematical implications also exist on both sides.
For example,
for many simple types of Lorentz violation,
a consistent dynamical spacetime cannot be accommodated by Riemann geometry 
and instead may require a generalization such as Finsler geometry 
\cite{Kostelecky:2003fs,Kostelecky:2011qz}.
The correspondence proposed here thus leads us to anticipate 
that Finsler geometry plays a role
in quasiparticle dynamics in Dirac and Weyl semimetals and,
conversely, 
that these systems can serve as laboratories for analogues
of fundamental particle dynamics governed by Finsler geometry, 
including scenarios that are experimentally accessible
but mathematically intractable.

The goal of the present work is to pave the way 
for future comprehensive studies of these ideas
and to illustrate some of the benefits of this thesis.
Here,
we focus specifically on implications of the SME framework 
for emergent Lorentz symmetry in the low-energy effective theory 
of various types of Dirac and Weyl semimetals. 
Special cases previously considered correspond
to particular terms in the SME formalism.
For Weyl semimetals,
certain violations can be parametrized by a single axial background gauge field,
and the electromagnetic response has been analyzed with field-theoretic methods
\cite{Grushin:2012mt,Liu:2012hk,Zyuzin:2012vn,Zyuzin:2012tv,Goswami:2012db,%
	Hannukainen:2020sif,Landsteiner:2015pdh,PhysRevX.6.041046,Elbistan:2016rla,%
	vanderWurff:2017fpv,Behrends_2019,Song:2019asj,Chernodub:2019blw,%
	Silva:2021fzh,Ji:2021aan,BitaghsirFadafan:2020lkh}.
Studies of related systems
have considered some additional types of Lorentz violations
\cite{Burkov:2011,Pozo:2018yzs,Gomez:2021aez,Rodgers:2021azg}.
These and other systems are described here using SME-based lattice models,
which provide a starting point for understanding higher-order contributions 
in the energy bands at energy scales relevant in real materials
and allow for nonperturbative effects in SME coefficients for Lorentz violation.
We thereby find novel phases of matter,
such as nodal Dirac semimetals 
arising from Lorentz-violating tensorial spin-orbit couplings. 
For these models,
we investigate the topological properties of the nodal lines and rings, 
and we characterize the surface bound states.
In high-energy physics,
the SME is typically viewed as a perturbative framework,
and in the present context this perspective is well suited 
to order-by-order computations of semimetal transport coefficients
in powers of the fine-structure constant 
and the SME coefficients for Lorentz violation.
We analyze the leading effects of all coefficients for Lorentz violation
on the electromagnetic response of Dirac materials,
revealing an unconventional response 
in the presence of tensorial spin-orbit couplings 
along with various velocity anisotropies.

The paper is organized as follows. 
Section~\ref{sec:SME} is dedicated to a brief introduction of the SME. 
In Sec.~\ref{sec:latt}, 
we formulate lattice Hamiltonians based on the SME 
and study their energy bands for the cases 
of a background axial gauge field 
and a background tensorial spin-orbit coupling. 
In Sec.~\ref{sec:topology},
we characterize the nontrivial topology of the bands 
in terms of their bulk topological invariants and their surface states. 
Section~\ref{sec:transport} presents perturbative calculations 
of transport coefficients in the SME, 
including for a background axial gauge field, 
a background tensorial spin-orbit coupling, 
and velocity anisotropies. 
Finally, 
Sec.~\ref{sec:sum} provides a summary of our findings 
and an outlook on some interesting open issues
for future investigation.

\section{SME Basics}\label{sec:SME}

Small departures from exact Lorentz invariance in nature
could arise in an underlying unified theory such as strings
\cite{Kostelecky:1988zi,Kostelecky:1991ak}.
A general description of the ensuing Lorentz violations 
appearing at attainable energy scales 
can be formulated using effective field theory 
\cite{Kostelecky:1994rn},
yielding the SME framework
\cite{Colladay:1996iz,Colladay_1998,Kostelecky:2003fs}.
The SME degrees of freedom include 
those of all known elementary particles and their interactions. 
The SME action consists of the Einstein-Hilbert action
for General Relativity coupled to the action for the Standard Model,
together with all possible terms formed from Lorentz-violating operators
that respect general coordinate invariance.
Each Lorentz-violating term is constructed using a background field
that remains unaffected by Lorentz transformations 
of the experimental system of interest. 
The background field is coupled to an operator formed from dynamical fields 
to yield a term in the action that is a scalar 
under general coordinate transformations. 
The components of the background field 
are called coefficients for Lorentz violation.
In a realistic effective field theory of this type,
any terms that break CPT symmetry,
which is the product of charge conjugation C,
parity inversion P, and time reversal T,
must also break Lorentz invariance 
\cite{Colladay:1996iz,Greenberg:2002uu}. 
The set of SME coefficients therefore controls CPT violation
as well as Lorentz violation.
Reviews of the SME can be found in Refs.\ 
\cite{Bluhm:2005uj,Will:2014kxa,Tasson:2014dfa,Hees:2016lyw}.

In the present work, 
we are interested in the behavior 
of electromagnetically coupled spin-$\frac{1}{2}$ quasiparticle excitations 
in condensed-matter systems, 
so the relevant SME limit is that 
of a single species of spin-$\frac{1}{2}$ Dirac fermions 
subject to a U(1) gauge interaction. 
For practical purposes, 
we further restrict our analysis to the minimal nongravitational SME 
\cite{Colladay:1996iz,Colladay_1998}.
Terms in the Lagrange density of this version of the SME 
contain only field operators of mass dimensions $d\leq 4$, 
a feature shared by established descriptions 
of electronic quasiparticle properties in Weyl and Dirac semimetals. 
More general SME contributions, 
including field operators with $d>4$ in the nonminimal SME 
\cite{Kostelecky:2009zp,Kostelecky:2011gq,Kostelecky:2013rta,Kostelecky:2018yfa}
and gravitational field operators
\cite{Kostelecky:2003fs,Kostelecky_2011,Kostelecky:2017zob,Kostelecky:2020hbb},
may well also be of interest for condensed-matter systems, 
but an investigation of their roles 
in this context lies beyond our present scope.

With the above considerations in mind, 
we are led to focus on the following flat-spacetime SME limit \cite{Kostelecky_2001}:
\begin{equation}\label{eq:SMEaction}
	S = \frac12 \int \mathrm{d}^4x\,\bar\psi\left(\mathrm{i}\Gamma^\mu D_\mu - M \right)  \psi + \text{h.c.}%\,.
\end{equation}
In this expression, 
$\psi$ denotes a four-component spinor field, 
and $\bar\psi \equiv \psi^{\dagger}\gamma^0$ is its Dirac conjugate,
as usual. 
The generalized Dirac and mass matrices, 
$\Gamma^{\mu}\equiv\gamma^{\mu}+\delta\Gamma^{\mu}$ and $M\equiv m+\delta M$, 
are composed of the ordinary Lorentz-symmetric pieces $\gamma^{\mu}$ and $m$ 
and Lorentz-violating contributions $\delta\Gamma^{\mu}$ and $\delta M$. 
Minimal coupling to the vector potential $A_{\mu}$ is implemented 
via the conventional U(1)-covariant derivative $D_{\mu}=\partial_{\mu}-\mathrm{i}qA_{\mu}$ 
with the particle charge $q$. 
Repeated spacetime indices are understood to be summed over, 
and our conventions for the Minkowski metric, 
the Levi-Civita symbol, 
and the Dirac matrices are $\eta^{\mu\nu} = \text{diag}(1,-1,-1,-1)$,
$\epsilon^{0123}=+1$, 
$\{ \gamma^\mu,\gamma^\nu \} = 2\eta^{\mu\nu}$, 
$\gamma_5=\mathrm{i}\gamma^0\gamma^1\gamma^2\gamma^3$,
and $\sigma^{\mu\nu} = \frac{\mathrm{i}}{2}[\gamma^\mu,\gamma^\nu]$. 
Unless stated otherwise, we work in natural units $\hbar=c=e=1$.

To expose the spacetime-transformation behavior 
of the various components of $\delta\Gamma^{\mu}$ and $\delta M$, 
it is customary to decompose these quantities 
in terms of the 16 Dirac matrices as follows:
\begin{subequations}
	\begin{align}
		\delta\Gamma^{\mu} &
		:= c^{\nu\mu}\gamma_{\nu}+d^{\nu\mu}\gamma_5\gamma_{\nu}
		+e^{\mu}+\mathrm{i}f^{\mu}\gamma_5+\tfrac{1}{2}g^{\kappa\lambda\mu}\sigma_{\kappa\lambda}\,,
		\label{GammaDecomp4}\\[1ex]
		\delta M &
		:= a^\mu \gamma_\mu + b^\mu \gamma_5\gamma_\mu + \tfrac12 H^{\mu\nu}\sigma_{\mu\nu}\,.
		\label{GammaDecomp3}
	\end{align}
\end{subequations}
Here, 
the SME coefficients $a_\mu$, 
$b_\mu$, 
$c_{\mu\nu}$, 
$d_{\mu\nu}$, 
$e_\mu$, 
$f_\mu$, 
$g_{\lambda\mu\nu}$, 
and $H_{\mu\nu}$ 
control the type and size of deviations from Lorentz symmetry, 
and in the present flat-spacetime context, 
they may consistently be assumed to be constant. 
Without loss of generality, 
$c_{\mu\nu}$ and $d_{\mu\nu}$ can be taken as traceless, 
$H_{\mu\nu}$ as antisymmetric, 
and $g_{\kappa\lambda\mu}$ as antisymmetric in its first two indices. 
The coefficients $a_\mu$, 
$b_\mu$, 
$e_\mu$, 
and $g_{\kappa\lambda\mu}$ 
parametrize CPT-odd behavior, 
while $c_{\mu\nu}$, 
$d_{\mu\nu}$, 
$f_\mu$, 
and $H_{\mu\nu}$ 
are associated with CPT-even physics.

We remark in passing 
that in certain limits of the SME, 
various coefficients for Lorentz violation
lead only to suppressed effects 
or become entirely undetectable. 
For example, 
in the present flat-spacetime, single-fermion situation, 
a judiciously chosen field redefinition 
removes $a_\mu$ from the action 
rendering these coefficients unobservable \cite{Colladay:1996iz,Colladay_1998,Kostelecky_2011}. 
Likewise, 
the $f_\mu$ coefficients can be completely absorbed into $c_{\mu\nu}$ 
by rescaling the Dirac matrices~\cite{Altschul:2006ts,Kostelecky_2010} 
and are thus superfluous.
For perturbatively small SME coefficients, 
additional leading-order transformations exist 
that either remove certain further SME coefficients
from the action~(\ref{eq:SMEaction}) 
or demonstrate their equivalence to other coefficients in the Lagrange density.

The set $\{\openone_4,\gamma_5,\gamma^{\mu},\gamma_5\gamma^{\mu},\sigma^{\mu\nu}\}$ 
spans the space of $(4\times4)$ matrices, 
so that the parametrizations~(\ref{GammaDecomp4}) and~(\ref{GammaDecomp3}) 
contain all possible nontrivial corrections 
to the free Dirac equation 
with fewer than two derivatives. 
It follows 
that general spin-$\frac{1}{2}$ quasiparticle excitations in condensed-matter systems, 
including ones not yet realized experimentally, 
are encompassed by the action~(\ref{eq:SMEaction}), 
with all generalizations to higher derivatives 
contained in the full SME. 
This broad scope, 
together with an abundance of existing theoretical SME explorations in high-energy physics, 
establishes the SME as a valuable framework 
for understanding, modeling, and predicting key features of Weyl and Dirac semimetals.

Examples of this assessment 
can readily be identified. 
Spin-independent and spin-dependent anisotropies in the Fermi velocity 
are associated with the SME coefficients $c^{\mu\nu}$ and $d^{\mu\nu}$, 
respectively. 
Furthermore, 
semimetals with Weyl nodes 
separated by $2b^{\mu}$ in four-momentum space 
are known to be governed by the $b^{\mu}$ contribution in the SME 
\cite{Behrends_2019}. 
Many aspects of the action in Eq.~\eqref{eq:SMEaction} 
have been investigated in high-energy physics,
such as its general plane-wave dispersion and propagation;
explicit eigenspinor solutions, spin sums, and propagators;
field redefinitions;
canonical field quantization;
classical-particle limit;
statistical physics; 
and its phenomenology~\cite{Colladay:1996iz,Bluhm:1997qb,Colladay_1998,Bluhm:1998rk,Kostelecky_2001,Bluhm:2001rw,Colladay:2002eh,Bluhm:2003un,Lehnert:2004ri,Colladay:2004qt,Altschul:2004wq,Lane:2005jv,Ferreira:2006kg,Altschul:2006ts,Altschul:2006uu,Lehnert:2006id,Ferreira:2007za,Ferreira:2007gnn,Altschul:2008qg,Altschul:2008ki,Colladay:2010ae,Altschul:2010na,Bocquet:2010ke,Kostelecky_2010,Kostelecky_2011,Fittante:2012ua,Noordmans:2014hxa,Schreck:2017isa,Aghababaei:2017bei,Escobar:2018hyo,Shao:2019tle}. 
Further results, 
which may also include SME terms of higher mass dimensions
or involve the QED extension of this action
can be found in Refs.~\cite{Colladay:2001wk,Kostelecky:2001jc,Kostelecky:2002ue,Altschul:2003ce,Altschul:2004gs,Altschul:2005za,Altschul:2006pv,Nascimento:2007rb,Kostelecky:2013rta,Cambiaso:2014eba,Gomes:2014kaa,SSantos:2015mzs,Kostelecky:2015nma,Mariz:2016ooa,Ding:2016lwt,Reis:2016hzu,Kostelecky:2018fmc,BaetaScarpelli:2018vsm,Ferrari:2018tps,Reis:2019jmm,Brito:2020eiy,Ding:2020aew,Ferrari:2021eam}. 
Therefore, 
Eq.~\eqref{eq:SMEaction} comes with a well developed tool kit 
for applications in condensed-matter systems. 
Part of this work will outline in more detail 
how known results for Weyl and Dirac semimetals 
closely mesh with SME physics of $b^{\mu}$ and $c^{\mu\nu}$.
	
The action~(\ref{eq:SMEaction}) also permits investigations of band structures 
that are possible in principle, 
but have not yet been established.
We will illustrate this capability of the SME 
in the context of its $g_{\kappa\lambda\mu}$ contribution. 
The associated dispersion 
can be extracted as usual 
via a plane-wave ansatz $\psi(x)\sim\exp(-\mathrm{i}k\cdot x)$, 
$k^{\mu}=(k^0,\vex{k})$, 
in the modified Dirac equation
emerging from the action~(\ref{eq:SMEaction}). 
This dispersion can only depend on $g_{\kappa\lambda\mu}$ 
through $K_{\kappa\lambda}{}:={} g_{\kappa\lambda\mu}k^{\mu}$ and $L_{\kappa}{}:={} g_{\kappa\lambda\mu}k^{\lambda}k^{\mu}$, 
and an explicit calculation yields
\begin{align}
	\label{FullDR}
	0& = (k^2-m^2)^2-4K^2(k^2+m^2)-4K^4
	\nonumber\\
	&\phantom{{}={}}{}+16 L^2+16K^{\kappa}{}_{\lambda}K^{\lambda}{}_{\mu}K^{\mu}{}_{\nu}K^{\nu}{}_{\kappa}\,,
\end{align}
where $K^2{}:={} K^{\mu\nu}K_{\mu\nu}$. 
In general, 
the latter represents a quartic equation in the energy variable $k^0$ 
for a given three-momentum $\vex{k}$. 
Its four roots correspond to particle and antiparticle states 
with two spin degrees of freedom each. 
Exact expressions for the roots can be given, 
but they are not particularly transparent. 
However, 
leading-order results, 
valid for $|g_{\kappa\lambda\mu}|\ll 1$, 
can be found from:
\begin{equation}
	\label{ApprDR}
	k^2-m^2 \simeq \pm 4\, \sqrt{\tfrac{1}{2}K^2m^2-L^2}\,.
\end{equation}
Here, 
it is understood that $K_{\kappa\lambda}$ and $L_{\kappa}$ 
are constructed with the zeroth-order roots $k_{\pm}^{\mu}\equiv(\pm\sqrt{\vex{k}^2+m^2},\vex{k})$.
The $\pm$ signs in Eq.~(\ref{ApprDR}) and in $k_{\pm}^{\mu}$ are uncorrelated, 
so that all the usual degeneracies are typically lifted.
	
The antisymmetric structure of $g_{\kappa\lambda\mu}$ in its first two indices 
results in 24 independent components. 
For many purposes, 
it is useful to decompose them into Lorentz-irreducible pieces:
\begin{subequations}
	\label{gDecomp}
	\begin{equation}
		g_{\kappa\lambda\mu}=
		g^{(M)}_{\kappa\lambda\mu}
		+\epsilon_{\kappa\lambda\mu}{}^{\nu}g^{(A)}_{\nu}
		-\tfrac{1}{3}\big(
		\eta_{\kappa\mu}g^{(T)}_{\lambda}
		-\eta_{\lambda\mu}g^{(T)}_{\kappa}
		\big)\,,
	\end{equation}
	where
	\begin{align}
		g^{(M)}_{\kappa\lambda\mu} & {}:={}
		{}+\tfrac{1}{3}(g_{\kappa\lambda\mu}+g_{\kappa\mu\lambda}+\eta_{\kappa\mu}g_{\lambda\nu}{}^{\nu})
		\nonumber\\
		&\phantom{{}:={}}-\tfrac{1}{3}(g_{\lambda\kappa\mu}+g_{\lambda\mu\kappa}+\eta_{\lambda\mu}g_{\kappa\nu}{}^{\nu})\,,	
		\label{eq:gM}
		\\
		g^{(A)}_{\nu} & {}:={}
		{}-\tfrac{1}{6}\epsilon_{\kappa\lambda\mu\nu}g^{\kappa\lambda\mu}\,,
		\label{eq:gA}
		\\
		g^{(T)}_{\kappa} & {}:={}
		{}+\eta^{\lambda\mu}g_{\kappa\lambda\mu}
		=g_{\kappa\lambda}{}^{\lambda}\,.
		\label{eq:gT}
	\end{align}
\end{subequations}
Here, 
$g^{(A)}_{\nu}$ and $g^{(T)}_{\kappa}$ each 
contain four independent components 
and denote the fully antisymmetric and trace pieces,
respectively. 
The mixed-symmetry piece $g^{(M)}_{\kappa\lambda\mu}$ is non-axial, 
$\epsilon^{\kappa\lambda\mu\nu}g^{(M)}_{\kappa\lambda\mu}=0$,
and traceless, 
$\eta^{\lambda\mu}g^{(M)}_{\kappa\lambda\mu}=0$. 
These eight constraints leave $g^{(M)}_{\kappa\lambda\mu}$ with 16 independent components.
At leading order, 
$m g^{(A)}_{\nu}$ can be absorbed into $b_{\nu}$, 
and $g^{(T)}_{\kappa}$ can be removed with a field redefinition.
For these reasons, 
we will focus on $g^{(M)}_{\kappa\lambda\mu}$ in this work.

\section{Lattice models}\label{sec:latt}

Lattice models 
whose low-energy theory contains desired terms in the SME 
can be constructed in various ways. 
Here, 
we illustrate this fact 
by taking the minimal SME as our starting point 
and constructing the corresponding lattice Bloch Hamiltonians 
via a Wilson map for every spatial direction $j$,
\begin{subequations}
\label{eq:wilson-map}
\begin{align}
-\mathrm{i}\partial_j &\mapsto  v \sin k_j =: v_j(\vex k), \\
\label{eq:wilson-map-mass}
m &\mapsto m + B \sum_{j=1}^3 [ 1- \cos (k_j)] =: \mu(\vex k).
\end{align}
\end{subequations}
Here, 
$\vex k = (k_1,k_2,k_3)$ is the lattice momentum in the Brillouin zone $[-\pi,\pi]^3$ 
where we choose the lattice spacing as the length unit. 
The electromagnetic vector potential 
is introduced on the lattice 
by the Peierls substitution, 
$\vex k \mapsto \vex k - \vex A$, 
where $\vex A = (A_1,A_2,A_3)$. 
The lattice parameters $m$ and $B$ 
control the gap structure of the model: 
for $\sgn(m)\sgn(m+B) > 0$, 
the energy bands are gapped, 
while for $m=0$ the gap closes at discrete points in the Brillouin zone, 
realizing a semimetal. 
The parameter $v$ is the Fermi velocity at $\vex k =0$.

Incorporating the temporal direction 
indicated by the index $\mu = 0$ in $\delta\Gamma^{\mu}$ of Eq.~\eqref{GammaDecomp4} 
needs some care, 
since it provides the link between the equations of motion 
obtained from the action~(\ref{eq:SMEaction}) 
and the Hamiltonian~\cite{Kostelecky_2001,Kostelecky_2011}. 
Here, 
we assume $\Gamma^0 = \gamma^0$ for simplicity, 
whereupon $\delta\Gamma^0=0$. 
This requirement restricts the number of nonzero elements of the Lorentz-violating background fields. 
Specifically, 
we will be setting $c_{\lambda 0} = g_{\kappa\lambda0} = d_{\lambda 0} = e_0 = f_0 = 0$ 
in our lattice models. 
Then, 
the generic Hamiltonian based on the Dirac-fermion sector of the minimal SME 
given by Eq.~(\ref{eq:SMEaction}) 
takes the form
\begin{equation}
\label{eq:sme-hamiltonian}
H (\vex k) = \gamma^0\Gamma^j v_j (\vex k - \vex A) + \gamma^0 M(\vex k - \vex A) + A_0\,,
\end{equation}
where, 
again, 
a sum over the repeated index $j$ is understood. 
For the explicit form of the free Hamiltonian, 
consult Ref.~\cite{Kostelecky:1999zh}.

In the following, 
we consider two cases in detail: 
the well-studied Weyl semimetal 
with only a nonzero $b_\mu$ background field 
and a novel Lorentz-violating Dirac semimetal with a nonzero $g_{\kappa\lambda\nu}$ background field.

\subsection{The $\boldsymbol b$ term: Weyl semimetals}\label{sec:latt-b}

An effective microscopic model for Weyl semimetals 
based on the $b$ term of Eq.~(\ref{eq:SMEaction}) 
has been considered in Refs.\ 
\cite{Liu:2012hk,Grushin:2012mt,Zyuzin:2012vn,Zyuzin:2012tv,Goswami:2012db,%
Hannukainen:2020sif,Landsteiner:2015pdh,PhysRevX.6.041046,Elbistan:2016rla,%
vanderWurff:2017fpv,Behrends_2019,Song:2019asj,Chernodub:2019blw,%
Silva:2021fzh,Ji:2021aan,BitaghsirFadafan:2020lkh}.
Note that the conventions for the $b$ term 
employed in the SME action of Eq.~(\ref{eq:SMEaction}) 
are different from those typically used in the condensed-matter context. 
First, 
the Lagrangian employed in Eq.~(\ref{eq:SMEaction}) 
is Hermitian by construction. 
Second, 
the $b$ term occurring in the SME 
comes with the opposite sign 
relative to the corresponding term in, 
e.g.,
Ref.~\cite{Grushin:2012mt}.

Keeping only the $b$ term nonzero 
among the Lorentz-violating background fields 
and setting $A_\mu=0$, 
we arrive at the following lattice Hamiltonian 
from Eqs.~(\ref{eq:wilson-map}), (\ref{eq:sme-hamiltonian}):
\begin{equation}\label{eq:Hb}
H_b(\vex k) = %\sum_{j=1}^3
	\gamma^0 \gamma^j [v_j(\vex k) - b_j \gamma_5] + \gamma^0 \mu(\vex k) - b_0 \gamma_5\,.
\end{equation}
As of now, 
we will suppress the explicit dependence of $\mathbf{v}$ and $\mu$ 
on $\vex k$ for brevity.
For a purely timelike background field $b_\mu = (b_0,0)$, 
the dispersion has the form
\begin{equation}\label{eq:Ebt}
E_b^\text{t}= \pm\sqrt{\left(|\vex v| \pm b_0 \right)^2 + \mu^2}\,,
\end{equation}
where $\vex v = (v_1,v_2,v_3)$ 
and the signs are chosen independently. 
For a purely spacelike background field $b_\mu = (0,\vex b)$, 
the dispersion reads,
\begin{equation}\label{eq:Ebs}
E_b^\text{s} = \pm \sqrt{|\vex v_{\perp b}|^2 + \left(\sqrt{|\vex v_{\parallel b}|^2 +\mu^2} \pm |\vex b| \right)^2 }\,,
\end{equation}
where $\vex v_{\parallel b} := (\vex v \cdot \vex b/|\vex b|) \vex b$ and $\vex v_{\perp b} := \vex v - \vex v_{\parallel b}$ 
are the components of $\vex v$ parallel and perpendicular to $\vex b$, 
respectively.

The closed form of the dispersion in the general case is cumbersome. 
However, 
for $\mu \equiv 0$, 
it simplifies to $E_b^\text{0}(\vex k) = \pm \left[b_0 + |\vex v \pm \vex b| \right]$.
This is useful 
in deducing the general dispersion for $m=0$ and small $b_\mu$ near $\vex k = 0$ as
\begin{equation}
E_b(\vex k) = \pm [b_0 + | v \vex k \pm \vex b|] + \mathcal{O}(B^2)\,,
\end{equation}
with $B$ employed in Eq.~(\ref{eq:wilson-map-mass}). 
The latter result shows the presence of Weyl nodes 
at energies $E_0^\pm = \pm b_0 + \mathcal{O}(B^2)$ 
and momenta $\vex k^\pm_0 = \pm \vex b / v + \mathcal{O}(B^2)$ \cite{Behrends_2019}.

\subsection{The $\boldsymbol g$ term: Dirac nodal semimetals}\label{sec:latt-g}

We will now focus 
on the $g$ term, 
defined in Eq.~(\ref{eq:SMEaction}), 
and its understanding within the context of semimetals.
For $A_\mu=0$, 
keeping only the $g$ term and having set $g_{\kappa\lambda 0}=0$ in the Hamiltonian formulation, 
we have
\begin{equation}
\label{eq:Hg}
H_g(\vex k) = \gamma^0\gamma^j v_j(\vex k) + \frac12 \gamma^0 \sigma^{\kappa\lambda} g_{\kappa\lambda j} v_j(\vex k) + \gamma^0 \mu(\vex k)\,,
\end{equation}
from Eqs.~(\ref{eq:wilson-map}) and (\ref{eq:sme-hamiltonian}). 
In the remainder of this paper, 
we will only consider the effect of the mixed component $g^{(M)}_{\kappa\lambda\mu}$ of Eq.~(\ref{gDecomp}) 
and set $g^{(A)}_\mu = g^{(T)}_\mu = 0$.

As we discard the four components $g_{\kappa\lambda 0}$, 
we can parametrize the remaining twelve components of the mixed piece 
in terms of two $(3\times 3)$ matrices $g_0$ and $g_1$ 
with spatial components
\begin{subequations}
\label{eq:definitions-g0-g1}
\begin{align}
(g_0)_{ij} &:= g_{0ij}\,, \label{eq:g0}\\[1ex]
(g_1)_{ij} &:= \frac12\epsilon_{ikl} g_{klj}\,, \label{eq:g1}
\end{align}
\end{subequations}
with the Levi-Civita symbol $\epsilon_{ikl}=\epsilon^{ikl}$ 
in three dimensions.
We note 
from Eq.~(\ref{eq:gT}) 
that $g_0^{(T)}=-\tr(g_0)$ and $g_k^{(T)}=-\epsilon_{kij}(g_1)_{ij}$, 
where $k$ is a spatial component. 
Similarly, 
from Eq.~(\ref{eq:gA}), 
$g_0^{(A)} = \tr(g_1)$ and $g_k^{(A)} = \epsilon_{kij} (g_0)_{ij}$. 
In what follows, 
we will set $g^{(T)}_{\mu}=g^{(A)}_{\mu}=0$, 
which renders the matrices $g_0$ and $g_1$ symmetric and traceless. 
In this case,
they have twelve independent coefficients in total.

Then, 
the Hamiltonian of Eq.~(\ref{eq:Hg}) takes the form
\begin{equation}\label{eq:Hg01}
H_g = \gamma^0\gamma^j v_j + i\gamma^j (g_0 \vex v)_j + \gamma^j\gamma^5 (g_1\vex v)_j + \gamma^0 \mu\,,
\end{equation}
where we again omit dependences on $\mathbf{k}$. 
The corresponding dispersion 
can be written in closed form 
for some special cases. 
For $g_1 = 0$, 
we have
\begin{equation}
\label{eq:Eg0}
E_{g0} = \pm\sqrt{|\vex v_{\parallel g_0}|^2 + \left(|\vex v_{\perp g_0}| \pm |g_0 \vex v|\right)^2 + \mu^2}\,,
\end{equation}
where $\vex v_{\parallel g_0} := ( \vex v^\trans g_0\vex v/|g_0 \vex v|^2) g_0\vex v$ 
and $\vex v_{\perp g_0} := \vex v - \vex v_{\parallel g_0}$ 
are the components of $\vex v$ 
parallel and perpendicular to $g_0 \vex v$, 
respectively. 
This spectrum is gapped everywhere 
except at the Dirac point 
where $\mu = |\vex v| = 0$.
For $g_0=0$, 
we find
\begin{equation}
\label{eq:Eg1}
E_{g1} = \pm\sqrt{|\vex v_{\parallel g_1}|^2 + \left(\sqrt{|\vex v_{\perp g_1}|^2 + \mu^2} \pm |g_1 \vex v| \right)^2 }\,,
\end{equation}
where $\vex v_{\parallel g_1} = ( \vex v^\trans g_1\vex v/|g_1 \vex v|^2) g_1\vex v$ 
and $\vex v_{\perp g_1} = \vex v - \vex v_{\parallel g_1}$ 
are the components of $\vex v$ parallel and perpendicular to $g_1 \vex v$, 
respectively. 
We note that $E_{g0}$ and $E_{g1}$ 
are closely analogous to $E^\text{t}_b$ of Eq.~(\ref{eq:Ebt}) 
and $E^\text{s}_b$ of Eq.~(\ref{eq:Ebs}), 
respectively, 
for the $b$ term, 
except: 
(i) the parallel and perpendicular directions of $\vex v$ are switched, 
and (ii) the $b$ term is fixed and finite, 
whereas the $g$ terms vanish 
at the Dirac point 
along with $\vex v$ itself.

We now analyze the spectrum 
in a system with periodic boundary conditions. 
We first give a geometric interpretation of $g_0$ and $g_1$: 
since they are traceless, symmetric matrices, 
they can be rotated to a set of orthogonal principal axes 
labeled with $(a,b,c)$, 
where they take the diagonal form $\text{diag}(g_a,g_b,g_c)$ 
with $g_a+g_b +g_c=0$. 
Therefore, 
each such matrix can be represented by a rotation 
to the principal coordinate system 
(with three free parameters) 
and a combination of reflections 
around two principal axes 
(with two free parameters). 
In the following, 
when needed we shall work in this principal coordinate system, 
in which we denote $\vex v = (v_a,v_b,v_c)$.

Note that the double degeneracy of the original Dirac energy bands 
is lifted by the $g$ term, 
except along certain lines where $g_0\vex v = 0$ or $g_1\vex v=0$, 
respectively, 
for the $g_1=0$ or $g_0=0$ case. 
Of course, 
such degenerate lines exist only 
if one of the eigenvalues of $g_0$ or $g_1$ vanishes.

Interestingly, 
the Dirac point may now be accompanied 
by nodal lines at $\vex k \neq 0$ 
where the two central energy bands 
are degenerate at $E_{g0} = \pm |\mu(\vex k)|$ and $E_{g1}=0$. 
Such nodal lines exist 
when $|\vex v_{\parallel g_0}| = 0$ and $|\vex v_{\perp g_0}| = |g_0 \vex v|$ 
for the $g_0$ term, 
or $|\vex v_{\parallel g_1}| = 0$ and $|\vex v_{\perp g_1}|^2 + \mu^2 = |g_1 \vex v|^2$ 
for the $g_1$ term. 
Since the condition for nodal lines of $g_0$ 
can be obtained from that of $g_1$ 
by setting $\mu =0$, 
we will study the more general case of $g_1$ nodal lines 
and drop the reference to $0$ and $1$ for brevity. 
Near the Dirac point with $m=0$, 
we can set $\mu = \mathcal{O}(k^2) \to 0$ 
compared to $|\vex v| = \mathcal{O}(k$), 
and the two conditions coincide.

Since $\vex v_{\parallel g}=0$, 
we may replace $\vex v_{\perp g} = \vex v$ 
and simplify the condition for nodal lines to
\begin{subequations}
\begin{align}
\vex v^\trans g\, \vex v &= 0\,, \label{eq:vgv} \\[1ex]
\vex v^\trans (g^2-\openone_3) \vex v &= \mu^2\,, \label{eq:vg2v}
\end{align}
\end{subequations}
where we have used $\vex v_{\parallel g} \propto \vex v^\trans g\, \vex v$ 
and $|g\vex v|^2 = \vex v^\trans g^2\; \vex v$ 
for a symmetric matrix $ g = g^\trans$. 
The latter equations describe quadric surfaces 
that can be brought to normal form 
by diagonalizing the matrices $g$ and $g^2-\openone_3$, 
respectively.

If one of the eigenvalues vanishes, 
say $g_a=0=g_b+g_c$, 
then Eq.~(\ref{eq:vgv}) yields $v_b^2 = v_c^2$. 
From Eq.~(\ref{eq:vg2v}), 
we have $2(g_b^2-1)v_b^2 = v_a^2$. 
The nodal lines then exist for $|g_c|>1$. 
Near the Dirac point, 
we can set $\mu\to 0$ 
compared to linear terms in $\vex v$ 
to find nodal lines 
along the four directions $(\pm\sqrt{2(g_c^2-1)},\pm 1, 1)$ in the principal basis. 
One expects 
that these nodal lines should exist also 
when $|g_a|\ll1$ in approximately the same direction. 
Indeed, 
we will show in the full solution below 
that this is the case. 
The argument works similarly 
for the other principal directions $b$ and $c$, 
as summarized in Table~\ref{tab:gnodes}.

%%%% Table 1: Nodal lines of g %%%%
\setlength{\tabcolsep}{4pt} % Default value: 6pt
\renewcommand{\arraystretch}{1.75} % Default value: 1
\begin{table}[t]
\begin{tabular}{c c}
\hline
Condition & Nodal direction in principal axes  \\
\hline \hline
$|g_a|\ll1$, $|g_b|\gtrsim1$ & $\left(\pm\sqrt{2(g_b^2-1)},\pm1,1\right)$  \\
\hline
$|g_b|\ll1$, $|g_c|\gtrsim1$ & $\left(1,\pm\sqrt{2(g_c^2-1)},\pm1\right)$  \\
\hline
$|g_c|\ll1$, $|g_a|\gtrsim1$ & $\left(\pm1,1,\pm\sqrt{2(g_a^2-1)}\right)$\\
\hline
\end{tabular}
\caption{Nodal lines of $E_{g0}$ and $E_{g1}$ 
stated in Eq.~(\ref{eq:Eg0}) and Eq.~(\ref{eq:Eg1}), 
respectively, 
near the Dirac point for $g_0$ and $g_1$ 
with a vanishing eigenvalue $g_a$, $g_c$, or $g_b$.}
\label{tab:gnodes}
\end{table}%

Let us now look at the general case 
assuming nonzero eigenvalues. 
Then we can solve Eq.~(\ref{eq:vgv}) 
for $v_c^2 = (g_av_a^2 + g_b v_b^2)/(g_a + g_b)$
and replace in Eq.~(\ref{eq:vg2v}) 
to find the set of equations
\begin{subequations}
\begin{align}
g_a v_a^2 + g_b v_b^2 &= -g_c v_c^2\,, \\[1ex]
G_b v_a^2 - G_a v_b^2 &= -g_c \mu^2\,, \label{eq:vab2}
\end{align}
\end{subequations}
where
\begin{subequations}
\begin{align}
G_a &= (g_b-g_c)(g_b g_c + 1)\,, \\[1ex]
G_b &= (g_c-g_a)(g_cg_a +1)\,.
\end{align}
\end{subequations}
Solving for $v_a^2$ and $v_b^2$, we find
\begin{subequations}
\begin{align}
v_a^2 &= \frac{G_a v_c^2 + g_b \mu^2}{G_c}\,, \label{eq:vac2} \\[1ex]
v_b^2 &= \frac{G_b v_c^2 - g_a \mu^2}{G_c}\,. \label{eq:vbc2}
\end{align}
with
\begin{equation}
G_c = (g_a-g_b)(g_ag_b +1)\,.
\end{equation}
\end{subequations}
Note that Eqs.~(\ref{eq:vac2}) and~(\ref{eq:vbc2}) 
can be obtained from Eq.~(\ref{eq:vab2}) 
by appropriate cyclic permutations of $(a,b,c)$. 
This makes sense 
since the choice of $v_a, v_c$ in Eq.~(\ref{eq:vab2}) is arbitrary.

\begin{figure}[t!]
\includegraphics[width=2.8in]{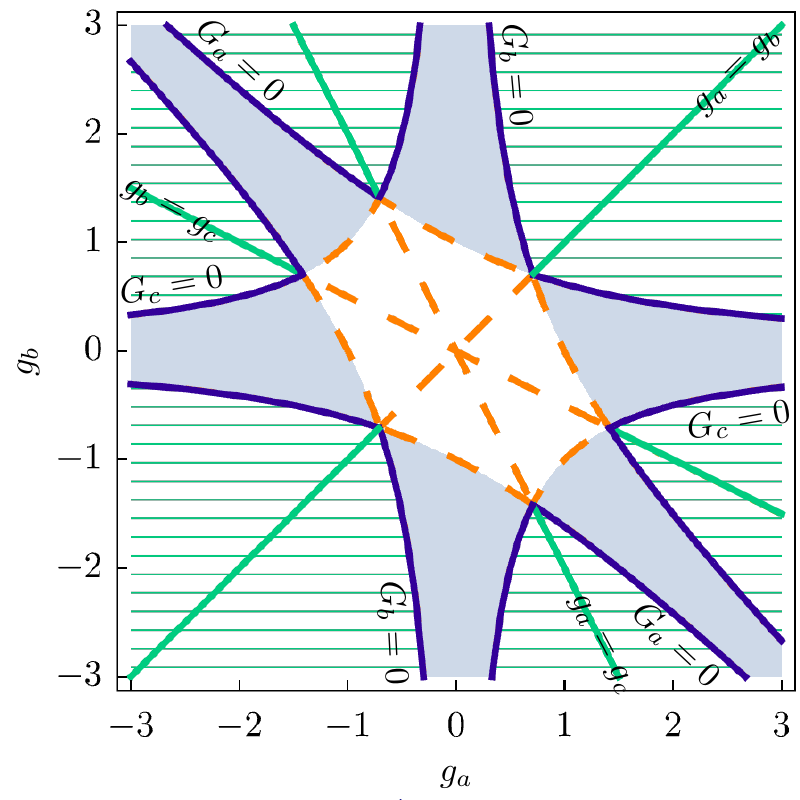}
\caption{Diagram of solutions for the nodal lines of $E_{g0}$ and $E_{g1}$ near the Dirac point, 
with $E_{g0},E_{g_1}$ taken from Eq.~(\ref{eq:Eg0}) and Eq.~(\ref{eq:Eg1}), 
respectively. 
In the solid (blue) regions, 
there are nodal lines near the Dirac point. 
The dark solid (blue) borders indicate 
that there are nodal lines 
through the Dirac point. 
When two eigenvalues of $g_0$ and $g_1$ are equal 
along the light solid (green) lines, 
there are nodal rings in $E_{g1}$. 
The hatched (green) regions are where nodal rings also form 
when the eigenvalues are not strictly equal. 
Analytical forms of the solutions are obtained perturbatively 
near light solid (green) lines 
as in Eq.~(\ref{eq:ka0c}). 
There are no nodal lines 
along the dashed curves and lines. 
The directions $g_a = 0$, $g_b = - g_a$, and $g_b=0$ 
bisecting the dark (blue) regions 
correspond to the cases in Table~\ref{tab:gnodes}.}
\label{fig:gnodes}
\end{figure}

The existence of nodal lines 
can be inferred from the relative signs of $G_a,G_b$ and $G_c$. 
Importantly, 
near the original Dirac point, 
we may set $\mu \to 0$ 
compared to the linear terms, 
and the nodal lines exist when $G_a$, $G_b$, and $G_c$ 
all have the same sign. 
These regions are shown in Fig.~\ref{fig:gnodes}. 
As expected, 
they include the cases 
with a single vanishingly small eigenvalue in Table~\ref{tab:gnodes}.

At the borders of these regions, 
say when $G_a=0$, 
we have $g_bg_c+1=0$. 
Then, 
$G_c = g_b^2 G_b$ 
so that $G_b$ and $G_c$ have the same sign. 
In the case of $E_{g1}$, $\mu\neq0$ away from the Dirac point 
and solutions for $v_a^2 = (g_b/G_c)\mu^2$ exist 
when  $g_b$ and $G_c$ have the same signs. 
Then, 
the second equation $G_b v_c^2 = G_c v_b^2 + g_a \mu^2$ has solutions 
that form \emph{lines} through the Dirac point, 
along which $G_b v_c^2 \approx G_c v_b^2$ near the Dirac point. 
In the case of $E_{g0}$, 
nodal lines also exist in the plane $v_a=0$ 
with $G_b v_c^2 = G_c v_b^2$ exactly. 
The analysis is similar for the other borders when $G_b=0$ or $G_c=0$. 
Borders along which solutions exist 
are shown with solid (blue) curves in Fig.~\ref{fig:gnodes}.

We may also have $G_a=0$ 
when $g_b=g_c=-g_a/2$. 
Then, 
$G_c = - G_b = 3g_b (2g_b^2-1)$ 
and the equations for the nodes simplify to $(v_b^2 + v_c^2)/2 = v_a^2 = (g_b/G_c)\mu^2$, 
which have solutions only for $\mu\neq 0$ 
when $G_c/g_b = 3(2g_b^2-1) > 0$, 
hence $|g_b| > 1/\sqrt2$. 
Thus, 
these solutions exist only in the case of $E_{g1}$ 
and form closed \emph{nodal rings} 
around the Dirac point at the intersection of the surfaces 
formed by $v_b^2 + v_c^2 = 2v_a^2$ and $v_a^2 = (g_b/G_c)\mu^2$. 
The nodal rings exist along the solid (green) lines in Fig.~\ref{fig:gnodes}.

%%%% Fig. 1: Nodal Line Solutions %%%%
%%

Clearly, 
the nodal rings do not appear or disappear discontinuously 
as we vary the eigenvalues. 
Instead, 
we expect 
that there be a region around the lines of equal eigenvalues in Fig.~\ref{fig:gnodes} 
for which nodal rings exists in the spectrum of $E_{g1}$. 
We can see this for $g_b \gtrsim 1/\sqrt{2}$, 
where we expect the rings to be close to the Dirac points.  
For $g_b\approx g_c$, 
we can use $\vex v \approx v \vex k$ and $\mu \approx B\vex k^2/2$ 
to find to the lowest-order in $\epsilon := (g_c-g_b)/g_b$ that, 
when projected to the $b$-$c$ plane, 
the nodal rings form an ellipse
\begin{subequations}
\begin{equation}\label{eq:ellipsering}
\zeta_b k_b^2+ \zeta_c k_c^2 = \zeta_a k_{a0}^2\,,
\end{equation}
where
\begin{align}
k_{a0}^2 &= (2g_b^2-1) \frac{4v^2}{3B^2}\,, \\
\zeta_a &= 2 + \epsilon\,, ~ \zeta_b = 1+\frac43\epsilon\,, ~
\zeta_c = 1 + \frac{4g_b^2-5}{3(2g_b^2-1)} \epsilon\,.
\end{align}
\end{subequations}
In the full $\vex k$ space, the nodes are found on the surface
\begin{subequations}
\begin{equation}\label{eq:ka0c}
k_a^2 = k_{a0}^2 + A(k_c) \epsilon\,,
\end{equation}
where
\begin{equation}\label{eq:Akc}
A(k_c) = \frac{5g_b^2-1}{3(2g_b^2-1)}k_c^2 - \frac43k_{a0}^2\,.
\end{equation}
\end{subequations}
When $g_b=g_c > 1/\sqrt2$, 
$\epsilon=0$ exactly 
and we find a circular nodal ring with radius $\sqrt2 |k_{a0}|$ 
in the $b$-$c$ planes at $k_a = \pm k_{a0}$. 
We have checked numerically 
that nodal rings exist 
for the hatched (green) regions in Fig.~\ref{fig:gnodes}.

\begin{figure}[t!]
\includegraphics[width=3.4in]{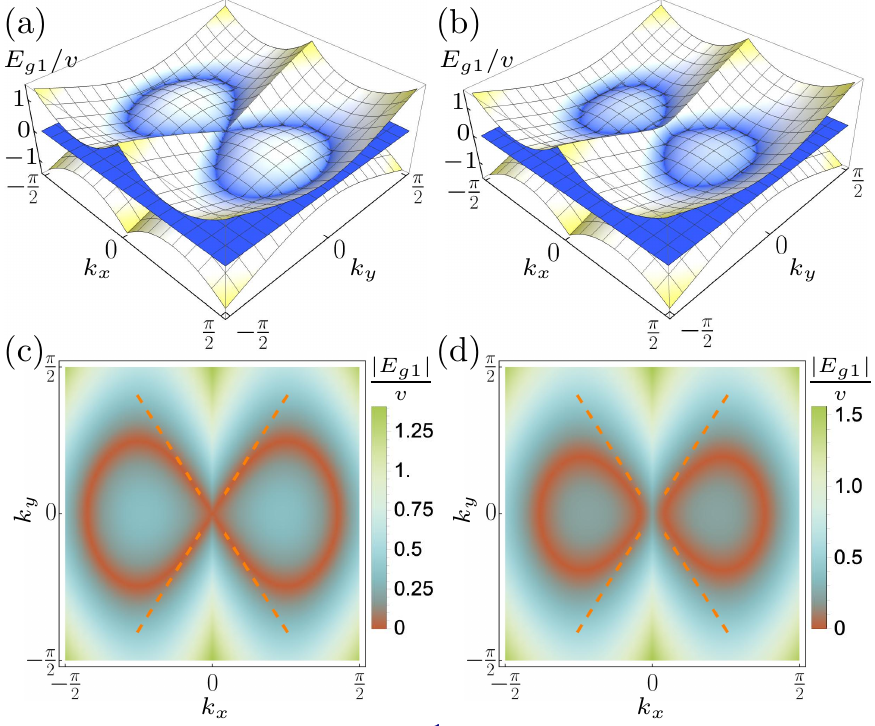}
\caption{The energy bands for $g_1 = \text{diag}(g_a,0,-g_a)$ with $g_a=1.5$ 
(in the dark blue region with $g_b=0$ in Fig.~\ref{fig:gnodes} and second row of Table~\ref{tab:gnodes}) 
with the principal axes $(a,b,c) = (x,y,z)$. 
The energy of the two central bands are shown in (a,b) in units of $v$, 
and the planes show the surface of zero energy. 
In (a,c), 
$m=0$ and the $\infty$-shaped nodal lines 
go through the Dirac point of $E_{g1}(k_x,k_y,\pm k_x)$. 
In (b,d),
$m=0.2v$ and the nodal lines are gapped out 
at the Dirac point
and turn into rings. 
The dashed lines in density plots (c,d) 
show the nodal directions $k_y = \pm\sqrt{2g_a^2-1}k_x$ 
through the Dirac point. 
We have assumed $B=v$ everywhere.}
\label{fig:nodespec}
\end{figure}

\begin{figure}[t!]
\includegraphics[width=3.4in]{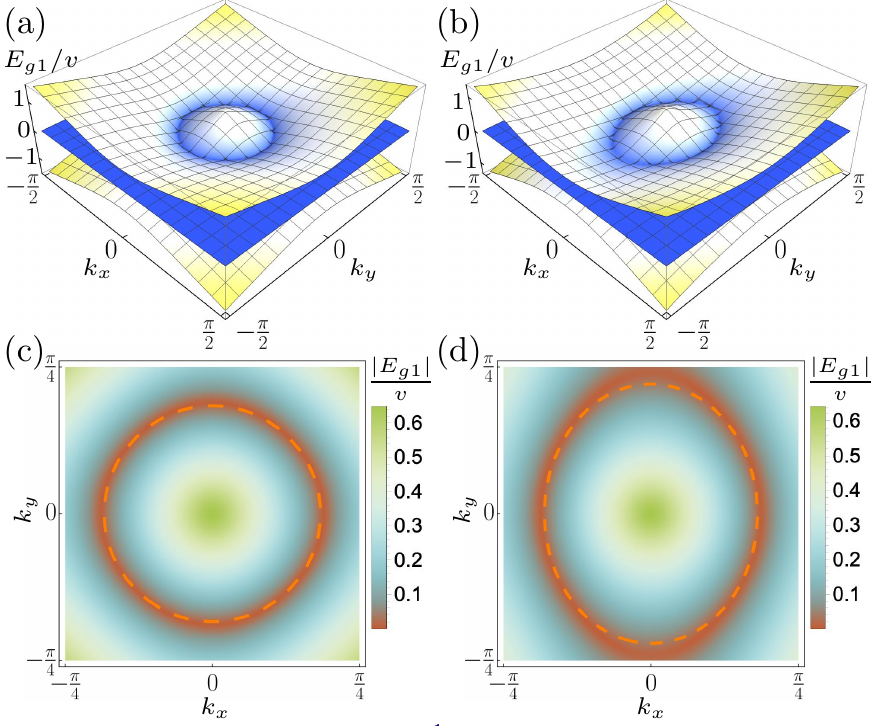}
\caption{The energy bands for $g_1 = g_b\, \text{diag}(-2-\epsilon,1,1+\epsilon)$ 
with $g_b=0.75$ and (a,c) $\epsilon=0$, (b,d) $\epsilon=0.04$, 
respectively, 
on and near the line $g_b=g_c$ in Fig.~\ref{fig:gnodes} 
with the principal axes $(a,b,c) = (x,y,z)$.
The energy of the two central bands in (a,b) 
are in units of $v$ 
and show nodal rings of $E_{g1}(k_x,k_y,k_z)$ 
(a) on the planes $k_x = \pm k_{x0}=\pm(2v/B)\sqrt{(2g_b^2-1)/3} = 0.41$ 
and (b) $k_x = \pm \sqrt{k_{x0}^2+A(k_z)\epsilon}$, 
see Eqs.~(\ref{eq:ka0c}-\ref{eq:Akc}).
The nodal rings shown on the density plots 
match well with the dashed (c) circle of radius $\sqrt2 |k_{x0}| = 0.58$ 
and (d) ellipse defined in Eq.~(\ref{eq:ellipsering}). 
We have assumed $m=0$ and $B=v$ 
for the lattice parameters.}
\label{fig:ringspec}
\end{figure}

On the lattice, 
the growing effect of $\mu(\vex k)$ 
away from the Dirac point 
can cause the nodal lines of $E_{g1}$ 
to close into an $\infty$ shape. 
The nodal lines of $E_{g0} = \pm |\mu|$, 
on the other hand, 
continue away from the Dirac point 
to the Brillouin zone edge. 
In Figs.~\ref{fig:nodespec} and~\ref{fig:ringspec}, 
we plot typical dispersions 
exemplifying the topology of nodal lines.

%%%% Fig. 2: Nodal lines and rings %%%%
%%

%%%% Fig. 3: Nodal rings %%%%
%%

\section{Topology of the band structure}\label{sec:topology}

\subsection{Topological invariants}\label{sec:topinv}

The topologically nontrivial nodal structure of the dispersion 
for a system with periodic boundary conditions 
can be characterized 
by bulk topological invariants. 
For example, 
for the $b$ term, 
the Chern number of the Hamiltonian~(\ref{eq:Hb}) 
with respect to the momenta perpendicular to $\vex b$ 
and as a function of the momentum component $q$ parallel to $\vex b$ 
is $C_b(q) = \pm\Theta( |q| < |\vex b|)$, 
where the step function $\Theta(s)$ is $1$ if $s$ is true 
and $0$ otherwise. 
The nonzero values of $C_b$ signify the topological nature of the Weyl semimetal.

For the $g$ term, 
we utilize the chiral symmetry of the Hamiltonian~(\ref{eq:Hg01}) when $g_0=0$ 
under the chiral operator $C=\mathrm{i}\gamma^0\gamma^5$,
\begin{equation}
\{ H_{g_1}, C \} = 0\,,
\end{equation}
to define an integer-valued winding number 
as a topological invariant. 
In the chiral eigenbasis, 
where we have $C= \sigma_z\otimes\openone_2$ and $H_g = \sigma_x \otimes h_x + \sigma_y \otimes h_y$, 
the winding number is defined as
\begin{equation}
W_C[H_{g_1}] := \frac1{2\pi \mathrm{i}} \oint \frac{\partial \ln\det h(q)}{\partial q} \dd q \in \mathbb{Z}\,,
\end{equation}
where $h = h_x - \mathrm{i} h_y$ and $q$ is a cyclic lattice momentum variable. 
The winding number is a function of $\vex p$, 
the momentum perpendicular to the cyclic momentum direction 
parametrized by $q$.  
For example, 
consider a two-dimensional system 
with momenta $(\vex p,q)$, 
where $\mathbf{p}$ is normal to $q$. 
If the system contains a pair of Dirac points 
with opposite chiralities at $(\pm \vex p_0, 0)$, 
the winding number reads $W_C(\vex p) = \pm \Theta( |\vex p| < |\vex p_0| )$. 
The sign here is determined 
by the orientation of the Dirac points 
with respect to the direction of integration over $q$.

We shall now demonstrate 
the topological characterization of the nodal lines and rings 
using this winding number under chiral symmetry. 
In Figs.~\ref{fig:wind}(a) and (b), 
we sketch the two cases corresponding to Figs.~\ref{fig:nodespec}(a) and~\ref{fig:ringspec}(a). 
Taking $\vex p$ to be the lattice momentum 
parallel to the green-shaded plane 
and the cyclic direction $q$ normal to it, 
we can see 
that as the (orange) plane containing $q$ and $\vex p$ 
scans the green plane 
(sampling different $\vex p$), 
two Dirac points emerge at the intersection 
with the nodal lines 
and move within the plane. 
Therefore, 
we expect the winding number $W_C(\vex p) = \pm1$ when $\vex p$ 
is sampling the area 
enclosed by the nodal lines, 
as sketched by the black and white shaded areas on the green plane. 
As the planes are rotated, 
the existence of other nodal lines and rings 
can lead to a partial cancellation of the winding number, 
since they contribute opposite signs 
to overlapping areas of $\vex p$.

We show the results of a numerical calculation of the winding numbers 
in Figs.~\ref{fig:wind}(c) and~\ref{fig:wind}(d) 
corresponding to the cases 
shown in panels (a) and (b), 
respectively, 
of the same figure. 
The parametrization of the plane of $\vex p$ in Fig.~\ref{fig:wind}(c) 
is the same as that used in Fig.~\ref{fig:nodespec}(a). 
In Fig.~\ref{fig:wind}(d), 
the plane of $\vex p$ is tilted by 45$^\circ$ 
in the $k_x$-$k_y$ plane 
compared to Fig.~\ref{fig:wind}(a) 
so as to resolve the two nodal rings in the spectrum. 
As expected, 
the partial overlap of the two rings at this angle 
leads to an area with winding number $+1-1 = 0$.

\subsection{Surface states}\label{sec:surfstates}

For a topological system with open boundaries, 
surface bound states may arise 
depending on the surface orientation 
and the nature of the bulk topology.
For example, 
for the Weyl semimetal with $\vex b\neq 0$ 
(see Sec.~\ref{sec:latt-b}), 
the generic dispersion for a boundary 
that is not orthogonal to $\vex b$ 
is an energy surface 
terminating at a contour 
that contains the projections of Weyl nodes on the boundary 
and whose curvature depends on the direction of the boundary. 
Thus, 
as is well-known, 
generic constant-energy contours 
for bound states on such open boundaries 
are open ``Fermi arcs'' \cite{Yan:2016euz} 
terminating on the contour 
containing the Weyl-node projections.

\begin{figure}[t!]
\includegraphics[width=3.3in]{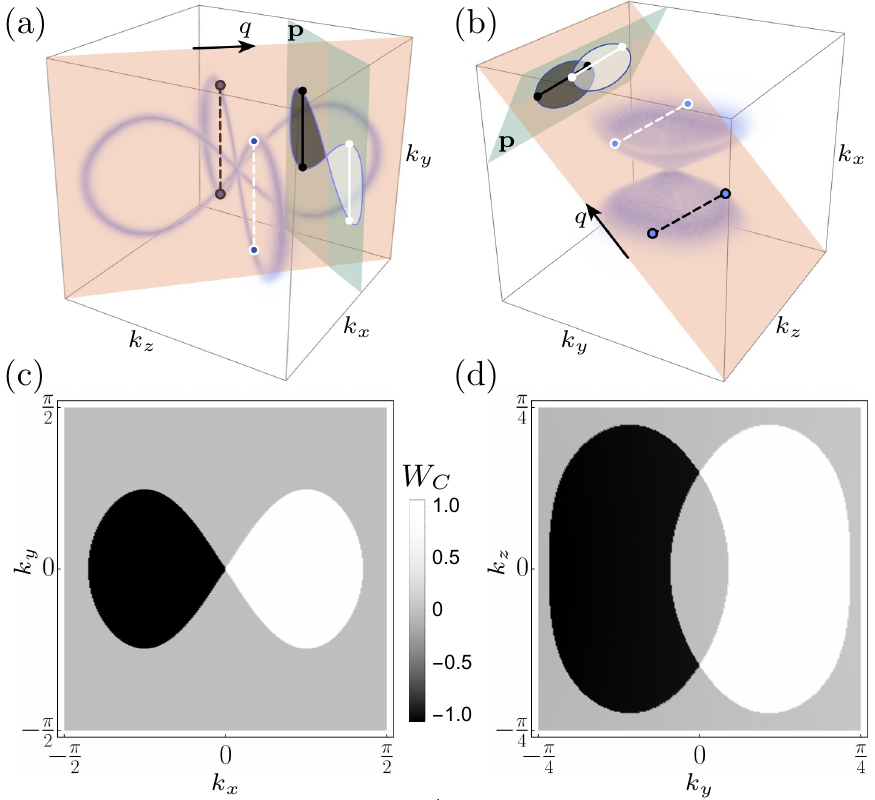}
\caption{The winding number $W_C(\vex p)$ for $g_1 = \text{diag}(g_a,0,-g_a)$ 
with $g_a = 1.5$ (a,c) and $g_1 = g_b\,\text{diag}(-2 - \epsilon, 1, 1+\epsilon)$ 
with $g_b = 0.75$, $\epsilon=0.04$ (c,d). 
The plots in (a) and (b) show the calculated nodal lines and rings, 
respectively, 
with the cyclic direction $q$ 
used to define the winding number 
contained in the (orange) plane 
normal to the (green) plane of $\vex p$. 
As the plane containing $q$ scans values of $\vex p$, 
the projection of the nodal lines 
forms regions with $W_C(\vex p)=\pm1$. 
In (d), 
we take $\epsilon=0$. 
The values of the other parameters are $m=0$ and $B=v$.}
\label{fig:wind}
\end{figure}

The topological winding number $W_C$ 
calculated in the previous section 
corresponds to zero-energy bound states 
on surfaces terminating the bulk 
normal to the direction of the cyclic momentum $q$. 
These bound states are eigenstates of the chiral operator 
with an eigenvalue equal to $\sgn(W_C)$ and, 
thus, 
their energy is pinned to zero 
by the chiral symmetry. 
Therefore, 
with open boundary conditions on surface terminations 
parallel to the green planes in Figs.~\ref{fig:wind}(a) and (b), 
we expect to obtain zero-energy surface bound states 
for all momenta $\vex p$ along the surface 
for which $W_C(\vex p) \neq 0$. 
Such surface bound states 
form a flat band over a finite area of $\vex p$ and, 
thus, 
have been called ``drumhead'' surface states.

The existence and the properties of such surface bound states 
can be studied in a number of ways 
that incorporate the physics near the boundary. 
In the continuum formulation, 
one needs to impose the boundary conditions judiciously 
so that the resulting Hamiltonian 
is self-adjoint~\cite{Ahari_2016,Seradjeh_2018}. 
Here, 
instead, 
we study surface spectra 
by terminating the lattice Hamiltonians appropriately 
to form open boundaries.

%%%% Fig. 4: g term winding number %%%%
%%

To simplify the choice of the boundary along lattice directions 
and still be able to resolve surface states 
with opposite chiral eigenvalues, 
we choose $g_0 = 0$ and
\begin{equation}\label{eq:g1Marco}
g_1 = g_a\begin{pmatrix} 1 & 0 & 0 \\ 0 & 0 & -1 \\ 0 & -1 & -1 \end{pmatrix}\,,
\end{equation}
where the columns and rows 
are along the lattice directions $j$. 
Then, 
$g_b = - \phi g_a$, 
$g_c = \phi^{-1} g_a$, 
with the golden ratio $\phi = \frac{\sqrt 5 +1}2$, 
and the eigenvectors of $g_1$ are
\begin{equation}
a: \begin{pmatrix} 1 \\ 0 \\ 0 \end{pmatrix}\,, \
b: \begin{pmatrix} 0 \\ \phi^{-1} \\ 1 \end{pmatrix}\,, \
c: \begin{pmatrix} 0 \\ -\phi \\ 1 \end{pmatrix}\,. \
\end{equation}
Thus, 
implementing the conditions in Fig.~\ref{fig:gnodes}, 
an $\infty$-shaped nodal line is expected 
for $\sqrt{\phi^{-1}} \approx 0.786 < |g_a| < 1$. 
For $|g_a|=1$, 
the two $\infty$-shaped nodal lines become tangent 
and for $|g_a|>1$ open into nodal rings. 
The directions of nodal lines 
passing through the Dirac point (for $m=0$) 
and the orientation of the nodal rings are now tilted 
relative to the lattice directions.

\begin{figure}[t!]
\includegraphics[width=3.4in]{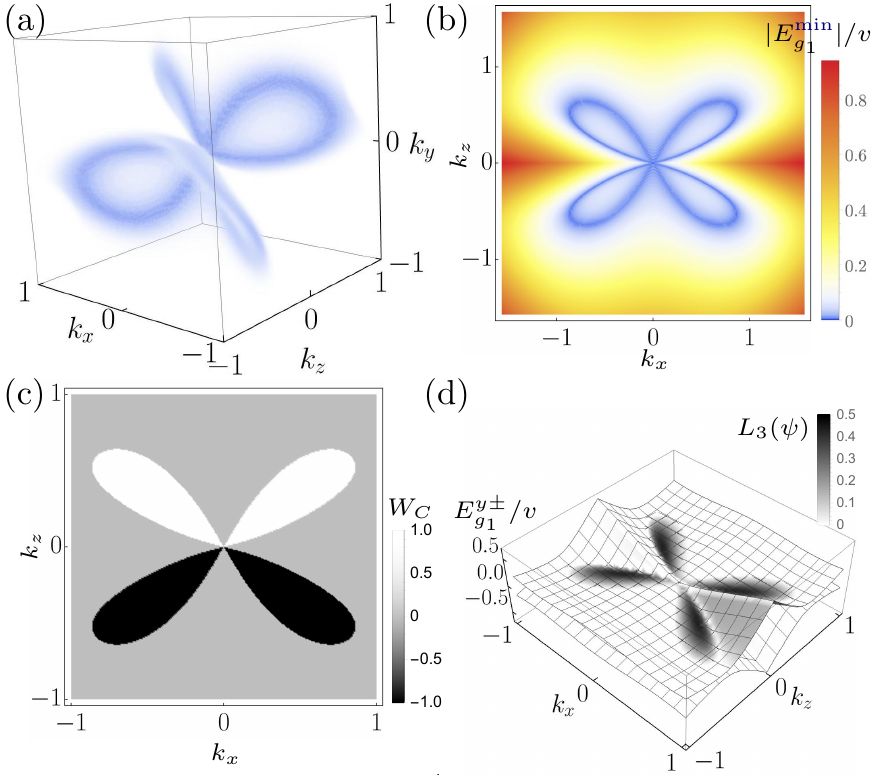}
\caption{Bulk-boundary correspondence for $H_{g_1}$ with $g_1$ given by Eq.~(\ref{eq:g1Marco}) and $g_a = 0.9$: (a) The contour of nodal lines; (b) The projection of nodal lines on the $k_x$-$k_z$ plane showing $E^\text{min}_{g_1}(k_x,k_z) \equiv \min_{k_y} E_{g_1}(k_x,k_y,k_z)$; (c) the winding number $W_C(k_x,k_z)$ with cyclic integration along $k_y$; (d) the two lowest (closest to zero) energy bands $E_{g_1}^{y\pm}$ in an open geometry along the $y$ direction with a lattice size $N_y=150$ and color showing the relative wavefunction weight $L_\delta(\psi)$ for $\delta = 3$ sites near the boundary, see Eq.~(\ref{eq:EL}).}
\label{fig:ss1}
\end{figure}

\begin{figure}[t!]
\includegraphics[width=3.4in]{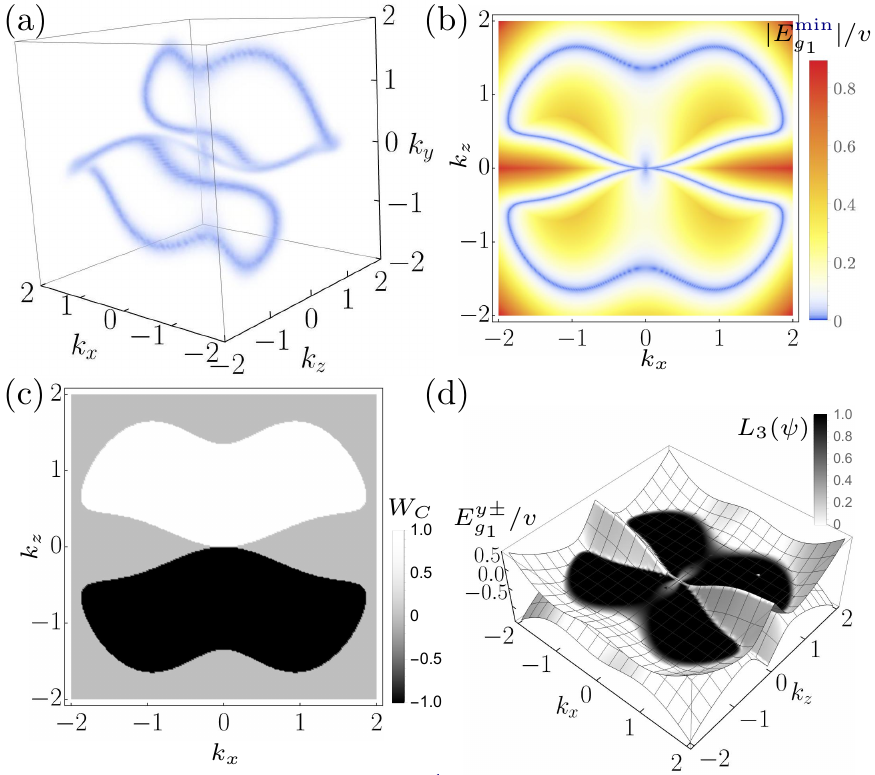}
\caption{Bulk-boundary correspondence for $H_{g_1}$ with $g_1$ 
given by Eq.~(\ref{eq:g1Marco}) and $g_a = 1.3$. 
Panels show the same data as in Fig.~\ref{fig:ss1}.}
\label{fig:ss2}
\end{figure}

In Figs.~\ref{fig:ss1} and~\ref{fig:ss2}, 
we present numerical results for $g_a=0.9$ and $g_a=1.3$, 
respectively, 
corresponding to the cases with $\infty$-shaped nodal lines and rings. 
In panel (a) of each figure, 
the contour of near-zero energy states in the bulk Brillouin zone $(k_x,k_y,k_z)$ 
are shown as a density plot of $|E_{g_1}|$ with $E_{g_1}$ of Eq.~\eqref{eq:Eg1}. 
In panel (b) of each figure, 
a two-dimensional projection of the minimum energy on the $k_x$-$k_z$ plane, 
$E^\text{min}_{g_1} \equiv \min_{k_y} E_{g_1}(k_x,k_y,k_z)$, 
is presented. 
This illustrates the expected path 
that pairs of Dirac points traverse 
as the plane containing the cyclic momentum $q\equiv k_y$ 
used for calculating the winding number 
scans values of $\vex p \equiv (k_x,k_z)$. 
This winding number is plotted in panel (c) of each figure.

%%%% Fig. 5: g term surface states, lines %%%%

Finally, 
in panel (d) of each figure 
we plot the lowest two energies 
(i.e., those closest to zero) 
in a geometry with open boundaries 
along the $y$ direction, 
forming bands $E_{g_1}^{y\pm}(k_x, k_z)$ as a function of momenta 
along the periodic directions. 
These bands contain both bulk and surface bound states. 
At momenta $\vex p = (k_x,k_z)$ 
for which there are states in the bulk gap, 
we would expect the lowest energies to be surface bound states. 
Such $\vex p$ should also correspond 
to nonzero values of $W_C(\vex p)$. 
In order to distinguish bulk and surface bound states, 
we calculate a measure of edge localization of a wavefunction $\psi$,
\begin{equation}\label{eq:EL}
L_\delta(\psi) = \sum_{0\leq |y - y_b| \leq \delta} |\psi(y)|^2 / \left\Vert \psi\right\Vert^2\,,
\end{equation}
where $y_b$ are the positions of the boundaries 
along the $y$ direction, 
and $\delta$ is the number of sites in the vicinity of the boundary. 
For a normalized state, 
$\left\Vert \psi \right\Vert^2=1$.

%%%% Fig. 6: g term surface states, rings %%%%

The results presented in Figs.~\ref{fig:ss1} and~\ref{fig:ss2} 
clearly show the expected bulk-boundary correspondence 
between the bulk topological invariant $W_C$ 
and the existence of surface bound states. 
For $g=0.9$ (Fig.~\ref{fig:ss1}), 
the states in the bulk gap are found close to the surface 
with half of their weight within a layer of thickness 
that is $2\%$ of the length of the system in the open direction. 
For $g=1.3$ (Fig.~\ref{fig:ss2}), 
the bound states have nearly all their weight in the same layer.

\section{Transport}\label{sec:transport}

Topological semimetals are known 
to exhibit novel transport phenomena. 
In particular, 
it has been shown 
that the Hall conductivity in Weyl semimetals, 
induced by an applied electromagnetic field, 
can be derived from the chiral anomaly 
related to the action in Eq.~(\ref{eq:SMEaction}) 
with nonzero $b$ coefficients~\cite{Zyuzin:2012vn,Zyuzin:2012tv,Goswami:2012db,PhysRevX.6.041046}. 
In this view, 
one may rotate away the $b$ term from the action at tree-level 
by an appropriate chiral rotation of the Dirac fields. 
As the partition function is not invariant under the same rotation, 
quantum corrections induce a nonconservation of chiral charge in the system. 
These observations provide a strong link 
between the microscopic theory of Weyl semimetals 
and the continuum effective action. 
In this section, 
we generalize the effective-action approach to transport phenomena 
and exhaust all possible induced fermion currents generated by quantum corrections 
at leading order in the coefficients in Eq.~(\ref{eq:SMEaction}). 
In this section, 
we follow the conventions outlined in Refs.~\cite{Peskin:1995ev,Kostelecky:2001jc}, 
i.e., 
natural units $\hbar=c=\varepsilon_0=1$ are used.

\subsection{Fermion current}

From the perspective of the continuum action, 
a systematic way to approach the calculation of induced currents 
starts with the effective action in $3+1$ Minkowski spacetime
\begin{equation}
	Z(A)=\int \mathcal{D}\bar{\psi}\mathcal{D}\psi\,\mathrm{e}^{\mathrm{i}S}\,,
\end{equation}
where $S$ is the action of the SME fermion sector 
stated in Eq.~(\ref{eq:SMEaction}), 
which is minimally coupled to an electromagnetic field. 
Furthermore, 
$\mathcal{D}$ indicates a path integral 
over appropriate Dirac spinor field configurations. 
The latter $Z(A)$ encodes the response of the system 
to an electromagnetic background field 
described by the four-potential $A_{\mu}$. 
By definition, 
the induced fermion current is then given by
\begin{align}
	\braket{j^\mu}&=\frac{\int\mathcal{D}\bar{\psi}\mathcal{D}\psi\left[\bar{\psi}\Gamma^\mu\psi\right] \mathrm{e}^{\mathrm{i}S}}{\int \mathcal{D}\bar{\psi}\mathcal{D}\psi\,\mathrm{e}^{\mathrm{i}S}} \notag \\
	&=\frac{1}{Z(A)}\left(-\mathrm{i}\frac{\delta}{\delta A_\mu(x)}\right)Z(A)\,.
	\label{eq:current}
\end{align}
In the latter formula, 
$\Gamma^{\mu}$ is defined as in Eq.~(\ref{eq:SMEaction}), 
$\frac{\delta}{\delta A_{\mu}}$ denotes the functional derivative for $A_{\mu}$, 
and we have set the charge $q=1$. 
The low-energy fluctuations around the vacuum 
are therefore obtained by integrating out the fermion fields:
\begin{align}
	Z(A) &= \int\mathcal{D}\bar{\psi}\mathcal{D}\psi\,\exp\left(\mathrm{i}\int\mathrm{d}^4x\,\bar{\psi}\Delta_A\psi\right) \notag \\
	&=\det(\Delta_A)\,,
\end{align}
with the modified Dirac operator $\Delta_A$ 
based on Eq.~(\ref{eq:SMEaction}) 
and minimally coupled to $A_{\mu}$.
In the limit of a weak electromagnetic coupling, 
this form of the effective action 
can be expanded as a series in powers of $A_{\mu}$ or, 
equivalently, 
in powers of the electromagnetic coupling:
\begin{subequations}
	\begin{equation}
		Z(A)=\det(\Delta)e^{\mathcal{S}(A)}\,,
	\end{equation}
	where $\Delta=\mathrm{i}\slashed{\partial}-m$ is the standard Dirac operator and
	\begin{equation}
		\mathcal{S}(A)=-\sum_{n=1}^{\infty}\frac{(-\mathrm{i})^{n}}{n}\int \mathrm{d}x_{1}\cdots \mathrm{d}x_{n}\,f(x_1,x_2\dots x_n)\,,
	\end{equation}
	with
	\begin{align}
		f(x_1,x_2\dots x_n)&=\mathrm{tr}[\slashed{A}(x_1)G_{F}(x_{2}-x_{1})\cdots \notag \\
		&\phantom{{}={}}\quad\times\slashed{A}(x_n)G_{F}(x_{1}-x_{n})]\,.
	\end{align}%%
\end{subequations}
Here, 
$\slashed{A}:=\gamma^{\mu}A_{\mu}$ and $G_{F}$ is the fermion propagator 
formally containing corrections 
from coefficients for Lorentz violation at all orders. 
The trace is computed 
with respect to the matrix structure in spinor space. 
Using this form of the effective action, 
the leading and sub-leading terms in the electric charge 
contributing to the induced current 
are given by
\begin{equation}
	\braket{j^\mu} = \Pi^{\mu\nu}A_{\nu} + \mathrm{V}^{\mu\lambda\kappa}A_{\lambda}A_{\kappa}\,,	
\end{equation}
where $\mathrm{i}\Pi^{\mu\nu}$ is the vacuum polarization at order $q^{2}$ 
and $\mathrm{i}\mathrm{V}^{\mu\lambda\kappa}$ is the three-photon vertex correction 
defined as the sum of all one-particle-irreducible diagrams 
contributing to the two- and three-point correlation functions 
in the effective theory, 
respectively. 
Note that 
we have ignored the first-order term in $Z(A)$, 
which is linear in $A_{\mu}$. 
In QED, 
this term vanishes due to Furry's theorem. 
In general, 
it does not vanish for each of the controlling coefficients in Eq.~(\ref{eq:SMEaction}). 
However, 
at leading order
it does so for all coefficients 
that can give a nonzero contribution to the induced current. 
For further details regarding the calculation of induced currents, 
see also Refs.~\cite{Grushin:2019uuu,McGinnis:2020tyj}.

Thus, 
the linear and quadratic response for any given coefficient in Eq.~(\ref{eq:SMEaction}) 
can be evaluated simply 
by computing the modified vacuum polarization or vertex function 
either by perturbative calculation or otherwise. 
In this work, 
we will focus only on the linear response 
and truncate at leading order in both 
the electromagnetic coupling and coefficients for Lorentz violation. 
In the perturbative approach, 
the response is given 
by the one-loop vacuum polarization 
with first-order corrections 
from coefficients for Lorentz violation. 
Incorporating the C, P, and T properties 
of each coefficient in Eq.~(\ref{eq:SMEaction}), 
the generalization of Furry's theorem reveals 
that the only coefficients 
which give a nonzero contribution to the induced linear response at this order 
are the $b$, $g$, and $c$ terms~\cite{Kostelecky:2001jc}. 
We note 
that by the same logic 
only the $d$ and $H$ terms are potentially relevant for nonlinear response. 
However, 
it was previously found by explicit calculation 
that these contributions to the modified vertex function vanish~\cite{Kostelecky:2002ue}. 
Since all other coefficients in Eq.~(\ref{eq:SMEaction}) 
may be neglected in Lorentz-violating QED due to field redefinitions, 
our results presented in this section 
encompass all possible effects 
in the low-energy response of the system 
at leading order in Lorentz violation. 
While the result for the $b$ term 
has already been widely explored in the context of Weyl semimetals, 
we establish the method of the effective action outlined here 
by reproducing known results 
which have been calculated by independent methods.

\begin{figure}[t]
	\centering
	\includegraphics[scale=1]{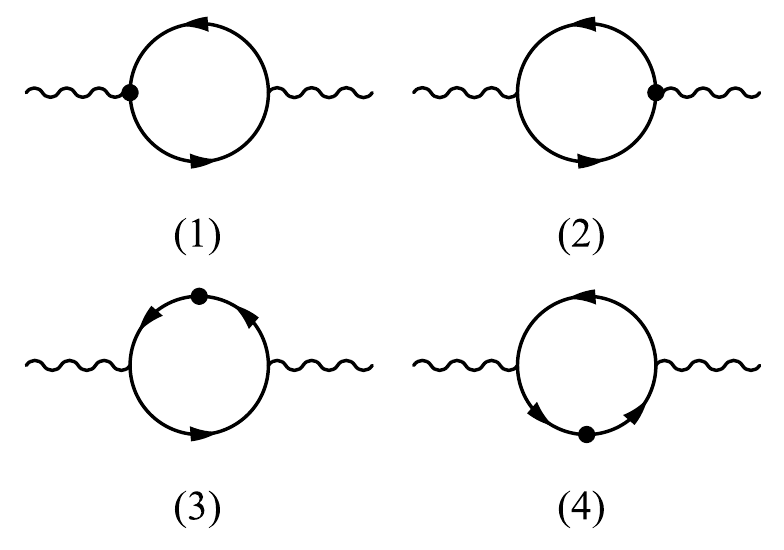}\\
	\caption{Diagrams contributing to the one-loop modified vacuum polarization at leading order in $g$. 
	Effective vertices are indicated by `$\bullet$' and denote a Lorentz-violating vertex insertion. 
	See Ref.~\cite{Kostelecky:2001jc} for details.}
	\label{fig:diags}
\end{figure}

\subsection{Chiral anomaly of the $\boldsymbol b$ term}

Recall from the previous section 
that the linear induced current 
is related to the vacuum polarization of the underlying theory. 
In the case of $b^{\mu}\neq 0$, 
the vacuum polarization has been previously calculated~\cite{Perez-Victoria:1999erb}. 
The Hall conductivity for spacelike $b^{\mu}$ with $\mathbf{b}^2<m^2$ 
is known to vanish \cite{Goswami:2012db}. 
Thus, 
we consider the regime $\mathbf{b}^2>m^2$ 
characterized by the following nonzero contribution:
\begin{equation}
	\label{eq:vacuum-polarization-b}
	\mathrm{i}\Pi^{\mu\nu}_{b}=\frac{1}{2\pi^{2}}\epsilon^{\mu\nu\alpha\beta}b_{\beta}p_{\alpha}\sqrt{1-\frac{m^{2}}{|b^{2}|}}\,.
\end{equation}
It is worth commenting 
that this result is ambiguous 
with respect to the chosen regularization scheme 
and could be shifted 
by an undetermined constant~\cite{Colladay_1998,Chung:1998jv,%
Jackiw:1999yp,Chung:1999gg,Perez-Victoria:1999erb,Chung:1999pt}. 
However, 
the additional microscopic details in the condensed-matter setting, 
in particular the requirement 
that the Hall current vanish for spacelike $b_{\mu}$, 
enforce this constant to be zero. 
For an extensive discussion, 
see~Ref.~\cite{Grushin:2012mt}.

Equation~(\ref{eq:vacuum-polarization-b}) immediately gives the induced current
\begin{align}
	\braket{j^\mu}&=\Pi_b^{\mu\nu}A_{\nu}(p) \notag \\
	&=\frac{1}{2\pi^2}\sqrt{1-\frac{m^2}{|b^2|}}\epsilon^{\mu\nu\alpha\beta}b_{\beta}(-\mathrm{i}p_{\alpha})A_\nu(p) \notag \\
	&=\frac{1}{4\pi^2}\sqrt{1-\frac{m^2}{|b^2|}}\epsilon^{\mu\nu\alpha\beta}b_{\beta}F_{\alpha\nu}\,,
	\label{eq:b_current}
\end{align}
where $F_{\mu\nu}=\partial_{\mu}A_{\nu}-\partial_{\nu}A_{\mu}$ is the electromagnetic field strength tensor in configuration space. 
For $m^2=0$, 
this agrees with the result obtained 
via calculation of the chiral anomaly 
using the Fujikawa method~\cite{Zyuzin:2012tv} 
where the latter is given in position space.
From Eq. (\ref{eq:b_current}), 
we can easily identify the conductivity of the medium 
generated by $b^{\mu}$ 
from the spatial components of the current 
via $\braket{j^{i}}=\sigma^{ij}E_{j}$. 
Using $\mathbf{E}=-\nabla\phi - \frac{\partial\mathbf{A}}{\partial t}$ 
and taking the $x$ component of the current, 
we obtain the well-known Hall conductivity
\begin{equation}
	\sigma^{xy}=\frac{1}{2\pi^2}{\sqrt{|b^2|-m^2}}\epsilon^{xyl}\hat{b}_l\,,
\end{equation}
where $\hat{b}_l$ is the $l$-th component of the unit vector $\hat{\mathbf{b}}\equiv \mathbf{b}/|\mathbf{b}|$ 
pointing along the spatial part $\mathbf{b}$.

\subsection{Novel effects of the $\boldsymbol g$ term}

We proceed now 
to the calculation of the induced fermion current 
associated with the $g$ term. 
Before presenting the main results, 
it is worth pointing out a few subtleties. 
Recall from Eq.~\eqref{gDecomp} in Sec.~\ref{sec:SME} 
that the $g$ term can be decomposed 
into three irreducible representations of the Lorentz group,
with the fully antisymmetric part $g^{(A)}_{\mu}$, 
the trace piece $g_{\mu}^{(T)}$, 
and the mixed-symmetry part $g^{(M)}_{\kappa\lambda\mu}$. 
The induced current generated by $g^{(A)}_{\mu}$ 
is redundant with that of the $b$ term. 
This can be understood from the fact 
that $mg^{(A)}_{\mu}$ can be rotated to a $b$-like contribution 
due to the field redefinitions mentioned in Sec.~\ref{sec:SME}. 
As was also described, 
the trace component $g^{(T)}_{\mu}$ is unphysical 
and can also be removed from Eq.~(\ref{eq:SMEaction}) 
without loss of generality. 
Thus, 
any new and physically relevant result 
must come from $g^{(M)}_{\kappa\lambda\mu}$, 
which was our initial motivation for setting $g^{(A)}_{\mu}=g^{(T)}_{\mu}=0$ 
starting from Eq.~\eqref{eq:definitions-g0-g1}.

The one-loop vacuum polarization is straightforward to compute 
in the perturbative regime via Feynman diagrams. 
In Fig.~\ref{fig:diags}, 
we show the diagrams needed for the modified vacuum polarization 
where vertices corresponding to an insertion of $g^{(M)}_{\kappa\lambda\mu}$ 
are denoted by the symbol `$\bullet$'. 
Note that while the first class of diagrams, 
$(1)$ and $(2)$, 
involve modified QED vertices, 
diagrams $(3)$ and $(4)$ involve a chirality flip 
on the internal fermion line. 
Thus, 
it is expected 
that the induced current be proportional to $m$ 
in contrast to the corresponding results generated by $b$. 
For a complete list of the associated Feynman rules, 
see Ref.~\cite{Kostelecky:2001jc}.

Proceeding with standard techniques to calculate the loop, 
we find
\begin{widetext}
	\begin{subequations}
		\begin{equation}
			\braket{j_{\mu}}=\Pi_{\mu\nu}A^{\nu}= -\frac{\mathrm{i}m}{2\pi^2}\mathcal{F}(p^2,m^2)\left[g^{(M)}_{\mu\nu\alpha}p^2 + p^{\lambda}\left(p_{\nu}g^{(M)}_{\lambda\mu\alpha} - p_{\mu}g^{(M)}_{\lambda\nu\alpha}\right)\right]p^{\alpha}A^\nu\,,
			\label{eq:current_gterm}
		\end{equation}
		where we introduced a dimensionless parameter $x:=\sqrt{p^{2}}/2m$ as well as a dimensionless function
		\begin{equation}
			\label{eq:auxiliary-function}
			p^2\mathcal{F}(p^2,m^2):=\hat{\mathcal{F}}(x)\,,
		\end{equation}
		such that
		\begin{equation}
			\hat{\mathcal{F}}(x)=1-\frac{\ln\left[1+2x(\sqrt{x^2-1}-x)\right]}{2x\sqrt{x^2-1}}
			= \begin{cases}
				1-\frac{\ln[1+2|x|(|x|+\sqrt{|x^{2}|+1})]}{2|x|\sqrt{|x^{2}|+1}}\,, & x^2 < 0 \\
				~\\
				1-\frac{\sin^{-1} (x)}{x\sqrt{1-x^2}}\,, & 0<x^2 < 1 \\
				~\\
				1-\frac{\mathrm{i}\pi + \ln |1+2x(\sqrt{x^{2}-1}-x)|}{2x\sqrt{x^{2}-1}}\,, & x^2 > 1\,.
			\end{cases}
		\end{equation}%%
	\end{subequations}
	Note that $\hat{\mathcal{F}}(|x|\to 0)=2|x|^2/3+\mathcal{O}(|x|^4)$ and $\hat{\mathcal{F}}(|x|\to \infty)=1 + \mathcal{O}[\ln|x|/|x|^2]$ for $x^2<0$ as well as $\hat{\mathcal{F}}(x \to 0) = -2x^2/3 + \mathcal{O}(x^4)$ and $\hat{\mathcal{F}}(x \to \infty) = 1 + \mathcal{O}[\ln(x)/x^2]$ for $x^2>0$.
\end{widetext}
It is worthwhile to note 
that this result is finite 
and thus there are no ambiguities 
related to divergences or the regularization scheme. 
This can be understood from the momentum structure 
appearing in the vacuum polarization. 
Note that each factor in Eq.~(\ref{eq:current_gterm}) 
appears with three factors of momenta. 
Thus, 
any divergence related to the vacuum polarization generated 
would need to be absorbed by a corresponding counterterm proportional to the $g$ term, 
three factors of $\partial_{\mu}$, and two factors of $A_{\mu}$. 
However, 
it is clear 
that there is no such counterterm 
that would be gauge-invariant and renormalizable. 
Thus, 
in the minimal SME, 
defined by Eq.~(\ref{eq:SMEaction}), 
which is gauge-invariant and renormalizable by construction, 
no such divergence can appear. 
Additionally, 
it should be understood 
that this result is valid for any irreducible component of $g_{\mu\nu\alpha}$, 
and we have simply assumed 
that $g^{(A)}_{\mu}=g^{(T)}_{\mu}=0$ in the final result 
so that $g_{\mu\nu\alpha}=g^{(M)}_{\mu\nu\alpha}$.

We note 
that while $\braket{j^{\mu}}$ is a gauge-invariant quantity, 
the same is not obvious of the right-hand side of Eq.~(\ref{eq:current_gterm}). 
Using the definition of the electromagnetic field-strength tensor $F^{\mu\nu}$ in momentum space, 
we can recast Eq.~(\ref{eq:current_gterm}) in a way 
where gauge invariance is manifest:
\begin{equation}
	\braket{j_{\mu}}= \frac{m}{2\pi^2p^2}\hat{\mathcal{F}}(x)\left(g^{(M)}_{\mu\nu\alpha}p_{\lambda} - \frac{1}{2}{g^{(M)}_{\lambda\nu\alpha}}p_{\mu}\right)p^{\alpha}F^{\lambda\nu}\,.
	\label{eq:current_gterm_v2}
\end{equation}%%	
Following the discussion of Sec.~\ref{sec:latt-b}, 
it is of interest to make the connection of this result 
to the components of $g^{\kappa\lambda\nu}$ 
which are relevant in the Hamiltonian formulation of the model, 
i.e., 
when $g^{(M)}_{\kappa\lambda0}=0$.

Parametrizing in terms of the matrices $g_{0}$ and $g_{1}$ of Eqs.~(\ref{eq:g0}) and~(\ref{eq:g1}) 
and denoting the momentum four-vector $p^\mu = (\omega,\vex p)$, 
the charge density $\braket{\rho} \equiv \braket{j^0} $ 
and the spatial current $\braket{\vex j}$ take the form
\begin{subequations}\label{eq:gjEB}
	\begin{align}
		\braket{\rho(\omega,\vex p)} &= \frac{m\omega\hat{\mathcal{F}}(x)}{2\pi^2p^2}\left[ (g_{0} \vex p) \cdot\vex E + (g_{1}\vex p)\cdot \vex B\right]\,, \\
		\braket{\vex j(\omega,\vex p)} &= \frac{m\hat{\mathcal{F}}(x)}{2\pi^2p^2} \vex p \left[ (g_{0} \vex p) \cdot\vex E + (g_{1}\vex p)\cdot \vex B\right]\,.
	\end{align}
\end{subequations}
Here, 
we have assumed 
that the applied electromagnetic background field 
obeys the source-free Maxwell equations.

Thus, 
$g_0$ and $g_1$ characterize, 
respectively, 
the electric conductivity tensor 
and the magnetoelectric response of the system. 
Remarkably, 
for $g_1\neq0$ the application of a magnetic field pulse 
can result in both a charge density 
and an electric current. 
For example, 
for static electromagnetic fields $E(\vex r)$ and $B(\vex r)$, 
we find $\braket{\rho} = 0$ and
\begin{subequations}\label{eq:gjEB0}
	\begin{equation}
		\braket{\vex j} = \frac{m}{2\pi^2} \left[\gvex\nabla (g_0 \gvex\nabla)\cdot {\vex  E}' + \gvex\nabla(g_1 \gvex\nabla)\cdot  {\vex B}' \right]\,,
	\end{equation}
	where the fields are obtained via ${\vex E}'(\vex r) = \int \mathrm{d}^3\vex r'\,\tilde{\mathcal{F}}(\vex r-\vex r') \vex E (\vex r')$ and similarly for ${\vex B}'(\vex r)$. The integration kernel follows from a Fourier transform of Eq.~(\ref{eq:auxiliary-function}). We employ the following asymptotic form of Eq.~\eqref{eq:auxiliary-function} that is valid for $x^2<0$ and has the correct asymptotic behavior for $|x|\to 0$ and $|x|\to\infty$:
	\begin{equation}
		\hat{\mathcal{F}}(x)|_{x^2<0} \sim \frac{|x|^2}{|x|^2+3/2}\,.
	\end{equation}
	Then,
	\begin{equation}
		\tilde{\mathcal{F}}(\vex r) = \int \frac{\mathrm{d}^3\vex p}{(2\pi)^3}\,\mathcal{F}(-|\vex p|^2,m^2)\mathrm{e}^{-\mathrm{i} \vex p\cdot \vex r} \approx \frac{\mathrm{e}^{-\sqrt6 m |\vex r|}}{4\pi |\vex r|}\,.
	\end{equation}
\end{subequations}
Thus, the kernel is approximately a screened Coulomb potential.

\subsection{The $\boldsymbol c$ term}

To complete the list of coefficients for Lorentz violation  
that give a nonzero induced current at leading order in Lorentz violation, 
we consider the $c$ term. 
In this case, 
the antisymmetric and trace components of $c$ 
can be removed via appropriate field redefinitions 
following a similar reasoning 
as that for the $g$ term. 
Thus, 
in our derivation 
we retain only the symmetric, traceless components of $c^{\mu\nu}$. 
The one-loop vacuum polarization at leading order in Lorentz violation 
is then given by
\begin{widetext}
	\begin{subequations}
		\begin{equation}
			\braket{j^{\mu}}=-\frac{1}{6\pi^2}\left\{\hat{\mathcal{K}}_1(x)\left[c^{\mu\nu}p^2-p^{\mu}(p\cdot c)^{\nu}-p^{\nu}(p\cdot c)^{\mu}\right]+{p\cdot c\cdot p}\left(\hat{\mathcal{K}}_2(x)\eta^{\mu\nu}+\hat{\mathcal{K}}_3(x)\frac{p^{\mu}p^{\nu}}{p^2}\right)\right\}A_{\nu}\,,
		\end{equation}
		with the dimensionless functions
		\begin{align}
			\hat{\mathcal{K}}_1(x)
			&= \left[\frac{5}{3}+\frac{1}{\epsilon}+\ln\left(\frac{\bar{\mu}^2}{m^2}\right)\right]+\left(1+\frac1{2x^2}\right)\hat{\mathcal{C}}(x)+\frac1{x^2}\,, \displaybreak[0]\\[2ex]
			\hat{\mathcal{K}}_2(x)
			&= \left[\frac{2}{3}+\frac{1}{\epsilon}+\ln\left(\frac{\bar{\mu}^2}{m^2}\right)\right] + \frac{4x^4-2x^2+1}{4x^2(x^2-1)} \hat{\mathcal{C}}(x) -\frac1{2x^2}\,, \displaybreak[0]\\[2ex]
			\hat{\mathcal{K}}_3(x)
			&= - \frac{3}{4x^2(x^2-1)} \hat{\mathcal{C}}(x) + 1 + \frac{3}{2x^2}\,,
		\end{align}
		where, 
		again, 
		$x=\sqrt{p^2}/2m$ 
		and we have used dimensional regularization 
		to evaluate the divergent pieces of the diagram, 
		thus introducing the unphysical mass scale $\bar{\mu}^{2}=\mu^{2}_r\mathrm{e}^{\gamma_{E}}/4\pi$, 
		with $\gamma_{E}$ the Euler-Mascheroni constant. 
		Furthermore, 
		$\mu_r$ is an arbitrary reference mass scale 
		needed for dimensional consistency. 
		For brevity, 
		we have also introduced the dimensionless function
		\begin{equation}
			\hat{\mathcal{C}}(x)=\frac{\sqrt{x^2-1}}{x}\ln\left[1+2x\left(\sqrt{x^2-1}-x\right)\right]\,.
		\end{equation}
	\end{subequations}
	Note that $\hat{\mathcal{C}}(x\to 0)=-2+2x^2/3+\mathcal{O}(x^4)$ and $\hat{\mathcal{C}}(x\to\infty)=\ln[-1/(4x^{2})] + \mathcal{O}(\ln[-1/(4x^2)]/x^{2})$.
\end{widetext}

In this case, 
there are divergent pieces 
that appear in the loop evaluation. 
The divergent pieces can be removed by appropriate counterterms 
in a gauge-invariant, renormalizable way~\cite{Kostelecky:2001jc}. 
One possible option would be 
to utilize the on-shell renormalization scheme, 
which has the advantages 
that the fermion mass $m$ 
appearing in the Lagrangian 
retains the meaning of the physical mass 
as opposed to the running mass 
and effectively removes the unphysical mass scale $\bar{\mu}^{2}$ 
from the vacuum polarization. 
In effect, 
the Lagrangian parameters will evolve 
with the choice of reference scale~\cite{Kostelecky:2001jc}. 
Irrespective of the choice of renormalization scheme 
the full expression we obtain 
is in agreement with the general considerations presented in Ref.~\cite{Colladay_1998}.

Finally, 
we come to the form of the induced current 
in which gauge invariance is manifest:
\begin{align}\nonumber
	\braket{j^{\mu}}&=-\frac{\mathrm{i}}{8\pi^2}\bigg[\hat{\mathcal{K}}_{1}(x) c^{\mu\alpha}p^{\nu}F_{\nu\alpha}
	+ \hat{\mathcal{K}}_{2}(x) c^{\alpha\beta}p_{\beta}F^{\mu}_{\phantom{\mu}\alpha}\\
	&\phantom{{}={}}\hspace{1.1cm}+\hat{\mathcal{K}}_{3}(x)c^{\alpha\beta}p_{\beta}p^{\mu}p^{\nu}F_{\alpha\nu}\bigg]\,.
\end{align}
Taking into account the components relevant to the Hamiltonian formulation, 
where $c^{\mu0}=0$, 
we find that the charge and current densities can be written as
\begin{subequations}
	\begin{align}
		\braket{\rho}&=-\mathrm{i}\frac{\hat{\mathcal{K}}_{2}(x)}{8\pi^2}c_{ij}p_{j}E_{j}\,, \displaybreak[0] \\
		\braket{j^{k}}&=-\mathrm{i}\frac{\hat{\mathcal{K}}_{2}(x)}{8\pi^{2}} \epsilon^{kil} c_{ij}p_{j} B_{l}\,,
	\end{align}
\end{subequations}
respectively.

\section{Summary and outlook}\label{sec:sum}

This paper proposes a correspondence between 
the field-theoretic description of emergent Lorentz symmetry 
in condensed-matter systems
and the SME framework,
which is the comprehensive effective field theory for Lorentz violation
appropriate for studies of fundamental theories of spacetime and matter.
The correspondence provides a foundation
for classifying and characterizing general quasiparticle excitations
using the SME,
and conversely it implies that features of emergent Lorentz symmetry 
in certain materials can yield insights into Lorentz-violating properties 
of spacetime and matter.

The body of this work focuses on emergent Lorentz invariance 
in three-dimensional Dirac materials 
as viewed from the general perspective offered by the SME. 
The correspondence with the SME enables the classification 
of field-theoretic terms in the action governing departures from emergent Lorentz symmetry,
according to the mass dimension and spinor structure 
of the operator.
This permits the construction of lattice Hamiltonians 
that incorporate all types of Lorentz violations 
around the original Lorentz-symmetric Dirac nodes of the material.

Part of our investigations involve the study of changes 
to the Dirac nodal structure arising from the presence 
of specific types of SME coefficients for Lorentz violation.
We discuss a modification of the Dirac operator,
known as the $b$ term and previously examined in the literature, 
which leads to a Weyl semimetal with two nodes of opposite chirality
separated in momentum and/or energy. 
We also consider another modification of the Dirac operator 
called the $g$ term,
previously unexplored in the condensed-matter context,
that describes Dirac nodal semimetals 
with intersecting Dirac lines and/or Dirac nodal rings.
The bulk topological invariants and the existence and properties
of surface bound states are explored.
The band structures associated with the $b$ and $g$ terms
are strikingly different,
a noteworthy feature given that the  C-, P-, and T-symmetry properties 
of the Hamiltonian components involving $(g_{0})_{ij}$ and $(g_{1})_{ij}$ 
of Eq.~\eqref{eq:definitions-g0-g1} 
are identical to those involving $b_{j}$ and $b_{0}$, 
respectively,
as can be verified from Table 1 of Ref.~\cite{Kostelecky:2001jc}. 
The general effective action describing 
P- or T-symmetry violation in semimetals
must therefore incorporate both terms,
which generate distinct band structures.

Another part of our investigation concerns the transport coefficients 
for various types of semimetals.
At leading order in perturbation theory,
we calculate the transport coefficients
for the $b$ and $g$ terms
and for another modification of the Dirac operator
called the $c$ term.
Interestingly, 
even at perturbatively small values, 
the $g$ term modifies the Maxwell equations 
in the material in unconventional ways. 
In particular, 
the current- and charge-density response of the material 
is determined by the second derivatives 
of nonlocal screened electromagnetic fields,
as given in Eqs.~(\ref{eq:gjEB}) and (\ref{eq:gjEB0}).

The correspondence proposed here 
between deviations from emergent Lorentz symmetry in materials
and the SME framework suggests various future research topics
spanning condensed-matter and high-energy physics.
Several of these arise directly 
from the approach and results obtained in the present work.
For example,
the lattice models for a few types of minimal SME coefficients 
remain to be investigated in detail,
including ones such as $c_{\lambda0}$ and $g_{\kappa\lambda0}$ 
that are disregarded here to minimize complications
in the Hamiltonian description. 
The associated band structures, 
topological invariants,
and bound surface states would be interesting to establish.
An intriguing open issue in the general case 
is the identification of appropriate boundary conditions 
and the resulting surface states 
in the presence of Lorentz-violating terms. 
For the $b$ term associated with Weyl semimetals, 
the most general boundary conditions 
can be found through self-adjoint extensions 
of the Hamiltonian~\cite{Seradjeh_2018},
and it would be valuable to generalize this method
to other SME coefficients.

Our study of transport coefficients could also be broadened.
While we have exhausted the list of minimal coefficients for Lorentz violation
that lead to a nonvanishing induced current at leading order, 
certain SME coefficients in the action (\ref{eq:SMEaction}) 
may generate a nonvanishing fermion current from the vacuum polarization 
when evaluated at second and higher orders in Lorentz violation. 
Arguments analogous to Furry's theorem suggest that
a nonzero current is to be expected from contributions 
such as the $d$ and $H$ terms in Eq.~(\ref{eq:SMEaction}).
Second-order effects from the $b$, $c$, and $g$ terms 
may also induce a nonlinear fermion current 
through the modified vertex function. 
It would be of interest to investigate the transport coefficients
nonperturbatively,
by including the full dispersion for large SME coefficients.

Intriguing open issues in the broader context
are also suggested by the correspondence
between the SME and condensed-matter systems.
The full SME incorporates additional terms with field operators
of mass dimensions $d>4$,
along with couplings to all known gauge fields and gravity
\cite{Kostelecky:2009zp,Kostelecky:2011gq,Kostelecky:2013rta,%
	Kostelecky:2018yfa,Kostelecky:2003fs,Kostelecky_2011,Kostelecky:2017zob,
	Kostelecky:2020hbb}.
These terms represent a large set of modifications to the Dirac operator 
that as yet remain unexplored in the condensed-matter context,
so there is considerable potential for the discovery
and possibly even the design of interesting and novel materials
emergent Lorentz symmetry.
In the other direction,
condensed-matter systems offer prospects for informing SME physics.
Issues in the SME such as the conditions for quantum stability,
the interpretation of scenarios with large coefficients for Lorentz violation,
and the understanding of ambiguities in radiative corrections
could be addressed in the context of materials 
with departures from emergent Lorentz symmetry,
both via methods from condensed-matter theory
and through material realizations of SME systems in the laboratory.

Another interesting angle to pursue is the connection to Finsler geometry.
In the SME,
the trajectory of the centroid of a fermion or scalar wavepacket
in the presence of Lorentz violation
is known to correspond to a geodesic in a Riemann-Finsler spacetime
\cite{Kostelecky:2011qz,AlanKostelecky:2012yjr,Russell:2015gwa,Foster:2015yta,%
	Schreck:2015seb,Reis:2017ayl,Colladay:2017bon,Schreck:2019mmr,Reis:2021ban}.
We therefore anticipate that Finsler geometry underlies the
motion of quasiparticles in Dirac and Weyl semimetals 
and other materials exhibiting departures from emergent Lorentz symmetry.
This situation offers the potential for 
interdisciplinary advances in several directions.
For example,
results in Finsler geometry can be expected to provide insights
into the physics of various materials with emergent Lorentz symmetry,
while these systems in turn provide analogue models
and laboratory realizations for challenging mathematical issues.
Overall,
the numerous topics open for investigation 
across these seemingly disparate subjects
offer rich prospects for future advances.

\begin{acknowledgments}

V.A.K.\ is supported in part by the U.S.\ Department of Energy 
under grant number {DE}-SC0010120. 
N.M.\ acknowledges partial support by the U.S.\ Department of Energy 
under contract No.\ DEAC02-06CH11357 
while at Argonne National Laboratory 
as well as the U.S.\ Department of Energy, Office of Science, 
Office of Workforce Development for Teachers and Scientists, 
Office of Science Graduate Student Research (SCGSR) program. 
The SCGSR program is administered 
by the Oak Ridge Institute for Science and Education (ORISE) for the DOE. 
ORISE is managed by ORAU under contract number {DE}-SC0014664. 
TRIUMF receives federal funding via a contribution agreement 
with the National Research Council of Canada. 
M.S.\ appreciates support by FAPEMA Universal 01149/17, 
FAPEMA Universal 00830/19, 
CNPq Universal 421566/2016-7, 
CNPq Produtividade 312201/2018-4, 
and CAPES/Finance Code 001. 
B.S.\ was supported in part by NSF CAREER award DMR-1350663 
and the U.S.\ Department of Energy, 
Office of Science, 
Basic Energy Sciences, 
under Award No.\ DE-SC0020343.

\end{acknowledgments}

\bibliography{WeylSME}

\begin{thebibliography}{129}%
\makeatletter
\providecommand \@ifxundefined [1]{%
 \@ifx{#1\undefined}
}%
\providecommand \@ifnum [1]{%
 \ifnum #1\expandafter \@firstoftwo
 \else \expandafter \@secondoftwo
 \fi
}%
\providecommand \@ifx [1]{%
 \ifx #1\expandafter \@firstoftwo
 \else \expandafter \@secondoftwo
 \fi
}%
\providecommand \natexlab [1]{#1}%
\providecommand \enquote  [1]{``#1''}%
\providecommand \bibnamefont  [1]{#1}%
\providecommand \bibfnamefont [1]{#1}%
\providecommand \citenamefont [1]{#1}%
\providecommand \href@noop [0]{\@secondoftwo}%
\providecommand \href [0]{\begingroup \@sanitize@url \@href}%
\providecommand \@href[1]{\@@startlink{#1}\@@href}%
\providecommand \@@href[1]{\endgroup#1\@@endlink}%
\providecommand \@sanitize@url [0]{\catcode `\\12\catcode `\$12\catcode
  `\&12\catcode `\#12\catcode `\^12\catcode `\_12\catcode `\%12\relax}%
\providecommand \@@startlink[1]{}%
\providecommand \@@endlink[0]{}%
\providecommand \url  [0]{\begingroup\@sanitize@url \@url }%
\providecommand \@url [1]{\endgroup\@href {#1}{\urlprefix }}%
\providecommand \urlprefix  [0]{URL }%
\providecommand \Eprint [0]{\href }%
\providecommand \doibase [0]{https://doi.org/}%
\providecommand \selectlanguage [0]{\@gobble}%
\providecommand \bibinfo  [0]{\@secondoftwo}%
\providecommand \bibfield  [0]{\@secondoftwo}%
\providecommand \translation [1]{[#1]}%
\providecommand \BibitemOpen [0]{}%
\providecommand \bibitemStop [0]{}%
\providecommand \bibitemNoStop [0]{.\EOS\space}%
\providecommand \EOS [0]{\spacefactor3000\relax}%
\providecommand \BibitemShut  [1]{\csname bibitem#1\endcsname}%
\let\auto@bib@innerbib\@empty
%</preamble>
\bibitem [{\citenamefont {Weinberg}(2010)}]{Weinberg:2009bg}%
  \BibitemOpen
  \bibfield  {author} {\bibinfo {author} {\bibfnamefont {S.}~\bibnamefont
  {Weinberg}},\ }\bibfield  {title} {\bibinfo {title} {{Effective field theory,
  past and future}},\ }in\ \href {https://doi.org/10.22323/1.086.0001} {\emph
  {\bibinfo {booktitle} {Proceedings of 6th International Workshop on Chiral
  Dynamics {\textemdash} PoS(CD09)}}},\ Vol.\ \bibinfo {volume} {086}\
  (\bibinfo {year} {2010})\ p.\ \bibinfo {pages} {001},\ \Eprint
  {https://arxiv.org/abs/0908.1964} {arXiv:0908.1964 [hep-th]} \BibitemShut
  {NoStop}%
\bibitem [{\citenamefont {Colladay}\ and\ \citenamefont
  {Kosteleck{\'{y}}}(1997)}]{Colladay:1996iz}%
  \BibitemOpen
  \bibfield  {author} {\bibinfo {author} {\bibfnamefont {D.}~\bibnamefont
  {Colladay}}\ and\ \bibinfo {author} {\bibfnamefont {V.~A.}\ \bibnamefont
  {Kosteleck{\'{y}}}},\ }\bibfield  {title} {\bibinfo {title} {{CPT violation
  and the standard model}},\ }\href {https://doi.org/10.1103/PhysRevD.55.6760}
  {\bibfield  {journal} {\bibinfo  {journal} {Phys. Rev. D}\ }\textbf {\bibinfo
  {volume} {55}},\ \bibinfo {pages} {6760} (\bibinfo {year} {1997})},\ \Eprint
  {https://arxiv.org/abs/hep-ph/9703464} {arXiv:hep-ph/9703464} \BibitemShut
  {NoStop}%
\bibitem [{\citenamefont {Colladay}\ and\ \citenamefont
  {Kosteleck{\'{y}}}(1998)}]{Colladay_1998}%
  \BibitemOpen
  \bibfield  {author} {\bibinfo {author} {\bibfnamefont {D.}~\bibnamefont
  {Colladay}}\ and\ \bibinfo {author} {\bibfnamefont {V.~A.}\ \bibnamefont
  {Kosteleck{\'{y}}}},\ }\bibfield  {title} {\bibinfo {title}
  {Lorentz-violating extension of the standard model},\ }\href
  {https://doi.org/10.1103/physrevd.58.116002} {\bibfield  {journal} {\bibinfo
  {journal} {Phys. Rev. D}\ }\textbf {\bibinfo {volume} {58}},\ \bibinfo
  {pages} {116002} (\bibinfo {year} {1998})},\ \Eprint
  {https://arxiv.org/abs/hep-ph/9809521} {arXiv:hep-ph/9809521} \BibitemShut
  {NoStop}%
\bibitem [{\citenamefont {Kosteleck\'y}(2004)}]{Kostelecky:2003fs}%
  \BibitemOpen
  \bibfield  {author} {\bibinfo {author} {\bibfnamefont {V.~A.}\ \bibnamefont
  {Kosteleck\'y}},\ }\bibfield  {title} {\bibinfo {title} {{Gravity, Lorentz
  violation, and the standard model}},\ }\href
  {https://doi.org/10.1103/PhysRevD.69.105009} {\bibfield  {journal} {\bibinfo
  {journal} {Phys. Rev. D}\ }\textbf {\bibinfo {volume} {69}},\ \bibinfo
  {pages} {105009} (\bibinfo {year} {2004})}\BibitemShut {NoStop}%
\bibitem [{\citenamefont {Kosteleck\'{y}}\ and\ \citenamefont
  {Russell}(2011)}]{Kostelecky:2008edit}%
  \BibitemOpen
  \bibfield  {author} {\bibinfo {author} {\bibfnamefont {V.~A.}\ \bibnamefont
  {Kosteleck\'{y}}}\ and\ \bibinfo {author} {\bibfnamefont {N.}~\bibnamefont
  {Russell}},\ }\bibfield  {title} {\bibinfo {title} {{Data tables for Lorentz
  and CPT violation}},\ }\href {https://doi.org/10.1103/RevModPhys.83.11}
  {\bibfield  {journal} {\bibinfo  {journal} {Rev. Mod. Phys.}\ }\textbf
  {\bibinfo {volume} {83}},\ \bibinfo {pages} {11} (\bibinfo {year} {2011})},\
  \bibinfo {note} {updated edition for 2021 available as arXiv:
  0801.0287v14}\BibitemShut {NoStop}%
\bibitem [{\citenamefont {Lv}\ \emph {et~al.}(2015)\citenamefont {Lv} \emph
  {et~al.}}]{Lv:2015pya}%
  \BibitemOpen
  \bibfield  {author} {\bibinfo {author} {\bibfnamefont {B.~Q.}\ \bibnamefont
  {Lv}} \emph {et~al.},\ }\bibfield  {title} {\bibinfo {title} {{Experimental
  Discovery of Weyl Semimetal TaAs}},\ }\href
  {https://doi.org/10.1103/PhysRevX.5.031013} {\bibfield  {journal} {\bibinfo
  {journal} {Phys. Rev. X}\ }\textbf {\bibinfo {volume} {5}},\ \bibinfo {pages}
  {031013} (\bibinfo {year} {2015})},\ \Eprint
  {https://arxiv.org/abs/1502.04684} {arXiv:1502.04684 [cond-mat.mtrl-sci]}
  \BibitemShut {NoStop}%
\bibitem [{\citenamefont {Tamai}\ \emph {et~al.}(2016)\citenamefont {Tamai}
  \emph {et~al.}}]{Tamai:2016}%
  \BibitemOpen
  \bibfield  {author} {\bibinfo {author} {\bibfnamefont {A.}~\bibnamefont
  {Tamai}} \emph {et~al.},\ }\bibfield  {title} {\bibinfo {title} {{Fermi Arcs
  and Their Topological Character in the Candidate Type-II Weyl Semimetal
  $\mathrm{MoTe_2}$}},\ }\href {https://doi.org/10.1103/PhysRevX.6.031021}
  {\bibfield  {journal} {\bibinfo  {journal} {Phys. Rev. X}\ }\textbf {\bibinfo
  {volume} {6}},\ \bibinfo {pages} {031021} (\bibinfo {year} {2016})},\ \Eprint
  {https://arxiv.org/abs/1604.08228} {arXiv:1604.08228 [cond-mat.mtrl-sci]}
  \BibitemShut {NoStop}%
\bibitem [{\citenamefont {Yan}\ and\ \citenamefont
  {Felser}(2017)}]{Yan:2016euz}%
  \BibitemOpen
  \bibfield  {author} {\bibinfo {author} {\bibfnamefont {B.}~\bibnamefont
  {Yan}}\ and\ \bibinfo {author} {\bibfnamefont {C.}~\bibnamefont {Felser}},\
  }\bibfield  {title} {\bibinfo {title} {{Topological materials: Weyl
  semimetals}},\ }\href
  {https://doi.org/10.1146/annurev-conmatphys-031016-025458} {\bibfield
  {journal} {\bibinfo  {journal} {Ann. Rev. Condensed Matter Phys.}\ }\textbf
  {\bibinfo {volume} {8}},\ \bibinfo {pages} {337} (\bibinfo {year} {2017})},\
  \Eprint {https://arxiv.org/abs/1611.04182} {arXiv:1611.04182
  [cond-mat.mtrl-sci]} \BibitemShut {NoStop}%
\bibitem [{\citenamefont {Armitage}\ \emph {et~al.}(2018)\citenamefont
  {Armitage}, \citenamefont {Mele},\ and\ \citenamefont
  {Vishwanath}}]{Armitage:2017cjs}%
  \BibitemOpen
  \bibfield  {author} {\bibinfo {author} {\bibfnamefont {N.~P.}\ \bibnamefont
  {Armitage}}, \bibinfo {author} {\bibfnamefont {E.~J.}\ \bibnamefont {Mele}},\
  and\ \bibinfo {author} {\bibfnamefont {A.}~\bibnamefont {Vishwanath}},\
  }\bibfield  {title} {\bibinfo {title} {{Weyl and Dirac semimetals in
  three-dimensional solids}},\ }\href
  {https://doi.org/10.1103/RevModPhys.90.015001} {\bibfield  {journal}
  {\bibinfo  {journal} {Rev. Mod. Phys.}\ }\textbf {\bibinfo {volume} {90}},\
  \bibinfo {pages} {015001} (\bibinfo {year} {2018})},\ \Eprint
  {https://arxiv.org/abs/1705.01111} {arXiv:1705.01111 [cond-mat.str-el]}
  \BibitemShut {NoStop}%
\bibitem [{\citenamefont {Gao}\ \emph {et~al.}(2019)\citenamefont {Gao},
  \citenamefont {Venderbos}, \citenamefont {Kim},\ and\ \citenamefont
  {Rappe}}]{Gao:2018xwm}%
  \BibitemOpen
  \bibfield  {author} {\bibinfo {author} {\bibfnamefont {H.}~\bibnamefont
  {Gao}}, \bibinfo {author} {\bibfnamefont {J.~W.~F.}\ \bibnamefont
  {Venderbos}}, \bibinfo {author} {\bibfnamefont {Y.}~\bibnamefont {Kim}},\
  and\ \bibinfo {author} {\bibfnamefont {A.~M.}\ \bibnamefont {Rappe}},\
  }\bibfield  {title} {\bibinfo {title} {{Topological semimetals from first
  principles}},\ }\href {https://doi.org/10.1146/annurev-matsci-070218-010049}
  {\bibfield  {journal} {\bibinfo  {journal} {Ann. Rev. Mat. Res.}\ }\textbf
  {\bibinfo {volume} {49}},\ \bibinfo {pages} {153} (\bibinfo {year} {2019})},\
  \Eprint {https://arxiv.org/abs/1810.08186} {arXiv:1810.08186
  [cond-mat.mes-hall]} \BibitemShut {NoStop}%
\bibitem [{\citenamefont {Lee}\ \emph {et~al.}(2021)\citenamefont {Lee} \emph
  {et~al.}}]{Lee:2021}%
  \BibitemOpen
  \bibfield  {author} {\bibinfo {author} {\bibfnamefont {S.~H.}\ \bibnamefont
  {Lee}} \emph {et~al.},\ }\bibfield  {title} {\bibinfo {title} {{Evidence for
  a Magnetic-Field-Induced Ideal Type-II Weyl State in Antiferromagnetic
  Topological Insulator $\mathrm{Mn(Bi_{1-x}Sb_x)Te_4}$}},\ }\href
  {https://doi.org/10.1103/PhysRevX.11.031032} {\bibfield  {journal} {\bibinfo
  {journal} {Phys. Rev. X}\ }\textbf {\bibinfo {volume} {11}},\ \bibinfo
  {pages} {031032} (\bibinfo {year} {2021})},\ \Eprint
  {https://arxiv.org/abs/2002.10683} {arXiv:2002.10683 [cond-mat.mtrl-sci]}
  \BibitemShut {NoStop}%
\bibitem [{\citenamefont {Grushin}(2012)}]{Grushin:2012mt}%
  \BibitemOpen
  \bibfield  {author} {\bibinfo {author} {\bibfnamefont {A.~G.}\ \bibnamefont
  {Grushin}},\ }\bibfield  {title} {\bibinfo {title} {{Consequences of a
  condensed matter realization of Lorentz-violating QED in Weyl semi-metals}},\
  }\href {https://doi.org/10.1103/PhysRevD.86.045001} {\bibfield  {journal}
  {\bibinfo  {journal} {Phys. Rev. D}\ }\textbf {\bibinfo {volume} {86}},\
  \bibinfo {pages} {045001} (\bibinfo {year} {2012})},\ \Eprint
  {https://arxiv.org/abs/1205.3722} {arXiv:1205.3722 [hep-th]} \BibitemShut
  {NoStop}%
\bibitem [{\citenamefont {Liu}\ \emph {et~al.}(2013)\citenamefont {Liu},
  \citenamefont {Ye},\ and\ \citenamefont {Qi}}]{Liu:2012hk}%
  \BibitemOpen
  \bibfield  {author} {\bibinfo {author} {\bibfnamefont {C.-X.}\ \bibnamefont
  {Liu}}, \bibinfo {author} {\bibfnamefont {P.}~\bibnamefont {Ye}},\ and\
  \bibinfo {author} {\bibfnamefont {X.-L.}\ \bibnamefont {Qi}},\ }\bibfield
  {title} {\bibinfo {title} {{Chiral gauge field and axial anomaly in a Weyl
  semimetal}},\ }\href {https://doi.org/10.1103/PhysRevB.87.235306} {\bibfield
  {journal} {\bibinfo  {journal} {Phys. Rev. B}\ }\textbf {\bibinfo {volume}
  {87}},\ \bibinfo {pages} {235306} (\bibinfo {year} {2013})},\ \bibinfo {note}
  {[Erratum: Phys. Rev. B \textbf{92}, 119904 (2015)]},\ \Eprint
  {https://arxiv.org/abs/1204.6551} {arXiv:1204.6551 [cond-mat.str-el]}
  \BibitemShut {NoStop}%
\bibitem [{\citenamefont {Zyuzin}\ \emph {et~al.}(2012)\citenamefont {Zyuzin},
  \citenamefont {Wu},\ and\ \citenamefont {Burkov}}]{Zyuzin:2012vn}%
  \BibitemOpen
  \bibfield  {author} {\bibinfo {author} {\bibfnamefont {A.~A.}\ \bibnamefont
  {Zyuzin}}, \bibinfo {author} {\bibfnamefont {S.}~\bibnamefont {Wu}},\ and\
  \bibinfo {author} {\bibfnamefont {A.~A.}\ \bibnamefont {Burkov}},\ }\bibfield
   {title} {\bibinfo {title} {{Weyl semimetal with broken time reversal and
  inversion symmetries}},\ }\href {https://doi.org/10.1103/PhysRevB.85.165110}
  {\bibfield  {journal} {\bibinfo  {journal} {Phys. Rev. B}\ }\textbf {\bibinfo
  {volume} {85}},\ \bibinfo {pages} {165110} (\bibinfo {year} {2012})},\
  \Eprint {https://arxiv.org/abs/1201.3624} {arXiv:1201.3624
  [cond-mat.mes-hall]} \BibitemShut {NoStop}%
\bibitem [{\citenamefont {Zyuzin}\ and\ \citenamefont
  {Burkov}(2012)}]{Zyuzin:2012tv}%
  \BibitemOpen
  \bibfield  {author} {\bibinfo {author} {\bibfnamefont {A.~A.}\ \bibnamefont
  {Zyuzin}}\ and\ \bibinfo {author} {\bibfnamefont {A.~A.}\ \bibnamefont
  {Burkov}},\ }\bibfield  {title} {\bibinfo {title} {{Topological response in
  Weyl semimetals and the chiral anomaly}},\ }\href
  {https://doi.org/10.1103/PhysRevB.86.115133} {\bibfield  {journal} {\bibinfo
  {journal} {Phys. Rev. B}\ }\textbf {\bibinfo {volume} {86}},\ \bibinfo
  {pages} {115133} (\bibinfo {year} {2012})},\ \Eprint
  {https://arxiv.org/abs/1206.1868} {arXiv:1206.1868 [cond-mat.mes-hall]}
  \BibitemShut {NoStop}%
\bibitem [{\citenamefont {Goswami}\ and\ \citenamefont
  {Tewari}(2013)}]{Goswami:2012db}%
  \BibitemOpen
  \bibfield  {author} {\bibinfo {author} {\bibfnamefont {P.}~\bibnamefont
  {Goswami}}\ and\ \bibinfo {author} {\bibfnamefont {S.}~\bibnamefont
  {Tewari}},\ }\bibfield  {title} {\bibinfo {title} {{Axionic field theory of
  (3+1)-dimensional Weyl semimetals}},\ }\href
  {https://doi.org/10.1103/PhysRevB.88.245107} {\bibfield  {journal} {\bibinfo
  {journal} {Phys. Rev. B}\ }\textbf {\bibinfo {volume} {88}},\ \bibinfo
  {pages} {245107} (\bibinfo {year} {2013})},\ \Eprint
  {https://arxiv.org/abs/1210.6352} {arXiv:1210.6352 [cond-mat.mes-hall]}
  \BibitemShut {NoStop}%
\bibitem [{\citenamefont {Hannukainen}\ \emph {et~al.}(2021)\citenamefont
  {Hannukainen}, \citenamefont {Cortijo}, \citenamefont {Bardarson},\ and\
  \citenamefont {Ferreiros}}]{Hannukainen:2020sif}%
  \BibitemOpen
  \bibfield  {author} {\bibinfo {author} {\bibfnamefont {J.~D.}\ \bibnamefont
  {Hannukainen}}, \bibinfo {author} {\bibfnamefont {A.}~\bibnamefont
  {Cortijo}}, \bibinfo {author} {\bibfnamefont {J.~H.}\ \bibnamefont
  {Bardarson}},\ and\ \bibinfo {author} {\bibfnamefont {Y.}~\bibnamefont
  {Ferreiros}},\ }\bibfield  {title} {\bibinfo {title} {{Electric manipulation
  of domain walls in magnetic Weyl semimetals via the axial anomaly}},\ }\href
  {https://doi.org/10.21468/SciPostPhys.10.5.102} {\bibfield  {journal}
  {\bibinfo  {journal} {SciPost Phys.}\ }\textbf {\bibinfo {volume} {10}},\
  \bibinfo {pages} {102} (\bibinfo {year} {2021})},\ \Eprint
  {https://arxiv.org/abs/2012.12785} {arXiv:2012.12785 [cond-mat.mes-hall]}
  \BibitemShut {NoStop}%
\bibitem [{\citenamefont {Baum}\ \emph {et~al.}(2015)\citenamefont {Baum},
  \citenamefont {Berg}, \citenamefont {Parameswaran},\ and\ \citenamefont
  {Stern}}]{Baum:2015}%
  \BibitemOpen
  \bibfield  {author} {\bibinfo {author} {\bibfnamefont {Y.}~\bibnamefont
  {Baum}}, \bibinfo {author} {\bibfnamefont {E.}~\bibnamefont {Berg}}, \bibinfo
  {author} {\bibfnamefont {S.~A.}\ \bibnamefont {Parameswaran}},\ and\ \bibinfo
  {author} {\bibfnamefont {A.}~\bibnamefont {Stern}},\ }\bibfield  {title}
  {\bibinfo {title} {{Current at a Distance and Resonant Transparency in Weyl
  Semimetals}},\ }\href {https://doi.org/10.1103/PhysRevX.5.041046} {\bibfield
  {journal} {\bibinfo  {journal} {Phys. Rev. X}\ }\textbf {\bibinfo {volume}
  {5}},\ \bibinfo {pages} {041046} (\bibinfo {year} {2015})},\ \Eprint
  {https://arxiv.org/abs/1508.03047} {arXiv:1508.03047 [cond-mat.mes-hall]}
  \BibitemShut {NoStop}%
\bibitem [{\citenamefont {Landsteiner}\ \emph {et~al.}(2016)\citenamefont
  {Landsteiner}, \citenamefont {Liu},\ and\ \citenamefont
  {Sun}}]{Landsteiner:2015pdh}%
  \BibitemOpen
  \bibfield  {author} {\bibinfo {author} {\bibfnamefont {K.}~\bibnamefont
  {Landsteiner}}, \bibinfo {author} {\bibfnamefont {Y.}~\bibnamefont {Liu}},\
  and\ \bibinfo {author} {\bibfnamefont {Y.-W.}\ \bibnamefont {Sun}},\
  }\bibfield  {title} {\bibinfo {title} {{Quantum Phase Transition between a
  Topological and a Trivial Semimetal from Holography}},\ }\href
  {https://doi.org/10.1103/PhysRevLett.116.081602} {\bibfield  {journal}
  {\bibinfo  {journal} {Phys. Rev. Lett.}\ }\textbf {\bibinfo {volume} {116}},\
  \bibinfo {pages} {081602} (\bibinfo {year} {2016})},\ \Eprint
  {https://arxiv.org/abs/1511.05505} {arXiv:1511.05505 [hep-th]} \BibitemShut
  {NoStop}%
\bibitem [{\citenamefont {Grushin}\ \emph {et~al.}(2016)\citenamefont
  {Grushin}, \citenamefont {Venderbos}, \citenamefont {Vishwanath},\ and\
  \citenamefont {Ilan}}]{PhysRevX.6.041046}%
  \BibitemOpen
  \bibfield  {author} {\bibinfo {author} {\bibfnamefont {A.~G.}\ \bibnamefont
  {Grushin}}, \bibinfo {author} {\bibfnamefont {J.~W.~F.}\ \bibnamefont
  {Venderbos}}, \bibinfo {author} {\bibfnamefont {A.}~\bibnamefont
  {Vishwanath}},\ and\ \bibinfo {author} {\bibfnamefont {R.}~\bibnamefont
  {Ilan}},\ }\bibfield  {title} {\bibinfo {title} {{Inhomogeneous Weyl and
  Dirac Semimetals: Transport in Axial Magnetic Fields and Fermi Arc Surface
  States from Pseudo-Landau Levels}},\ }\href
  {https://doi.org/10.1103/PhysRevX.6.041046} {\bibfield  {journal} {\bibinfo
  {journal} {Phys. Rev. X}\ }\textbf {\bibinfo {volume} {6}},\ \bibinfo {pages}
  {041046} (\bibinfo {year} {2016})},\ \Eprint
  {https://arxiv.org/abs/1607.04268} {arXiv:1607.04268 [cond-mat.mes-hall]}
  \BibitemShut {NoStop}%
\bibitem [{\citenamefont {Elbistan}(2017)}]{Elbistan:2016rla}%
  \BibitemOpen
  \bibfield  {author} {\bibinfo {author} {\bibfnamefont {M.}~\bibnamefont
  {Elbistan}},\ }\bibfield  {title} {\bibinfo {title} {{Weyl semimetal and
  topological numbers}},\ }\href {https://doi.org/10.1142/S0217979217502216}
  {\bibfield  {journal} {\bibinfo  {journal} {Int. J. Mod. Phys. B}\ }\textbf
  {\bibinfo {volume} {31}},\ \bibinfo {pages} {1750221} (\bibinfo {year}
  {2017})},\ \Eprint {https://arxiv.org/abs/1605.00759} {arXiv:1605.00759
  [hep-th]} \BibitemShut {NoStop}%
\bibitem [{\citenamefont {van~der Wurff}\ and\ \citenamefont
  {Stoof}(2017)}]{vanderWurff:2017fpv}%
  \BibitemOpen
  \bibfield  {author} {\bibinfo {author} {\bibfnamefont {E.~C.~I.}\
  \bibnamefont {van~der Wurff}}\ and\ \bibinfo {author} {\bibfnamefont
  {H.~T.~C.}\ \bibnamefont {Stoof}},\ }\bibfield  {title} {\bibinfo {title}
  {{Anisotropic chiral magnetic effect from tilted Weyl cones}},\ }\href
  {https://doi.org/10.1103/PhysRevB.96.121116} {\bibfield  {journal} {\bibinfo
  {journal} {Phys. Rev. B}\ }\textbf {\bibinfo {volume} {96}},\ \bibinfo
  {pages} {121116(R)} (\bibinfo {year} {2017})},\ \Eprint
  {https://arxiv.org/abs/1707.00598} {arXiv:1707.00598 [cond-mat.mes-hall]}
  \BibitemShut {NoStop}%
\bibitem [{\citenamefont {Behrends}\ \emph {et~al.}(2019)\citenamefont
  {Behrends}, \citenamefont {Roy}, \citenamefont {Kolodrubetz}, \citenamefont
  {Bardarson},\ and\ \citenamefont {Grushin}}]{Behrends_2019}%
  \BibitemOpen
  \bibfield  {author} {\bibinfo {author} {\bibfnamefont {J.}~\bibnamefont
  {Behrends}}, \bibinfo {author} {\bibfnamefont {S.}~\bibnamefont {Roy}},
  \bibinfo {author} {\bibfnamefont {M.~H.}\ \bibnamefont {Kolodrubetz}},
  \bibinfo {author} {\bibfnamefont {J.~H.}\ \bibnamefont {Bardarson}},\ and\
  \bibinfo {author} {\bibfnamefont {A.~G.}\ \bibnamefont {Grushin}},\
  }\bibfield  {title} {\bibinfo {title} {{Landau levels, Bardeen polynomials,
  and Fermi arcs in Weyl semimetals: Lattice-based approach to the chiral
  anomaly}},\ }\href {https://doi.org/10.1103/physrevb.99.140201} {\bibfield
  {journal} {\bibinfo  {journal} {Phys. Rev. B}\ }\textbf {\bibinfo {volume}
  {99}},\ \bibinfo {pages} {140201(R)} (\bibinfo {year} {2019})},\ \Eprint
  {https://arxiv.org/abs/1807.06615} {arXiv:1807.06615 [cond-mat.mes-hall]}
  \BibitemShut {NoStop}%
\bibitem [{\citenamefont {Song}\ \emph {et~al.}(2019)\citenamefont {Song},
  \citenamefont {Rong},\ and\ \citenamefont {Sin}}]{Song:2019asj}%
  \BibitemOpen
  \bibfield  {author} {\bibinfo {author} {\bibfnamefont {G.}~\bibnamefont
  {Song}}, \bibinfo {author} {\bibfnamefont {J.}~\bibnamefont {Rong}},\ and\
  \bibinfo {author} {\bibfnamefont {S.-J.}\ \bibnamefont {Sin}},\ }\bibfield
  {title} {\bibinfo {title} {{Stability of topology in interacting Weyl
  semi-metal and topological dipole in holography}},\ }\href
  {https://doi.org/10.1007/JHEP10(2019)109} {\bibfield  {journal} {\bibinfo
  {journal} {{JHEP}}\ }\textbf {\bibinfo {volume} {10}},\ \bibinfo {pages}
  {109} (\bibinfo {year} {2019})},\ \Eprint {https://arxiv.org/abs/1904.09349}
  {arXiv:1904.09349 [hep-th]} \BibitemShut {NoStop}%
\bibitem [{\citenamefont {Chernodub}\ and\ \citenamefont
  {Vozmediano}(2019)}]{Chernodub:2019blw}%
  \BibitemOpen
  \bibfield  {author} {\bibinfo {author} {\bibfnamefont {M.~N.}\ \bibnamefont
  {Chernodub}}\ and\ \bibinfo {author} {\bibfnamefont {M.~A.~H.}\ \bibnamefont
  {Vozmediano}},\ }\bibfield  {title} {\bibinfo {title} {{Direct measurement of
  a beta function and an indirect check of the Schwinger effect near the
  boundary in Dirac semimetals}},\ }\href
  {https://doi.org/10.1103/PhysRevResearch.1.032002} {\bibfield  {journal}
  {\bibinfo  {journal} {Phys. Rev. Research}\ }\textbf {\bibinfo {volume}
  {1}},\ \bibinfo {pages} {032002(R)} (\bibinfo {year} {2019})},\ \Eprint
  {https://arxiv.org/abs/1902.02694} {arXiv:1902.02694 [cond-mat.str-el]}
  \BibitemShut {NoStop}%
\bibitem [{\citenamefont {Silva}\ \emph {et~al.}(2021)\citenamefont {Silva},
  \citenamefont {Lisboa-Santos}, \citenamefont {Ferreira},\ and\ \citenamefont
  {Schreck}}]{Silva:2021fzh}%
  \BibitemOpen
  \bibfield  {author} {\bibinfo {author} {\bibfnamefont {P.~D.~S.}\
  \bibnamefont {Silva}}, \bibinfo {author} {\bibfnamefont {L.}~\bibnamefont
  {Lisboa-Santos}}, \bibinfo {author} {\bibfnamefont {M.~M.}\ \bibnamefont
  {Ferreira}, \bibfnamefont {Jr.}},\ and\ \bibinfo {author} {\bibfnamefont
  {M.}~\bibnamefont {Schreck}},\ }\bibfield  {title} {\bibinfo {title}
  {{Effects of CPT-odd terms of dimensions three and five on electromagnetic
  propagation in continuous matter}},\ }\Eprint
  {https://arxiv.org/abs/2109.04659} {arXiv:2109.04659 [hep-th]}  (\bibinfo
  {year} {2021})\BibitemShut {NoStop}%
\bibitem [{\citenamefont {Ji}\ \emph {et~al.}(2021)\citenamefont {Ji},
  \citenamefont {Liu}, \citenamefont {Sun},\ and\ \citenamefont
  {Zhang}}]{Ji:2021aan}%
  \BibitemOpen
  \bibfield  {author} {\bibinfo {author} {\bibfnamefont {X.}~\bibnamefont
  {Ji}}, \bibinfo {author} {\bibfnamefont {Y.}~\bibnamefont {Liu}}, \bibinfo
  {author} {\bibfnamefont {Y.-W.}\ \bibnamefont {Sun}},\ and\ \bibinfo {author}
  {\bibfnamefont {Y.-L.}\ \bibnamefont {Zhang}},\ }\bibfield  {title} {\bibinfo
  {title} {{A Weyl-Z$_{2}$ semimetal from holography}},\ }\href
  {https://doi.org/10.1007/JHEP12(2021)066} {\bibfield  {journal} {\bibinfo
  {journal} {{JHEP}}\ }\textbf {\bibinfo {volume} {12}},\ \bibinfo {pages}
  {066} (\bibinfo {year} {2021})},\ \Eprint {https://arxiv.org/abs/2109.05993}
  {arXiv:2109.05993 [hep-th]} \BibitemShut {NoStop}%
\bibitem [{\citenamefont {Bitaghsir~Fadafan}\ \emph {et~al.}(2021)\citenamefont
  {Bitaghsir~Fadafan}, \citenamefont {O'Bannon}, \citenamefont {Rodgers},\ and\
  \citenamefont {Russell}}]{BitaghsirFadafan:2020lkh}%
  \BibitemOpen
  \bibfield  {author} {\bibinfo {author} {\bibfnamefont {K.}~\bibnamefont
  {Bitaghsir~Fadafan}}, \bibinfo {author} {\bibfnamefont {A.}~\bibnamefont
  {O'Bannon}}, \bibinfo {author} {\bibfnamefont {R.}~\bibnamefont {Rodgers}},\
  and\ \bibinfo {author} {\bibfnamefont {M.}~\bibnamefont {Russell}},\
  }\bibfield  {title} {\bibinfo {title} {{A Weyl semimetal from AdS/CFT with
  flavour}},\ }\href {https://doi.org/10.1007/JHEP04(2021)162} {\bibfield
  {journal} {\bibinfo  {journal} {{JHEP}}\ }\textbf {\bibinfo {volume} {04}},\
  \bibinfo {pages} {162} (\bibinfo {year} {2021})},\ \Eprint
  {https://arxiv.org/abs/2012.11434} {arXiv:2012.11434 [hep-th]} \BibitemShut
  {NoStop}%
\bibitem [{\citenamefont {Burkov}\ \emph {et~al.}(2011)\citenamefont {Burkov},
  \citenamefont {Hook},\ and\ \citenamefont {Balents}}]{Burkov:2011}%
  \BibitemOpen
  \bibfield  {author} {\bibinfo {author} {\bibfnamefont {A.~A.}\ \bibnamefont
  {Burkov}}, \bibinfo {author} {\bibfnamefont {M.~D.}\ \bibnamefont {Hook}},\
  and\ \bibinfo {author} {\bibfnamefont {L.}~\bibnamefont {Balents}},\
  }\bibfield  {title} {\bibinfo {title} {{Topological nodal semimetals}},\
  }\href {https://doi.org/10.1103/PhysRevB.84.235126} {\bibfield  {journal}
  {\bibinfo  {journal} {Phys. Rev. B}\ }\textbf {\bibinfo {volume} {84}},\
  \bibinfo {pages} {235126} (\bibinfo {year} {2011})},\ \Eprint
  {https://arxiv.org/abs/1110.1089} {arXiv:1110.1089 [cond-mat.mes-hall]}
  \BibitemShut {NoStop}%
\bibitem [{\citenamefont {{\'{O}. Pozo, Y. Ferreiros, and M. A. H.
  Vozmediano}}(2018)}]{Pozo:2018yzs}%
  \BibitemOpen
  \bibfield  {author} {\bibinfo {author} {\bibnamefont {{\'{O}. Pozo, Y.
  Ferreiros, and M. A. H. Vozmediano}}},\ }\bibfield  {title} {\bibinfo {title}
  {{Anisotropic fixed points in Dirac and Weyl semimetals}},\ }\href
  {https://doi.org/10.1103/PhysRevB.98.115122} {\bibfield  {journal} {\bibinfo
  {journal} {Phys. Rev. B}\ }\textbf {\bibinfo {volume} {98}},\ \bibinfo
  {pages} {115122} (\bibinfo {year} {2018})},\ \Eprint
  {https://arxiv.org/abs/1802.02632} {arXiv:1802.02632 [cond-mat.str-el]}
  \BibitemShut {NoStop}%
\bibitem [{\citenamefont {G\'omez}\ and\ \citenamefont
  {Urrutia}(2021)}]{Gomez:2021aez}%
  \BibitemOpen
  \bibfield  {author} {\bibinfo {author} {\bibfnamefont {A.}~\bibnamefont
  {G\'omez}}\ and\ \bibinfo {author} {\bibfnamefont {L.}~\bibnamefont
  {Urrutia}},\ }\bibfield  {title} {\bibinfo {title} {{The axial anomaly in
  Lorentz violating theories: Towards the electromagnetic response of weakly
  tilted Weyl semimetals}},\ }\href {https://doi.org/10.3390/sym13071181}
  {\bibfield  {journal} {\bibinfo  {journal} {Symmetry}\ }\textbf {\bibinfo
  {volume} {13}},\ \bibinfo {pages} {1181} (\bibinfo {year} {2021})},\ \Eprint
  {https://arxiv.org/abs/2106.15062} {arXiv:2106.15062 [cond-mat.mes-hall]}
  \BibitemShut {NoStop}%
\bibitem [{\citenamefont {Rodgers}\ \emph {et~al.}(2021)\citenamefont
  {Rodgers}, \citenamefont {Mauri}, \citenamefont {G\"ursoy},\ and\
  \citenamefont {Stoof}}]{Rodgers:2021azg}%
  \BibitemOpen
  \bibfield  {author} {\bibinfo {author} {\bibfnamefont {R.}~\bibnamefont
  {Rodgers}}, \bibinfo {author} {\bibfnamefont {E.}~\bibnamefont {Mauri}},
  \bibinfo {author} {\bibfnamefont {U.}~\bibnamefont {G\"ursoy}},\ and\
  \bibinfo {author} {\bibfnamefont {H.~T.~C.}\ \bibnamefont {Stoof}},\
  }\bibfield  {title} {\bibinfo {title} {{Thermodynamics and transport of
  holographic nodal line semimetals}},\ }\href
  {https://doi.org/10.1007/JHEP11(2021)191} {\bibfield  {journal} {\bibinfo
  {journal} {{JHEP}}\ }\textbf {\bibinfo {volume} {11}},\ \bibinfo {pages}
  {191} (\bibinfo {year} {2021})},\ \Eprint {https://arxiv.org/abs/2109.07187}
  {arXiv:2109.07187 [hep-th]} \BibitemShut {NoStop}%
\bibitem [{\citenamefont {Van~Harlingen}(1995)}]{Van-Harlingen_1995}%
  \BibitemOpen
  \bibfield  {author} {\bibinfo {author} {\bibfnamefont {D.~J.}\ \bibnamefont
  {Van~Harlingen}},\ }\bibfield  {title} {\bibinfo {title} {Phase-sensitive
  tests of the symmetry of the pairing state in the high-temperature
  superconductors: Evidence for $d_{x^{2}-y^{2}}$ symmetry},\ }\href
  {https://doi.org/10.1103/RevModPhys.67.515} {\bibfield  {journal} {\bibinfo
  {journal} {Rev. Mod. Phys.}\ }\textbf {\bibinfo {volume} {67}},\ \bibinfo
  {pages} {515} (\bibinfo {year} {1995})}\BibitemShut {NoStop}%
\bibitem [{\citenamefont {Tsuei}\ and\ \citenamefont
  {Kirtley}(2000)}]{Tsuei_Kirtley_2000}%
  \BibitemOpen
  \bibfield  {author} {\bibinfo {author} {\bibfnamefont {C.~C.}\ \bibnamefont
  {Tsuei}}\ and\ \bibinfo {author} {\bibfnamefont {J.~R.}\ \bibnamefont
  {Kirtley}},\ }\bibfield  {title} {\bibinfo {title} {Pairing symmetry in
  cuprate superconductors},\ }\href {https://doi.org/10.1103/RevModPhys.72.969}
  {\bibfield  {journal} {\bibinfo  {journal} {Rev. Mod. Phys.}\ }\textbf
  {\bibinfo {volume} {72}},\ \bibinfo {pages} {969} (\bibinfo {year}
  {2000})}\BibitemShut {NoStop}%
\bibitem [{\citenamefont {Franz}\ \emph {et~al.}(2002)\citenamefont {Franz},
  \citenamefont {Te\v{s}anovi\'{c}},\ and\ \citenamefont {Vafek}}]{Franz_2002}%
  \BibitemOpen
  \bibfield  {author} {\bibinfo {author} {\bibfnamefont {M.}~\bibnamefont
  {Franz}}, \bibinfo {author} {\bibfnamefont {Z.}~\bibnamefont
  {Te\v{s}anovi\'{c}}},\ and\ \bibinfo {author} {\bibfnamefont
  {O.}~\bibnamefont {Vafek}},\ }\bibfield  {title} {\bibinfo {title} {{QED$_3$
  theory of pairing pseudogap in cuprates: From d-wave superconductor to
  antiferromagnet via an algebraic Fermi liquid}},\ }\href
  {https://doi.org/10.1103/PhysRevB.66.054535} {\bibfield  {journal} {\bibinfo
  {journal} {Phys. Rev. B}\ }\textbf {\bibinfo {volume} {66}},\ \bibinfo
  {pages} {054535} (\bibinfo {year} {2002})}\BibitemShut {NoStop}%
\bibitem [{\citenamefont {Herbut}(2002)}]{Herbut_2002}%
  \BibitemOpen
  \bibfield  {author} {\bibinfo {author} {\bibfnamefont {I.~F.}\ \bibnamefont
  {Herbut}},\ }\bibfield  {title} {\bibinfo {title} {{QED$_3$ theory of
  underdoped high-temperature superconductors}},\ }\href
  {https://doi.org/10.1103/PhysRevB.66.094504} {\bibfield  {journal} {\bibinfo
  {journal} {Phys. Rev. B}\ }\textbf {\bibinfo {volume} {66}},\ \bibinfo
  {pages} {094504} (\bibinfo {year} {2002})}\BibitemShut {NoStop}%
\bibitem [{\citenamefont {Seradjeh}\ and\ \citenamefont
  {Herbut}(2002)}]{Seradjeh_Herbut_2002}%
  \BibitemOpen
  \bibfield  {author} {\bibinfo {author} {\bibfnamefont {B.~H.}\ \bibnamefont
  {Seradjeh}}\ and\ \bibinfo {author} {\bibfnamefont {I.~F.}\ \bibnamefont
  {Herbut}},\ }\bibfield  {title} {\bibinfo {title} {{Fine structure of chiral
  symmetry breaking in the QED$_3$ theory of underdoped high-$T_c$
  superconductors}},\ }\href {https://doi.org/10.1103/PhysRevB.66.184507}
  {\bibfield  {journal} {\bibinfo  {journal} {Phys. Rev. B}\ }\textbf {\bibinfo
  {volume} {66}},\ \bibinfo {pages} {184507} (\bibinfo {year}
  {2002})}\BibitemShut {NoStop}%
\bibitem [{\citenamefont {Hasan}\ and\ \citenamefont
  {Kane}(2010)}]{Hasan_Kane_2010}%
  \BibitemOpen
  \bibfield  {author} {\bibinfo {author} {\bibfnamefont {M.~Z.}\ \bibnamefont
  {Hasan}}\ and\ \bibinfo {author} {\bibfnamefont {C.~L.}\ \bibnamefont
  {Kane}},\ }\bibfield  {title} {\bibinfo {title} {{\it Colloquium}:
  Topological insulators},\ }\href {https://doi.org/10.1103/RevModPhys.82.3045}
  {\bibfield  {journal} {\bibinfo  {journal} {Rev. Mod. Phys.}\ }\textbf
  {\bibinfo {volume} {82}},\ \bibinfo {pages} {3045} (\bibinfo {year}
  {2010})}\BibitemShut {NoStop}%
\bibitem [{\citenamefont {Ryu}\ \emph {et~al.}(2010)\citenamefont {Ryu},
  \citenamefont {Schnyder}, \citenamefont {Furusaki},\ and\ \citenamefont
  {Ludwig}}]{Ryu_2010}%
  \BibitemOpen
  \bibfield  {author} {\bibinfo {author} {\bibfnamefont {S.}~\bibnamefont
  {Ryu}}, \bibinfo {author} {\bibfnamefont {A.~P.}\ \bibnamefont {Schnyder}},
  \bibinfo {author} {\bibfnamefont {A.}~\bibnamefont {Furusaki}},\ and\
  \bibinfo {author} {\bibfnamefont {A.~W.~W.}\ \bibnamefont {Ludwig}},\
  }\bibfield  {title} {\bibinfo {title} {Topological insulators and
  superconductors: Tenfold way and dimensional hierarchy},\ }\href
  {https://doi.org/10.1088/1367-2630/12/6/065010} {\bibfield  {journal}
  {\bibinfo  {journal} {New J. Phys.}\ }\textbf {\bibinfo {volume} {12}},\
  \bibinfo {pages} {065010} (\bibinfo {year} {2010})}\BibitemShut {NoStop}%
\bibitem [{\citenamefont {Ando}\ and\ \citenamefont {Fu}(2015)}]{Ando_Fu_2015}%
  \BibitemOpen
  \bibfield  {author} {\bibinfo {author} {\bibfnamefont {Y.}~\bibnamefont
  {Ando}}\ and\ \bibinfo {author} {\bibfnamefont {L.}~\bibnamefont {Fu}},\
  }\bibfield  {title} {\bibinfo {title} {Topological crystalline insulators and
  topological superconductors: From concepts to materials},\ }\href
  {https://doi.org/10.1146/annurev-conmatphys-031214-014501} {\bibfield
  {journal} {\bibinfo  {journal} {Annu. Rev. Condens. Matter Phys.}\ }\textbf
  {\bibinfo {volume} {6}},\ \bibinfo {pages} {361} (\bibinfo {year}
  {2015})}\BibitemShut {NoStop}%
\bibitem [{\citenamefont {Xu}\ \emph {et~al.}(2016)\citenamefont {Xu},
  \citenamefont {Li}, \citenamefont {Acosta}, \citenamefont {Li}, \citenamefont
  {Swartzentruber}, \citenamefont {Zheng}, \citenamefont {Sinitsyn},
  \citenamefont {Htoon}, \citenamefont {Wang},\ and\ \citenamefont
  {Zhang}}]{Xu_2016}%
  \BibitemOpen
  \bibfield  {author} {\bibinfo {author} {\bibfnamefont {E.}~\bibnamefont
  {Xu}}, \bibinfo {author} {\bibfnamefont {Z.}~\bibnamefont {Li}}, \bibinfo
  {author} {\bibfnamefont {J.~A.}\ \bibnamefont {Acosta}}, \bibinfo {author}
  {\bibfnamefont {N.}~\bibnamefont {Li}}, \bibinfo {author} {\bibfnamefont
  {B.}~\bibnamefont {Swartzentruber}}, \bibinfo {author} {\bibfnamefont
  {S.}~\bibnamefont {Zheng}}, \bibinfo {author} {\bibfnamefont
  {N.}~\bibnamefont {Sinitsyn}}, \bibinfo {author} {\bibfnamefont
  {H.}~\bibnamefont {Htoon}}, \bibinfo {author} {\bibfnamefont
  {J.}~\bibnamefont {Wang}},\ and\ \bibinfo {author} {\bibfnamefont
  {S.}~\bibnamefont {Zhang}},\ }\bibfield  {title} {\bibinfo {title} {{Enhanced
  thermoelectric properties of topological crystalline insulator PbSnTe
  nanowires grown by vapor transport}},\ }\href
  {https://doi.org/10.1007/s12274-015-0961-1} {\bibfield  {journal} {\bibinfo
  {journal} {Nano Res.}\ }\textbf {\bibinfo {volume} {9}},\ \bibinfo {pages}
  {820} (\bibinfo {year} {2016})}\BibitemShut {NoStop}%
\bibitem [{\citenamefont {Reja}\ \emph {et~al.}(2017)\citenamefont {Reja},
  \citenamefont {Fertig}, \citenamefont {Brey},\ and\ \citenamefont
  {Zhang}}]{Reja_2017}%
  \BibitemOpen
  \bibfield  {author} {\bibinfo {author} {\bibfnamefont {S.}~\bibnamefont
  {Reja}}, \bibinfo {author} {\bibfnamefont {H.~A.}\ \bibnamefont {Fertig}},
  \bibinfo {author} {\bibfnamefont {L.}~\bibnamefont {Brey}},\ and\ \bibinfo
  {author} {\bibfnamefont {S.}~\bibnamefont {Zhang}},\ }\bibfield  {title}
  {\bibinfo {title} {Surface magnetism in topological crystalline insulators},\
  }\href {https://doi.org/10.1103/PhysRevB.96.201111} {\bibfield  {journal}
  {\bibinfo  {journal} {Phys. Rev. B}\ }\textbf {\bibinfo {volume} {96}},\
  \bibinfo {pages} {201111(R)} (\bibinfo {year} {2017})}\BibitemShut {NoStop}%
\bibitem [{\citenamefont {Novoselov}\ \emph {et~al.}(2004)\citenamefont
  {Novoselov}, \citenamefont {Geim}, \citenamefont {Morozov}, \citenamefont
  {Jiang}, \citenamefont {Zhang}, \citenamefont {Dubonos}, \citenamefont
  {Grigorieva},\ and\ \citenamefont {Firsov}}]{Novoselov:2004}%
  \BibitemOpen
  \bibfield  {author} {\bibinfo {author} {\bibfnamefont {K.~S.}\ \bibnamefont
  {Novoselov}}, \bibinfo {author} {\bibfnamefont {A.~K.}\ \bibnamefont {Geim}},
  \bibinfo {author} {\bibfnamefont {S.~V.}\ \bibnamefont {Morozov}}, \bibinfo
  {author} {\bibfnamefont {D.}~\bibnamefont {Jiang}}, \bibinfo {author}
  {\bibfnamefont {Y.}~\bibnamefont {Zhang}}, \bibinfo {author} {\bibfnamefont
  {S.~V.}\ \bibnamefont {Dubonos}}, \bibinfo {author} {\bibfnamefont {I.~V.}\
  \bibnamefont {Grigorieva}},\ and\ \bibinfo {author} {\bibfnamefont {A.~A.}\
  \bibnamefont {Firsov}},\ }\bibfield  {title} {\bibinfo {title} {{Electric
  Field Effect in Atomically Thin Carbon Films}},\ }\href
  {https://doi.org/10.1126/science.1102896} {\bibfield  {journal} {\bibinfo
  {journal} {Science}\ }\textbf {\bibinfo {volume} {306}},\ \bibinfo {pages}
  {666} (\bibinfo {year} {2004})},\ \Eprint
  {https://arxiv.org/abs/cond-mat/0410550} {arXiv:cond-mat/0410550
  [cond-mat.mtrl-sci]} \BibitemShut {NoStop}%
\bibitem [{\citenamefont {Castro~Neto}\ \emph {et~al.}(2009)\citenamefont
  {Castro~Neto}, \citenamefont {Guinea}, \citenamefont {Peres}, \citenamefont
  {Novoselov},\ and\ \citenamefont {Geim}}]{Castro-Neto_2009}%
  \BibitemOpen
  \bibfield  {author} {\bibinfo {author} {\bibfnamefont {A.~H.}\ \bibnamefont
  {Castro~Neto}}, \bibinfo {author} {\bibfnamefont {F.}~\bibnamefont {Guinea}},
  \bibinfo {author} {\bibfnamefont {N.~M.~R.}\ \bibnamefont {Peres}}, \bibinfo
  {author} {\bibfnamefont {K.~S.}\ \bibnamefont {Novoselov}},\ and\ \bibinfo
  {author} {\bibfnamefont {A.~K.}\ \bibnamefont {Geim}},\ }\bibfield  {title}
  {\bibinfo {title} {The electronic properties of graphene},\ }\href
  {https://doi.org/10.1103/RevModPhys.81.109} {\bibfield  {journal} {\bibinfo
  {journal} {Rev. Mod. Phys.}\ }\textbf {\bibinfo {volume} {81}},\ \bibinfo
  {pages} {109} (\bibinfo {year} {2009})}\BibitemShut {NoStop}%
\bibitem [{\citenamefont {Kosteleck{\'{y}}}\ and\ \citenamefont
  {Lehnert}(2001)}]{Kostelecky_2001}%
  \BibitemOpen
  \bibfield  {author} {\bibinfo {author} {\bibfnamefont {V.~A.}\ \bibnamefont
  {Kosteleck{\'{y}}}}\ and\ \bibinfo {author} {\bibfnamefont {R.}~\bibnamefont
  {Lehnert}},\ }\bibfield  {title} {\bibinfo {title} {{Stability, causality,
  and Lorentz and CPT violation}},\ }\href
  {https://doi.org/10.1103/physrevd.63.065008} {\bibfield  {journal} {\bibinfo
  {journal} {Phys. Rev. D}\ }\textbf {\bibinfo {volume} {63}},\ \bibinfo
  {pages} {065008} (\bibinfo {year} {2001})},\ \Eprint
  {https://arxiv.org/abs/hep-th/0012060} {arXiv:hep-th/0012060} \BibitemShut
  {NoStop}%
\bibitem [{\citenamefont {Chung}\ and\ \citenamefont
  {Oh}(1999)}]{Chung:1998jv}%
  \BibitemOpen
  \bibfield  {author} {\bibinfo {author} {\bibfnamefont {J.-M.}\ \bibnamefont
  {Chung}}\ and\ \bibinfo {author} {\bibfnamefont {P.}~\bibnamefont {Oh}},\
  }\bibfield  {title} {\bibinfo {title} {{Lorentz and CPT violating
  Chern-Simons term in the derivative expansion of QED}},\ }\href
  {https://doi.org/10.1103/PhysRevD.60.067702} {\bibfield  {journal} {\bibinfo
  {journal} {Phys. Rev. D}\ }\textbf {\bibinfo {volume} {60}},\ \bibinfo
  {pages} {067702} (\bibinfo {year} {1999})},\ \Eprint
  {https://arxiv.org/abs/hep-th/9812132} {arXiv:hep-th/9812132} \BibitemShut
  {NoStop}%
\bibitem [{\citenamefont {Jackiw}\ and\ \citenamefont
  {Kosteleck\'{y}}(1999)}]{Jackiw:1999yp}%
  \BibitemOpen
  \bibfield  {author} {\bibinfo {author} {\bibfnamefont {R.}~\bibnamefont
  {Jackiw}}\ and\ \bibinfo {author} {\bibfnamefont {V.~A.}\ \bibnamefont
  {Kosteleck\'{y}}},\ }\bibfield  {title} {\bibinfo {title} {{Radiatively
  Induced Lorentz and CPT Violation in Electrodynamics}},\ }\href
  {https://doi.org/10.1103/PhysRevLett.82.3572} {\bibfield  {journal} {\bibinfo
   {journal} {Phys. Rev. Lett.}\ }\textbf {\bibinfo {volume} {82}},\ \bibinfo
  {pages} {3572} (\bibinfo {year} {1999})},\ \Eprint
  {https://arxiv.org/abs/hep-ph/9901358} {arXiv:hep-ph/9901358} \BibitemShut
  {NoStop}%
\bibitem [{\citenamefont {P\'{e}rez-Victoria}(1999)}]{Perez-Victoria:1999erb}%
  \BibitemOpen
  \bibfield  {author} {\bibinfo {author} {\bibfnamefont {M.}~\bibnamefont
  {P\'{e}rez-Victoria}},\ }\bibfield  {title} {\bibinfo {title} {{Exact
  Calculation of the Radiatively Induced Lorentz and CPT Violation in QED}},\
  }\href {https://doi.org/10.1103/PhysRevLett.83.2518} {\bibfield  {journal}
  {\bibinfo  {journal} {Phys. Rev. Lett.}\ }\textbf {\bibinfo {volume} {83}},\
  \bibinfo {pages} {2518} (\bibinfo {year} {1999})},\ \Eprint
  {https://arxiv.org/abs/hep-th/9905061} {arXiv:hep-th/9905061} \BibitemShut
  {NoStop}%
\bibitem [{\citenamefont {Chung}(1999{\natexlab{a}})}]{Chung:1999gg}%
  \BibitemOpen
  \bibfield  {author} {\bibinfo {author} {\bibfnamefont {J.-M.}\ \bibnamefont
  {Chung}},\ }\bibfield  {title} {\bibinfo {title} {{Lorentz- and CPT-violating
  Chern-Simons term in the functional integral formalism}},\ }\href
  {https://doi.org/10.1103/PhysRevD.60.127901} {\bibfield  {journal} {\bibinfo
  {journal} {Phys. Rev. D}\ }\textbf {\bibinfo {volume} {60}},\ \bibinfo
  {pages} {127901} (\bibinfo {year} {1999}{\natexlab{a}})},\ \Eprint
  {https://arxiv.org/abs/hep-th/9904037} {arXiv:hep-th/9904037} \BibitemShut
  {NoStop}%
\bibitem [{\citenamefont {Chung}(1999{\natexlab{b}})}]{Chung:1999pt}%
  \BibitemOpen
  \bibfield  {author} {\bibinfo {author} {\bibfnamefont {J.-M.}\ \bibnamefont
  {Chung}},\ }\bibfield  {title} {\bibinfo {title} {{Radiatively-induced
  Lorentz and CPT violating Chern-Simons term in QED}},\ }\href
  {https://doi.org/10.1016/S0370-2693(99)00822-9} {\bibfield  {journal}
  {\bibinfo  {journal} {Phys. Lett. B}\ }\textbf {\bibinfo {volume} {461}},\
  \bibinfo {pages} {138} (\bibinfo {year} {1999}{\natexlab{b}})},\ \Eprint
  {https://arxiv.org/abs/hep-th/9905095} {arXiv:hep-th/9905095} \BibitemShut
  {NoStop}%
\bibitem [{\citenamefont {Altschul}(2004{\natexlab{a}})}]{Altschul:2003ce}%
  \BibitemOpen
  \bibfield  {author} {\bibinfo {author} {\bibfnamefont {B.}~\bibnamefont
  {Altschul}},\ }\bibfield  {title} {\bibinfo {title} {{Failure of gauge
  invariance in the nonperturbative formulation of massless Lorentz-violating
  QED}},\ }\href {https://doi.org/10.1103/PhysRevD.69.125009} {\bibfield
  {journal} {\bibinfo  {journal} {Phys. Rev. D}\ }\textbf {\bibinfo {volume}
  {69}},\ \bibinfo {pages} {125009} (\bibinfo {year} {2004}{\natexlab{a}})},\
  \Eprint {https://arxiv.org/abs/hep-th/0311200} {arXiv:hep-th/0311200}
  \BibitemShut {NoStop}%
\bibitem [{\citenamefont {Altschul}(2004{\natexlab{b}})}]{Altschul:2004gs}%
  \BibitemOpen
  \bibfield  {author} {\bibinfo {author} {\bibfnamefont {B.}~\bibnamefont
  {Altschul}},\ }\bibfield  {title} {\bibinfo {title} {{Gauge invariance and
  the Pauli-Villars regulator in Lorentz- and CPT-violating electrodynamics}},\
  }\href {https://doi.org/10.1103/PhysRevD.70.101701} {\bibfield  {journal}
  {\bibinfo  {journal} {Phys. Rev. D}\ }\textbf {\bibinfo {volume} {70}},\
  \bibinfo {pages} {101701(R)} (\bibinfo {year} {2004}{\natexlab{b}})},\
  \Eprint {https://arxiv.org/abs/hep-th/0407172} {arXiv:hep-th/0407172}
  \BibitemShut {NoStop}%
\bibitem [{\citenamefont {Kosteleck\'y}(2011)}]{Kostelecky:2011qz}%
  \BibitemOpen
  \bibfield  {author} {\bibinfo {author} {\bibfnamefont {V.~A.}\ \bibnamefont
  {Kosteleck\'y}},\ }\bibfield  {title} {\bibinfo {title} {{Riemann-Finsler
  geometry and Lorentz-violating kinematics}},\ }\href
  {https://doi.org/10.1016/j.physletb.2011.05.041} {\bibfield  {journal}
  {\bibinfo  {journal} {Phys. Lett. B}\ }\textbf {\bibinfo {volume} {701}},\
  \bibinfo {pages} {137} (\bibinfo {year} {2011})},\ \Eprint
  {https://arxiv.org/abs/1104.5488} {arXiv:1104.5488 [hep-th]} \BibitemShut
  {NoStop}%
\bibitem [{\citenamefont {Kosteleck\'y}\ and\ \citenamefont
  {Samuel}(1989)}]{Kostelecky:1988zi}%
  \BibitemOpen
  \bibfield  {author} {\bibinfo {author} {\bibfnamefont {V.~A.}\ \bibnamefont
  {Kosteleck\'y}}\ and\ \bibinfo {author} {\bibfnamefont {S.}~\bibnamefont
  {Samuel}},\ }\bibfield  {title} {\bibinfo {title} {{Spontaneous breaking of
  Lorentz symmetry in string theory}},\ }\href
  {https://doi.org/10.1103/PhysRevD.39.683} {\bibfield  {journal} {\bibinfo
  {journal} {Phys. Rev. D}\ }\textbf {\bibinfo {volume} {39}},\ \bibinfo
  {pages} {683} (\bibinfo {year} {1989})}\BibitemShut {NoStop}%
\bibitem [{\citenamefont {Kosteleck\'y}\ and\ \citenamefont
  {Potting}(1991)}]{Kostelecky:1991ak}%
  \BibitemOpen
  \bibfield  {author} {\bibinfo {author} {\bibfnamefont {V.~A.}\ \bibnamefont
  {Kosteleck\'y}}\ and\ \bibinfo {author} {\bibfnamefont {R.}~\bibnamefont
  {Potting}},\ }\bibfield  {title} {\bibinfo {title} {{CPT and strings}},\
  }\href {https://doi.org/10.1016/0550-3213(91)90071-5} {\bibfield  {journal}
  {\bibinfo  {journal} {Nucl. Phys. B}\ }\textbf {\bibinfo {volume} {359}},\
  \bibinfo {pages} {545} (\bibinfo {year} {1991})}\BibitemShut {NoStop}%
\bibitem [{\citenamefont {Kosteleck\'y}\ and\ \citenamefont
  {Potting}(1995)}]{Kostelecky:1994rn}%
  \BibitemOpen
  \bibfield  {author} {\bibinfo {author} {\bibfnamefont {V.~A.}\ \bibnamefont
  {Kosteleck\'y}}\ and\ \bibinfo {author} {\bibfnamefont {R.}~\bibnamefont
  {Potting}},\ }\bibfield  {title} {\bibinfo {title} {{CPT, strings, and meson
  factories}},\ }\href {https://doi.org/10.1103/PhysRevD.51.3923} {\bibfield
  {journal} {\bibinfo  {journal} {Phys. Rev. D}\ }\textbf {\bibinfo {volume}
  {51}},\ \bibinfo {pages} {3923} (\bibinfo {year} {1995})},\ \Eprint
  {https://arxiv.org/abs/hep-ph/9501341} {arXiv:hep-ph/9501341} \BibitemShut
  {NoStop}%
\bibitem [{\citenamefont {Greenberg}(2002)}]{Greenberg:2002uu}%
  \BibitemOpen
  \bibfield  {author} {\bibinfo {author} {\bibfnamefont {O.~W.}\ \bibnamefont
  {Greenberg}},\ }\bibfield  {title} {\bibinfo {title} {{CPT Violation Implies
  Violation of Lorentz Invariance}},\ }\href
  {https://doi.org/10.1103/PhysRevLett.89.231602} {\bibfield  {journal}
  {\bibinfo  {journal} {Phys. Rev. Lett.}\ }\textbf {\bibinfo {volume} {89}},\
  \bibinfo {pages} {231602} (\bibinfo {year} {2002})},\ \Eprint
  {https://arxiv.org/abs/hep-ph/0201258} {arXiv:hep-ph/0201258} \BibitemShut
  {NoStop}%
\bibitem [{\citenamefont {Bluhm}(2006)}]{Bluhm:2005uj}%
  \BibitemOpen
  \bibfield  {author} {\bibinfo {author} {\bibfnamefont {R.}~\bibnamefont
  {Bluhm}},\ }\bibfield  {title} {\bibinfo {title} {{Overview of the Standard
  Model Extension: Implications and Phenomenology of Lorentz Violation}},\ }in\
  \href {https://doi.org/10.1007/b11758914} {\emph {\bibinfo {booktitle}
  {{Special Relativity {\textemdash} Will it Survive the Next 101 Years?}}}},\
  \bibinfo {editor} {edited by\ \bibinfo {editor} {\bibfnamefont
  {J.}~\bibnamefont {Ehlers}}\ and\ \bibinfo {editor} {\bibfnamefont
  {C.}~\bibnamefont {Lammerzahl}}}\ (\bibinfo  {publisher} {Springer},\
  \bibinfo {address} {Heidelberg, Germany},\ \bibinfo {year} {2006})\
  Chap.~\bibinfo {chapter} {3}, pp.\ \bibinfo {pages} {191--226},\ \Eprint
  {https://arxiv.org/abs/hep-ph/0506054} {arXiv:hep-ph/0506054} \BibitemShut
  {NoStop}%
\bibitem [{\citenamefont {Will}(2014)}]{Will:2014kxa}%
  \BibitemOpen
  \bibfield  {author} {\bibinfo {author} {\bibfnamefont {C.~M.}\ \bibnamefont
  {Will}},\ }\bibfield  {title} {\bibinfo {title} {{The confrontation between
  general relativity and experiment}},\ }\href
  {https://doi.org/10.12942/lrr-2014-4} {\bibfield  {journal} {\bibinfo
  {journal} {Living Rev. Rel.}\ }\textbf {\bibinfo {volume} {17}},\ \bibinfo
  {pages} {4} (\bibinfo {year} {2014})},\ \Eprint
  {https://arxiv.org/abs/1403.7377} {arXiv:1403.7377 [gr-qc]} \BibitemShut
  {NoStop}%
\bibitem [{\citenamefont {Tasson}(2014)}]{Tasson:2014dfa}%
  \BibitemOpen
  \bibfield  {author} {\bibinfo {author} {\bibfnamefont {J.~D.}\ \bibnamefont
  {Tasson}},\ }\bibfield  {title} {\bibinfo {title} {{What do we know about
  Lorentz invariance?}},\ }\href
  {https://doi.org/10.1088/0034-4885/77/6/062901} {\bibfield  {journal}
  {\bibinfo  {journal} {Rept. Prog. Phys.}\ }\textbf {\bibinfo {volume} {77}},\
  \bibinfo {pages} {062901} (\bibinfo {year} {2014})},\ \Eprint
  {https://arxiv.org/abs/1403.7785} {arXiv:1403.7785 [hep-ph]} \BibitemShut
  {NoStop}%
\bibitem [{\citenamefont {Hees}\ \emph {et~al.}(2016)\citenamefont {Hees},
  \citenamefont {Bailey}, \citenamefont {Bourgoin}, \citenamefont {Bars},
  \citenamefont {Guerlin},\ and\ \citenamefont
  {Le~Poncin-Lafitte}}]{Hees:2016lyw}%
  \BibitemOpen
  \bibfield  {author} {\bibinfo {author} {\bibfnamefont {A.}~\bibnamefont
  {Hees}}, \bibinfo {author} {\bibfnamefont {Q.~G.}\ \bibnamefont {Bailey}},
  \bibinfo {author} {\bibfnamefont {A.}~\bibnamefont {Bourgoin}}, \bibinfo
  {author} {\bibfnamefont {H.~P.-L.}\ \bibnamefont {Bars}}, \bibinfo {author}
  {\bibfnamefont {C.}~\bibnamefont {Guerlin}},\ and\ \bibinfo {author}
  {\bibfnamefont {C.}~\bibnamefont {Le~Poncin-Lafitte}},\ }\bibfield  {title}
  {\bibinfo {title} {{Tests of Lorentz symmetry in the gravitational sector}},\
  }\href {https://doi.org/10.3390/universe2040030} {\bibfield  {journal}
  {\bibinfo  {journal} {Universe}\ }\textbf {\bibinfo {volume} {2}},\ \bibinfo
  {pages} {30} (\bibinfo {year} {2016})},\ \Eprint
  {https://arxiv.org/abs/1610.04682} {arXiv:1610.04682 [gr-qc]} \BibitemShut
  {NoStop}%
\bibitem [{\citenamefont {Kosteleck\'y}\ and\ \citenamefont
  {Mewes}(2009)}]{Kostelecky:2009zp}%
  \BibitemOpen
  \bibfield  {author} {\bibinfo {author} {\bibfnamefont {V.~A.}\ \bibnamefont
  {Kosteleck\'y}}\ and\ \bibinfo {author} {\bibfnamefont {M.}~\bibnamefont
  {Mewes}},\ }\bibfield  {title} {\bibinfo {title} {{Electrodynamics with
  Lorentz-violating operators of arbitrary dimension}},\ }\href
  {https://doi.org/10.1103/PhysRevD.80.015020} {\bibfield  {journal} {\bibinfo
  {journal} {Phys. Rev. D}\ }\textbf {\bibinfo {volume} {80}},\ \bibinfo
  {pages} {015020} (\bibinfo {year} {2009})},\ \Eprint
  {https://arxiv.org/abs/0905.0031} {arXiv:0905.0031 [hep-ph]} \BibitemShut
  {NoStop}%
\bibitem [{\citenamefont {Kosteleck\'y}\ and\ \citenamefont
  {Mewes}(2012)}]{Kostelecky:2011gq}%
  \BibitemOpen
  \bibfield  {author} {\bibinfo {author} {\bibfnamefont {V.~A.}\ \bibnamefont
  {Kosteleck\'y}}\ and\ \bibinfo {author} {\bibfnamefont {M.}~\bibnamefont
  {Mewes}},\ }\bibfield  {title} {\bibinfo {title} {{Neutrinos with
  Lorentz-violating operators of arbitrary dimension}},\ }\href
  {https://doi.org/10.1103/PhysRevD.85.096005} {\bibfield  {journal} {\bibinfo
  {journal} {Phys. Rev. D}\ }\textbf {\bibinfo {volume} {85}},\ \bibinfo
  {pages} {096005} (\bibinfo {year} {2012})},\ \Eprint
  {https://arxiv.org/abs/1112.6395} {arXiv:1112.6395 [hep-ph]} \BibitemShut
  {NoStop}%
\bibitem [{\citenamefont {Kosteleck\'y}\ and\ \citenamefont
  {Mewes}(2013)}]{Kostelecky:2013rta}%
  \BibitemOpen
  \bibfield  {author} {\bibinfo {author} {\bibfnamefont {V.~A.}\ \bibnamefont
  {Kosteleck\'y}}\ and\ \bibinfo {author} {\bibfnamefont {M.}~\bibnamefont
  {Mewes}},\ }\bibfield  {title} {\bibinfo {title} {{Fermions with
  Lorentz-violating operators of arbitrary dimension}},\ }\href
  {https://doi.org/10.1103/PhysRevD.88.096006} {\bibfield  {journal} {\bibinfo
  {journal} {Phys. Rev. D}\ }\textbf {\bibinfo {volume} {88}},\ \bibinfo
  {pages} {096006} (\bibinfo {year} {2013})},\ \Eprint
  {https://arxiv.org/abs/1308.4973} {arXiv:1308.4973 [hep-ph]} \BibitemShut
  {NoStop}%
\bibitem [{\citenamefont {Kosteleck\'y}\ and\ \citenamefont
  {Li}(2019)}]{Kostelecky:2018yfa}%
  \BibitemOpen
  \bibfield  {author} {\bibinfo {author} {\bibfnamefont {V.~A.}\ \bibnamefont
  {Kosteleck\'y}}\ and\ \bibinfo {author} {\bibfnamefont {Z.}~\bibnamefont
  {Li}},\ }\bibfield  {title} {\bibinfo {title} {{Gauge field theories with
  Lorentz-violating operators of arbitrary dimension}},\ }\href
  {https://doi.org/10.1103/PhysRevD.99.056016} {\bibfield  {journal} {\bibinfo
  {journal} {Phys. Rev. D}\ }\textbf {\bibinfo {volume} {99}},\ \bibinfo
  {pages} {056016} (\bibinfo {year} {2019})},\ \Eprint
  {https://arxiv.org/abs/1812.11672} {arXiv:1812.11672 [hep-ph]} \BibitemShut
  {NoStop}%
\bibitem [{\citenamefont {Kosteleck{\'{y}}}\ and\ \citenamefont
  {Tasson}(2011)}]{Kostelecky_2011}%
  \BibitemOpen
  \bibfield  {author} {\bibinfo {author} {\bibfnamefont {V.~A.}\ \bibnamefont
  {Kosteleck{\'{y}}}}\ and\ \bibinfo {author} {\bibfnamefont {J.~D.}\
  \bibnamefont {Tasson}},\ }\bibfield  {title} {\bibinfo {title}
  {{Matter-gravity couplings and Lorentz violation}},\ }\href
  {https://doi.org/10.1103/physrevd.83.016013} {\bibfield  {journal} {\bibinfo
  {journal} {Phys. Rev. D}\ }\textbf {\bibinfo {volume} {83}},\ \bibinfo
  {pages} {016013} (\bibinfo {year} {2011})},\ \Eprint
  {https://arxiv.org/abs/1006.4106} {arXiv:1006.4106 [gr-qc]} \BibitemShut
  {NoStop}%
\bibitem [{\citenamefont {Kosteleck\'y}\ and\ \citenamefont
  {Mewes}(2018)}]{Kostelecky:2017zob}%
  \BibitemOpen
  \bibfield  {author} {\bibinfo {author} {\bibfnamefont {V.~A.}\ \bibnamefont
  {Kosteleck\'y}}\ and\ \bibinfo {author} {\bibfnamefont {M.}~\bibnamefont
  {Mewes}},\ }\bibfield  {title} {\bibinfo {title} {{Lorentz and diffeomorphism
  violations in linearized gravity}},\ }\href
  {https://doi.org/10.1016/j.physletb.2018.01.082} {\bibfield  {journal}
  {\bibinfo  {journal} {Phys. Lett. B}\ }\textbf {\bibinfo {volume} {779}},\
  \bibinfo {pages} {136} (\bibinfo {year} {2018})},\ \Eprint
  {https://arxiv.org/abs/1712.10268} {arXiv:1712.10268 [gr-qc]} \BibitemShut
  {NoStop}%
\bibitem [{\citenamefont {Kosteleck\'y}\ and\ \citenamefont
  {Li}(2021)}]{Kostelecky:2020hbb}%
  \BibitemOpen
  \bibfield  {author} {\bibinfo {author} {\bibfnamefont {V.~A.}\ \bibnamefont
  {Kosteleck\'y}}\ and\ \bibinfo {author} {\bibfnamefont {Z.}~\bibnamefont
  {Li}},\ }\bibfield  {title} {\bibinfo {title} {{Backgrounds in gravitational
  effective field theory}},\ }\href
  {https://doi.org/10.1103/PhysRevD.103.024059} {\bibfield  {journal} {\bibinfo
   {journal} {Phys. Rev. D}\ }\textbf {\bibinfo {volume} {103}},\ \bibinfo
  {pages} {024059} (\bibinfo {year} {2021})},\ \Eprint
  {https://arxiv.org/abs/2008.12206} {arXiv:2008.12206 [gr-qc]} \BibitemShut
  {NoStop}%
\bibitem [{\citenamefont {Altschul}(2006{\natexlab{a}})}]{Altschul:2006ts}%
  \BibitemOpen
  \bibfield  {author} {\bibinfo {author} {\bibfnamefont {B.}~\bibnamefont
  {Altschul}},\ }\bibfield  {title} {\bibinfo {title} {{Eliminating the CPT-odd
  $f$ coefficient from the Lorentz-violating standard model extension}},\
  }\href {https://doi.org/10.1088/0305-4470/39/44/010} {\bibfield  {journal}
  {\bibinfo  {journal} {J. Phys. A}\ }\textbf {\bibinfo {volume} {39}},\
  \bibinfo {pages} {13757} (\bibinfo {year} {2006}{\natexlab{a}})},\ \Eprint
  {https://arxiv.org/abs/hep-th/0602235} {arXiv:hep-th/0602235} \BibitemShut
  {NoStop}%
\bibitem [{\citenamefont {Kosteleck{\'{y}}}\ and\ \citenamefont
  {Russell}(2010)}]{Kostelecky_2010}%
  \BibitemOpen
  \bibfield  {author} {\bibinfo {author} {\bibfnamefont {V.~A.}\ \bibnamefont
  {Kosteleck{\'{y}}}}\ and\ \bibinfo {author} {\bibfnamefont {N.}~\bibnamefont
  {Russell}},\ }\bibfield  {title} {\bibinfo {title} {Classical kinematics for
  lorentz violation},\ }\href {https://doi.org/10.1016/j.physletb.2010.08.069}
  {\bibfield  {journal} {\bibinfo  {journal} {Physics Letters B}\ }\textbf
  {\bibinfo {volume} {693}},\ \bibinfo {pages} {443} (\bibinfo {year}
  {2010})}\BibitemShut {NoStop}%
\bibitem [{\citenamefont {Bluhm}\ \emph {et~al.}(1998)\citenamefont {Bluhm},
  \citenamefont {Kosteleck\'{y}},\ and\ \citenamefont
  {Russell}}]{Bluhm:1997qb}%
  \BibitemOpen
  \bibfield  {author} {\bibinfo {author} {\bibfnamefont {R.}~\bibnamefont
  {Bluhm}}, \bibinfo {author} {\bibfnamefont {V.~A.}\ \bibnamefont
  {Kosteleck\'{y}}},\ and\ \bibinfo {author} {\bibfnamefont {N.}~\bibnamefont
  {Russell}},\ }\bibfield  {title} {\bibinfo {title} {{CPT and Lorentz tests in
  Penning traps}},\ }\href {https://doi.org/10.1103/PhysRevD.57.3932}
  {\bibfield  {journal} {\bibinfo  {journal} {Phys. Rev. D}\ }\textbf {\bibinfo
  {volume} {57}},\ \bibinfo {pages} {3932} (\bibinfo {year} {1998})},\ \Eprint
  {https://arxiv.org/abs/hep-ph/9809543} {arXiv:hep-ph/9809543} \BibitemShut
  {NoStop}%
\bibitem [{\citenamefont {Bluhm}\ \emph {et~al.}(1999)\citenamefont {Bluhm},
  \citenamefont {Kosteleck\'{y}},\ and\ \citenamefont
  {Russell}}]{Bluhm:1998rk}%
  \BibitemOpen
  \bibfield  {author} {\bibinfo {author} {\bibfnamefont {R.}~\bibnamefont
  {Bluhm}}, \bibinfo {author} {\bibfnamefont {V.~A.}\ \bibnamefont
  {Kosteleck\'{y}}},\ and\ \bibinfo {author} {\bibfnamefont {N.}~\bibnamefont
  {Russell}},\ }\bibfield  {title} {\bibinfo {title} {{CPT and Lorentz Tests in
  Hydrogen and Antihydrogen}},\ }\href
  {https://doi.org/10.1103/PhysRevLett.82.2254} {\bibfield  {journal} {\bibinfo
   {journal} {Phys. Rev. Lett.}\ }\textbf {\bibinfo {volume} {82}},\ \bibinfo
  {pages} {2254} (\bibinfo {year} {1999})},\ \Eprint
  {https://arxiv.org/abs/hep-ph/9810269} {arXiv:hep-ph/9810269} \BibitemShut
  {NoStop}%
\bibitem [{\citenamefont {Bluhm}\ \emph {et~al.}(2002)\citenamefont {Bluhm},
  \citenamefont {Kosteleck\'{y}}, \citenamefont {Lane},\ and\ \citenamefont
  {Russell}}]{Bluhm:2001rw}%
  \BibitemOpen
  \bibfield  {author} {\bibinfo {author} {\bibfnamefont {R.}~\bibnamefont
  {Bluhm}}, \bibinfo {author} {\bibfnamefont {V.~A.}\ \bibnamefont
  {Kosteleck\'{y}}}, \bibinfo {author} {\bibfnamefont {C.~D.}\ \bibnamefont
  {Lane}},\ and\ \bibinfo {author} {\bibfnamefont {N.}~\bibnamefont
  {Russell}},\ }\bibfield  {title} {\bibinfo {title} {{Clock-Comparison Tests
  of Lorentz and CPT Symmetry in Space}},\ }\href
  {https://doi.org/10.1103/PhysRevLett.88.090801} {\bibfield  {journal}
  {\bibinfo  {journal} {Phys. Rev. Lett.}\ }\textbf {\bibinfo {volume} {88}},\
  \bibinfo {pages} {090801} (\bibinfo {year} {2002})},\ \Eprint
  {https://arxiv.org/abs/hep-ph/0111141} {arXiv:hep-ph/0111141} \BibitemShut
  {NoStop}%
\bibitem [{\citenamefont {Colladay}\ and\ \citenamefont
  {McDonald}(2002)}]{Colladay:2002eh}%
  \BibitemOpen
  \bibfield  {author} {\bibinfo {author} {\bibfnamefont {D.}~\bibnamefont
  {Colladay}}\ and\ \bibinfo {author} {\bibfnamefont {P.}~\bibnamefont
  {McDonald}},\ }\bibfield  {title} {\bibinfo {title} {{Redefining spinors in
  Lorentz-violating quantum electrodynamics}},\ }\href
  {https://doi.org/10.1063/1.1477938} {\bibfield  {journal} {\bibinfo
  {journal} {J. Math. Phys.}\ }\textbf {\bibinfo {volume} {43}},\ \bibinfo
  {pages} {3554} (\bibinfo {year} {2002})},\ \Eprint
  {https://arxiv.org/abs/hep-ph/0202066} {arXiv:hep-ph/0202066} \BibitemShut
  {NoStop}%
\bibitem [{\citenamefont {Bluhm}\ \emph {et~al.}(2003)\citenamefont {Bluhm},
  \citenamefont {Kosteleck\'{y}}, \citenamefont {Lane},\ and\ \citenamefont
  {Russell}}]{Bluhm:2003un}%
  \BibitemOpen
  \bibfield  {author} {\bibinfo {author} {\bibfnamefont {R.}~\bibnamefont
  {Bluhm}}, \bibinfo {author} {\bibfnamefont {V.~A.}\ \bibnamefont
  {Kosteleck\'{y}}}, \bibinfo {author} {\bibfnamefont {C.~D.}\ \bibnamefont
  {Lane}},\ and\ \bibinfo {author} {\bibfnamefont {N.}~\bibnamefont
  {Russell}},\ }\bibfield  {title} {\bibinfo {title} {{Probing Lorentz and CPT
  violation with space-based experiments}},\ }\href
  {https://doi.org/10.1103/PhysRevD.68.125008} {\bibfield  {journal} {\bibinfo
  {journal} {Phys. Rev. D}\ }\textbf {\bibinfo {volume} {68}},\ \bibinfo
  {pages} {125008} (\bibinfo {year} {2003})},\ \Eprint
  {https://arxiv.org/abs/hep-ph/0306190} {arXiv:hep-ph/0306190} \BibitemShut
  {NoStop}%
\bibitem [{\citenamefont {Lehnert}(2004)}]{Lehnert:2004ri}%
  \BibitemOpen
  \bibfield  {author} {\bibinfo {author} {\bibfnamefont {R.}~\bibnamefont
  {Lehnert}},\ }\bibfield  {title} {\bibinfo {title} {{Dirac theory within the
  standard-model extension}},\ }\href {https://doi.org/10.1063/1.1769105}
  {\bibfield  {journal} {\bibinfo  {journal} {J. Math. Phys.}\ }\textbf
  {\bibinfo {volume} {45}},\ \bibinfo {pages} {3399} (\bibinfo {year}
  {2004})},\ \Eprint {https://arxiv.org/abs/hep-ph/0401084}
  {arXiv:hep-ph/0401084} \BibitemShut {NoStop}%
\bibitem [{\citenamefont {Colladay}\ and\ \citenamefont
  {McDonald}(2004)}]{Colladay:2004qt}%
  \BibitemOpen
  \bibfield  {author} {\bibinfo {author} {\bibfnamefont {D.}~\bibnamefont
  {Colladay}}\ and\ \bibinfo {author} {\bibfnamefont {P.}~\bibnamefont
  {McDonald}},\ }\bibfield  {title} {\bibinfo {title} {{Statistical mechanics
  and Lorentz violation}},\ }\href {https://doi.org/10.1103/PhysRevD.70.125007}
  {\bibfield  {journal} {\bibinfo  {journal} {Phys. Rev. D}\ }\textbf {\bibinfo
  {volume} {70}},\ \bibinfo {pages} {125007} (\bibinfo {year} {2004})},\
  \Eprint {https://arxiv.org/abs/hep-ph/0407354} {arXiv:hep-ph/0407354}
  \BibitemShut {NoStop}%
\bibitem [{\citenamefont {Altschul}\ and\ \citenamefont
  {Colladay}(2005)}]{Altschul:2004wq}%
  \BibitemOpen
  \bibfield  {author} {\bibinfo {author} {\bibfnamefont {B.}~\bibnamefont
  {Altschul}}\ and\ \bibinfo {author} {\bibfnamefont {D.}~\bibnamefont
  {Colladay}},\ }\bibfield  {title} {\bibinfo {title} {{Velocity in
  Lorentz-violating fermion theories}},\ }\href
  {https://doi.org/10.1103/PhysRevD.71.125015} {\bibfield  {journal} {\bibinfo
  {journal} {Phys. Rev. D}\ }\textbf {\bibinfo {volume} {71}},\ \bibinfo
  {pages} {125015} (\bibinfo {year} {2005})},\ \Eprint
  {https://arxiv.org/abs/hep-th/0412112} {arXiv:hep-th/0412112} \BibitemShut
  {NoStop}%
\bibitem [{\citenamefont {Lane}(2005)}]{Lane:2005jv}%
  \BibitemOpen
  \bibfield  {author} {\bibinfo {author} {\bibfnamefont {C.~D.}\ \bibnamefont
  {Lane}},\ }\bibfield  {title} {\bibinfo {title} {{Probing Lorentz violation
  with Doppler-shift experiments}},\ }\href
  {https://doi.org/10.1103/PhysRevD.72.016005} {\bibfield  {journal} {\bibinfo
  {journal} {Phys. Rev. D}\ }\textbf {\bibinfo {volume} {72}},\ \bibinfo
  {pages} {016005} (\bibinfo {year} {2005})},\ \Eprint
  {https://arxiv.org/abs/hep-ph/0505130} {arXiv:hep-ph/0505130} \BibitemShut
  {NoStop}%
\bibitem [{\citenamefont {Ferreira}\ and\ \citenamefont
  {Moucherek}(2006)}]{Ferreira:2006kg}%
  \BibitemOpen
  \bibfield  {author} {\bibinfo {author} {\bibfnamefont {M.~M.}\ \bibnamefont
  {Ferreira}, \bibfnamefont {Jr.}}\ and\ \bibinfo {author} {\bibfnamefont
  {F.~M.~O.}\ \bibnamefont {Moucherek}},\ }\bibfield  {title} {\bibinfo {title}
  {{Influence of Lorentz- and CPT-violating terms on the Dirac equation}},\
  }\href {https://doi.org/10.1142/S0217751X06033842} {\bibfield  {journal}
  {\bibinfo  {journal} {Int. J. Mod. Phys. A}\ }\textbf {\bibinfo {volume}
  {21}},\ \bibinfo {pages} {6211} (\bibinfo {year} {2006})},\ \Eprint
  {https://arxiv.org/abs/hep-th/0601018} {arXiv:hep-th/0601018} \BibitemShut
  {NoStop}%
\bibitem [{\citenamefont {Altschul}(2007)}]{Altschul:2006uu}%
  \BibitemOpen
  \bibfield  {author} {\bibinfo {author} {\bibfnamefont {B.}~\bibnamefont
  {Altschul}},\ }\bibfield  {title} {\bibinfo {title} {{Limits on neutron
  Lorentz violation from pulsar timing}},\ }\href
  {https://doi.org/10.1103/PhysRevD.75.023001} {\bibfield  {journal} {\bibinfo
  {journal} {Phys. Rev. D}\ }\textbf {\bibinfo {volume} {75}},\ \bibinfo
  {pages} {023001} (\bibinfo {year} {2007})},\ \Eprint
  {https://arxiv.org/abs/hep-ph/0608094} {arXiv:hep-ph/0608094} \BibitemShut
  {NoStop}%
\bibitem [{\citenamefont {Lehnert}(2006)}]{Lehnert:2006id}%
  \BibitemOpen
  \bibfield  {author} {\bibinfo {author} {\bibfnamefont {R.}~\bibnamefont
  {Lehnert}},\ }\bibfield  {title} {\bibinfo {title} {{Nonlocal on-shell field
  redefinition for the standard-model extension}},\ }\href
  {https://doi.org/10.1103/PhysRevD.74.125001} {\bibfield  {journal} {\bibinfo
  {journal} {Phys. Rev. D}\ }\textbf {\bibinfo {volume} {74}},\ \bibinfo
  {pages} {125001} (\bibinfo {year} {2006})},\ \Eprint
  {https://arxiv.org/abs/hep-th/0609162} {arXiv:hep-th/0609162} \BibitemShut
  {NoStop}%
\bibitem [{\citenamefont {Ferreira}\ \emph {et~al.}(2007)\citenamefont
  {Ferreira}, \citenamefont {Gomes},\ and\ \citenamefont
  {Lopes}}]{Ferreira:2007za}%
  \BibitemOpen
  \bibfield  {author} {\bibinfo {author} {\bibfnamefont {M.~M.}\ \bibnamefont
  {Ferreira}, \bibfnamefont {Jr.}}, \bibinfo {author} {\bibfnamefont {A.~R.}\
  \bibnamefont {Gomes}},\ and\ \bibinfo {author} {\bibfnamefont {R.~C.~C.}\
  \bibnamefont {Lopes}},\ }\bibfield  {title} {\bibinfo {title} {{Influence of
  Lorentz-violating terms on a two-level system}},\ }\href
  {https://doi.org/10.1103/PhysRevD.76.105031} {\bibfield  {journal} {\bibinfo
  {journal} {Phys. Rev. D}\ }\textbf {\bibinfo {volume} {76}},\ \bibinfo
  {pages} {105031} (\bibinfo {year} {2007})},\ \Eprint
  {https://arxiv.org/abs/0707.4660} {arXiv:0707.4660 [hep-th]} \BibitemShut
  {NoStop}%
\bibitem [{\citenamefont {Ferreira}\ and\ \citenamefont
  {Moucherek}(2007)}]{Ferreira:2007gnn}%
  \BibitemOpen
  \bibfield  {author} {\bibinfo {author} {\bibfnamefont {M.~M.}\ \bibnamefont
  {Ferreira}, \bibfnamefont {Jr.}}\ and\ \bibinfo {author} {\bibfnamefont
  {F.~M.~O.}\ \bibnamefont {Moucherek}},\ }\bibfield  {title} {\bibinfo {title}
  {{Influence of Lorentz violation on the hydrogen spectrum}},\ }\href
  {https://doi.org/10.1016/j.nuclphysa.2007.03.108} {\bibfield  {journal}
  {\bibinfo  {journal} {Nucl. Phys. A}\ }\textbf {\bibinfo {volume} {790}},\
  \bibinfo {pages} {635c} (\bibinfo {year} {2007})}\BibitemShut {NoStop}%
\bibitem [{\citenamefont {Altschul}(2008)}]{Altschul:2008qg}%
  \BibitemOpen
  \bibfield  {author} {\bibinfo {author} {\bibfnamefont {B.}~\bibnamefont
  {Altschul}},\ }\bibfield  {title} {\bibinfo {title} {{Limits on neutron
  Lorentz violation from the stability of primary cosmic ray protons}},\ }\href
  {https://doi.org/10.1103/PhysRevD.78.085018} {\bibfield  {journal} {\bibinfo
  {journal} {Phys. Rev. D}\ }\textbf {\bibinfo {volume} {78}},\ \bibinfo
  {pages} {085018} (\bibinfo {year} {2008})},\ \Eprint
  {https://arxiv.org/abs/0805.0781} {arXiv:0805.0781 [hep-ph]} \BibitemShut
  {NoStop}%
\bibitem [{\citenamefont {Altschul}(2009)}]{Altschul:2008ki}%
  \BibitemOpen
  \bibfield  {author} {\bibinfo {author} {\bibfnamefont {B.}~\bibnamefont
  {Altschul}},\ }\bibfield  {title} {\bibinfo {title} {{Lorentz violation and
  $\alpha$ decay}},\ }\href {https://doi.org/10.1103/PhysRevD.79.016004}
  {\bibfield  {journal} {\bibinfo  {journal} {Phys. Rev. D}\ }\textbf {\bibinfo
  {volume} {79}},\ \bibinfo {pages} {016004} (\bibinfo {year} {2009})},\
  \Eprint {https://arxiv.org/abs/0812.2236} {arXiv:0812.2236 [hep-ph]}
  \BibitemShut {NoStop}%
\bibitem [{\citenamefont {Colladay}\ \emph {et~al.}(2010)\citenamefont
  {Colladay}, \citenamefont {McDonald},\ and\ \citenamefont
  {Mullins}}]{Colladay:2010ae}%
  \BibitemOpen
  \bibfield  {author} {\bibinfo {author} {\bibfnamefont {D.}~\bibnamefont
  {Colladay}}, \bibinfo {author} {\bibfnamefont {P.}~\bibnamefont {McDonald}},\
  and\ \bibinfo {author} {\bibfnamefont {D.}~\bibnamefont {Mullins}},\
  }\bibfield  {title} {\bibinfo {title} {{Factoring the dispersion relation in
  the presence of Lorentz violation}},\ }\href
  {https://doi.org/10.1088/1751-8113/43/27/275202} {\bibfield  {journal}
  {\bibinfo  {journal} {J. Phys. A}\ }\textbf {\bibinfo {volume} {43}},\
  \bibinfo {pages} {275202} (\bibinfo {year} {2010})},\ \Eprint
  {https://arxiv.org/abs/1001.3839} {arXiv:1001.3839 [hep-ph]} \BibitemShut
  {NoStop}%
\bibitem [{\citenamefont {Altschul}(2010)}]{Altschul:2010na}%
  \BibitemOpen
  \bibfield  {author} {\bibinfo {author} {\bibfnamefont {B.}~\bibnamefont
  {Altschul}},\ }\bibfield  {title} {\bibinfo {title} {{Laboratory bounds on
  electron Lorentz violation}},\ }\href
  {https://doi.org/10.1103/PhysRevD.82.016002} {\bibfield  {journal} {\bibinfo
  {journal} {Phys. Rev. D}\ }\textbf {\bibinfo {volume} {82}},\ \bibinfo
  {pages} {016002} (\bibinfo {year} {2010})},\ \Eprint
  {https://arxiv.org/abs/1005.2994} {arXiv:1005.2994 [hep-ph]} \BibitemShut
  {NoStop}%
\bibitem [{\citenamefont {Bocquet}\ \emph {et~al.}(2010)\citenamefont {Bocquet}
  \emph {et~al.}}]{Bocquet:2010ke}%
  \BibitemOpen
  \bibfield  {author} {\bibinfo {author} {\bibfnamefont {J.-P.}\ \bibnamefont
  {Bocquet}} \emph {et~al.},\ }\bibfield  {title} {\bibinfo {title} {{Limits on
  Light-Speed Anisotropies from Compton Scattering of High-Energy Electrons}},\
  }\href {https://doi.org/10.1103/PhysRevLett.104.241601} {\bibfield  {journal}
  {\bibinfo  {journal} {Phys. Rev. Lett.}\ }\textbf {\bibinfo {volume} {104}},\
  \bibinfo {pages} {241601} (\bibinfo {year} {2010})},\ \Eprint
  {https://arxiv.org/abs/1005.5230} {arXiv:1005.5230 [hep-ex]} \BibitemShut
  {NoStop}%
\bibitem [{\citenamefont {Fittante}\ and\ \citenamefont
  {Russell}(2012)}]{Fittante:2012ua}%
  \BibitemOpen
  \bibfield  {author} {\bibinfo {author} {\bibfnamefont {A.}~\bibnamefont
  {Fittante}}\ and\ \bibinfo {author} {\bibfnamefont {N.}~\bibnamefont
  {Russell}},\ }\bibfield  {title} {\bibinfo {title} {{Fermion observables for
  Lorentz violation}},\ }\href {https://doi.org/10.1088/0954-3899/39/12/125004}
  {\bibfield  {journal} {\bibinfo  {journal} {J. Phys. G}\ }\textbf {\bibinfo
  {volume} {39}},\ \bibinfo {pages} {125004} (\bibinfo {year} {2012})},\
  \Eprint {https://arxiv.org/abs/1210.2003} {arXiv:1210.2003 [hep-ph]}
  \BibitemShut {NoStop}%
\bibitem [{\citenamefont {Noordmans}\ \emph {et~al.}(2016)\citenamefont
  {Noordmans}, \citenamefont {Onderwater}, \citenamefont {Wilschut},\ and\
  \citenamefont {Timmermans}}]{Noordmans:2014hxa}%
  \BibitemOpen
  \bibfield  {author} {\bibinfo {author} {\bibfnamefont {J.~P.}\ \bibnamefont
  {Noordmans}}, \bibinfo {author} {\bibfnamefont {C.~J.~G.}\ \bibnamefont
  {Onderwater}}, \bibinfo {author} {\bibfnamefont {H.~W.}\ \bibnamefont
  {Wilschut}},\ and\ \bibinfo {author} {\bibfnamefont {R.~G.~E.}\ \bibnamefont
  {Timmermans}},\ }\bibfield  {title} {\bibinfo {title} {{Question of Lorentz
  violation in muon decay}},\ }\href
  {https://doi.org/10.1103/PhysRevD.93.116001} {\bibfield  {journal} {\bibinfo
  {journal} {Phys. Rev. D}\ }\textbf {\bibinfo {volume} {93}},\ \bibinfo
  {pages} {116001} (\bibinfo {year} {2016})},\ \Eprint
  {https://arxiv.org/abs/1412.3257} {arXiv:1412.3257 [hep-ph]} \BibitemShut
  {NoStop}%
\bibitem [{\citenamefont {Schreck}(2017)}]{Schreck:2017isa}%
  \BibitemOpen
  \bibfield  {author} {\bibinfo {author} {\bibfnamefont {M.}~\bibnamefont
  {Schreck}},\ }\bibfield  {title} {\bibinfo {title} {{Vacuum Cherenkov
  radiation for Lorentz-violating fermions}},\ }\href
  {https://doi.org/10.1103/PhysRevD.96.095026} {\bibfield  {journal} {\bibinfo
  {journal} {Phys. Rev. D}\ }\textbf {\bibinfo {volume} {96}},\ \bibinfo
  {pages} {095026} (\bibinfo {year} {2017})},\ \Eprint
  {https://arxiv.org/abs/1702.03171} {arXiv:1702.03171 [hep-ph]} \BibitemShut
  {NoStop}%
\bibitem [{\citenamefont {Aghababaei}\ \emph {et~al.}(2017)\citenamefont
  {Aghababaei}, \citenamefont {Haghighat},\ and\ \citenamefont
  {Motie}}]{Aghababaei:2017bei}%
  \BibitemOpen
  \bibfield  {author} {\bibinfo {author} {\bibfnamefont {S.}~\bibnamefont
  {Aghababaei}}, \bibinfo {author} {\bibfnamefont {M.}~\bibnamefont
  {Haghighat}},\ and\ \bibinfo {author} {\bibfnamefont {I.}~\bibnamefont
  {Motie}},\ }\bibfield  {title} {\bibinfo {title} {{Muon anomalous magnetic
  moment in the standard model extension}},\ }\href
  {https://doi.org/10.1103/PhysRevD.96.115028} {\bibfield  {journal} {\bibinfo
  {journal} {Phys. Rev. D}\ }\textbf {\bibinfo {volume} {96}},\ \bibinfo
  {pages} {115028} (\bibinfo {year} {2017})},\ \Eprint
  {https://arxiv.org/abs/1712.09028} {arXiv:1712.09028 [hep-ph]} \BibitemShut
  {NoStop}%
\bibitem [{\citenamefont {Escobar}\ \emph {et~al.}(2018)\citenamefont
  {Escobar}, \citenamefont {Noordmans},\ and\ \citenamefont
  {Potting}}]{Escobar:2018hyo}%
  \BibitemOpen
  \bibfield  {author} {\bibinfo {author} {\bibfnamefont {C.~A.}\ \bibnamefont
  {Escobar}}, \bibinfo {author} {\bibfnamefont {J.~P.}\ \bibnamefont
  {Noordmans}},\ and\ \bibinfo {author} {\bibfnamefont {R.}~\bibnamefont
  {Potting}},\ }\bibfield  {title} {\bibinfo {title} {{Cosmic-ray fermion decay
  through tau-antitau emission with Lorentz violation}},\ }\href
  {https://doi.org/10.1103/PhysRevD.97.115030} {\bibfield  {journal} {\bibinfo
  {journal} {Phys. Rev. D}\ }\textbf {\bibinfo {volume} {97}},\ \bibinfo
  {pages} {115030} (\bibinfo {year} {2018})},\ \Eprint
  {https://arxiv.org/abs/1804.07586} {arXiv:1804.07586 [hep-ph]} \BibitemShut
  {NoStop}%
\bibitem [{\citenamefont {Shao}(2019)}]{Shao:2019tle}%
  \BibitemOpen
  \bibfield  {author} {\bibinfo {author} {\bibfnamefont {L.}~\bibnamefont
  {Shao}},\ }\bibfield  {title} {\bibinfo {title} {{Lorentz-violating
  matter-gravity couplings in small-eccentricity binary pulsars}},\ }\href
  {https://doi.org/10.3390/sym11091098} {\bibfield  {journal} {\bibinfo
  {journal} {Symmetry}\ }\textbf {\bibinfo {volume} {11}},\ \bibinfo {pages}
  {1098} (\bibinfo {year} {2019})},\ \Eprint {https://arxiv.org/abs/1908.10019}
  {arXiv:1908.10019 [hep-ph]} \BibitemShut {NoStop}%
\bibitem [{\citenamefont {Colladay}\ and\ \citenamefont
  {Kosteleck\'{y}}(2001)}]{Colladay:2001wk}%
  \BibitemOpen
  \bibfield  {author} {\bibinfo {author} {\bibfnamefont {D.}~\bibnamefont
  {Colladay}}\ and\ \bibinfo {author} {\bibfnamefont {V.~A.}\ \bibnamefont
  {Kosteleck\'{y}}},\ }\bibfield  {title} {\bibinfo {title} {{Cross sections
  and lorentz violation}},\ }\href
  {https://doi.org/10.1016/S0370-2693(01)00649-9} {\bibfield  {journal}
  {\bibinfo  {journal} {Phys. Lett. B}\ }\textbf {\bibinfo {volume} {511}},\
  \bibinfo {pages} {209} (\bibinfo {year} {2001})},\ \Eprint
  {https://arxiv.org/abs/hep-ph/0104300} {arXiv:hep-ph/0104300} \BibitemShut
  {NoStop}%
\bibitem [{\citenamefont {Kosteleck\'{y}}\ \emph {et~al.}(2002)\citenamefont
  {Kosteleck\'{y}}, \citenamefont {Lane},\ and\ \citenamefont
  {Pickering}}]{Kostelecky:2001jc}%
  \BibitemOpen
  \bibfield  {author} {\bibinfo {author} {\bibfnamefont {V.~A.}\ \bibnamefont
  {Kosteleck\'{y}}}, \bibinfo {author} {\bibfnamefont {C.~D.}\ \bibnamefont
  {Lane}},\ and\ \bibinfo {author} {\bibfnamefont {A.~G.~M.}\ \bibnamefont
  {Pickering}},\ }\bibfield  {title} {\bibinfo {title} {{One-loop
  renormalization of Lorentz-violating electrodynamics}},\ }\href
  {https://doi.org/10.1103/PhysRevD.65.056006} {\bibfield  {journal} {\bibinfo
  {journal} {Phys. Rev. D}\ }\textbf {\bibinfo {volume} {65}},\ \bibinfo
  {pages} {056006} (\bibinfo {year} {2002})},\ \Eprint
  {https://arxiv.org/abs/hep-th/0111123} {arXiv:hep-th/0111123} \BibitemShut
  {NoStop}%
\bibitem [{\citenamefont {Kosteleck\'{y}}\ and\ \citenamefont
  {Pickering}(2003)}]{Kostelecky:2002ue}%
  \BibitemOpen
  \bibfield  {author} {\bibinfo {author} {\bibfnamefont {V.~A.}\ \bibnamefont
  {Kosteleck\'{y}}}\ and\ \bibinfo {author} {\bibfnamefont {A.~G.~M.}\
  \bibnamefont {Pickering}},\ }\bibfield  {title} {\bibinfo {title} {{Vacuum
  Photon Splitting in Lorentz-Violating Quantum Electrodynamics}},\ }\href
  {https://doi.org/10.1103/PhysRevLett.91.031801} {\bibfield  {journal}
  {\bibinfo  {journal} {Phys. Rev. Lett.}\ }\textbf {\bibinfo {volume} {91}},\
  \bibinfo {pages} {031801} (\bibinfo {year} {2003})},\ \Eprint
  {https://arxiv.org/abs/hep-ph/0212382} {arXiv:hep-ph/0212382} \BibitemShut
  {NoStop}%
\bibitem [{\citenamefont {Altschul}(2005)}]{Altschul:2005za}%
  \BibitemOpen
  \bibfield  {author} {\bibinfo {author} {\bibfnamefont {B.}~\bibnamefont
  {Altschul}},\ }\bibfield  {title} {\bibinfo {title} {{Lorentz violation and
  synchrotron radiation}},\ }\href {https://doi.org/10.1103/PhysRevD.72.085003}
  {\bibfield  {journal} {\bibinfo  {journal} {Phys. Rev. D}\ }\textbf {\bibinfo
  {volume} {72}},\ \bibinfo {pages} {085003} (\bibinfo {year} {2005})},\
  \Eprint {https://arxiv.org/abs/hep-th/0507258} {arXiv:hep-th/0507258}
  \BibitemShut {NoStop}%
\bibitem [{\citenamefont {Altschul}(2006{\natexlab{b}})}]{Altschul:2006pv}%
  \BibitemOpen
  \bibfield  {author} {\bibinfo {author} {\bibfnamefont {B.}~\bibnamefont
  {Altschul}},\ }\bibfield  {title} {\bibinfo {title} {{Synchrotron and inverse
  Compton constraints on Lorentz violations for electrons}},\ }\href
  {https://doi.org/10.1103/PhysRevD.74.083003} {\bibfield  {journal} {\bibinfo
  {journal} {Phys. Rev. D}\ }\textbf {\bibinfo {volume} {74}},\ \bibinfo
  {pages} {083003} (\bibinfo {year} {2006}{\natexlab{b}})},\ \Eprint
  {https://arxiv.org/abs/hep-ph/0608332} {arXiv:hep-ph/0608332} \BibitemShut
  {NoStop}%
\bibitem [{\citenamefont {Nascimento}\ \emph {et~al.}(2007)\citenamefont
  {Nascimento}, \citenamefont {Passos}, \citenamefont {Petrov},\ and\
  \citenamefont {Brito}}]{Nascimento:2007rb}%
  \BibitemOpen
  \bibfield  {author} {\bibinfo {author} {\bibfnamefont {J.~R.}\ \bibnamefont
  {Nascimento}}, \bibinfo {author} {\bibfnamefont {E.}~\bibnamefont {Passos}},
  \bibinfo {author} {\bibfnamefont {A.~Y.}\ \bibnamefont {Petrov}},\ and\
  \bibinfo {author} {\bibfnamefont {F.~A.}\ \bibnamefont {Brito}},\ }\bibfield
  {title} {\bibinfo {title} {{Lorentz-CPT violation, radiative corrections and
  finite temperature}},\ }\href {https://doi.org/10.1088/1126-6708/2007/06/016}
  {\bibfield  {journal} {\bibinfo  {journal} {{JHEP}}\ }\textbf {\bibinfo
  {volume} {06}},\ \bibinfo {pages} {016} (\bibinfo {year} {2007})},\ \Eprint
  {https://arxiv.org/abs/0705.1338} {arXiv:0705.1338 [hep-th]} \BibitemShut
  {NoStop}%
\bibitem [{\citenamefont {Cambiaso}\ \emph {et~al.}(2014)\citenamefont
  {Cambiaso}, \citenamefont {Lehnert},\ and\ \citenamefont
  {Potting}}]{Cambiaso:2014eba}%
  \BibitemOpen
  \bibfield  {author} {\bibinfo {author} {\bibfnamefont {M.}~\bibnamefont
  {Cambiaso}}, \bibinfo {author} {\bibfnamefont {R.}~\bibnamefont {Lehnert}},\
  and\ \bibinfo {author} {\bibfnamefont {R.}~\bibnamefont {Potting}},\
  }\bibfield  {title} {\bibinfo {title} {{Asymptotic states and renormalization
  in Lorentz-violating quantum field theory}},\ }\href
  {https://doi.org/10.1103/PhysRevD.90.065003} {\bibfield  {journal} {\bibinfo
  {journal} {Phys. Rev. D}\ }\textbf {\bibinfo {volume} {90}},\ \bibinfo
  {pages} {065003} (\bibinfo {year} {2014})},\ \Eprint
  {https://arxiv.org/abs/1401.7317} {arXiv:1401.7317 [hep-th]} \BibitemShut
  {NoStop}%
\bibitem [{\citenamefont {Gomes}\ \emph {et~al.}(2014)\citenamefont {Gomes},
  \citenamefont {Kosteleck\'y},\ and\ \citenamefont {Vargas}}]{Gomes:2014kaa}%
  \BibitemOpen
  \bibfield  {author} {\bibinfo {author} {\bibfnamefont {A.~H.}\ \bibnamefont
  {Gomes}}, \bibinfo {author} {\bibfnamefont {A.}~\bibnamefont
  {Kosteleck\'y}},\ and\ \bibinfo {author} {\bibfnamefont {A.~J.}\ \bibnamefont
  {Vargas}},\ }\bibfield  {title} {\bibinfo {title} {{Laboratory tests of
  Lorentz and CPT symmetry with muons}},\ }\href
  {https://doi.org/10.1103/PhysRevD.90.076009} {\bibfield  {journal} {\bibinfo
  {journal} {Phys. Rev. D}\ }\textbf {\bibinfo {volume} {90}},\ \bibinfo
  {pages} {076009} (\bibinfo {year} {2014})},\ \Eprint
  {https://arxiv.org/abs/1407.7748} {arXiv:1407.7748 [hep-ph]} \BibitemShut
  {NoStop}%
\bibitem [{\citenamefont {S.~Santos}\ and\ \citenamefont
  {Sobreiro}(2016)}]{SSantos:2015mzs}%
  \BibitemOpen
  \bibfield  {author} {\bibinfo {author} {\bibfnamefont {T.~R.}\ \bibnamefont
  {S.~Santos}}\ and\ \bibinfo {author} {\bibfnamefont {R.~F.}\ \bibnamefont
  {Sobreiro}},\ }\bibfield  {title} {\bibinfo {title} {{Remarks on the
  renormalization properties of Lorentz- and CPT-violating quantum
  electrodynamics}},\ }\href {https://doi.org/10.1007/s13538-016-0423-6}
  {\bibfield  {journal} {\bibinfo  {journal} {Braz. J. Phys.}\ }\textbf
  {\bibinfo {volume} {46}},\ \bibinfo {pages} {437} (\bibinfo {year} {2016})},\
  \Eprint {https://arxiv.org/abs/1502.06881} {arXiv:1502.06881 [hep-th]}
  \BibitemShut {NoStop}%
\bibitem [{\citenamefont {Kosteleck\'y}\ and\ \citenamefont
  {Vargas}(2015)}]{Kostelecky:2015nma}%
  \BibitemOpen
  \bibfield  {author} {\bibinfo {author} {\bibfnamefont {V.~A.}\ \bibnamefont
  {Kosteleck\'y}}\ and\ \bibinfo {author} {\bibfnamefont {A.~J.}\ \bibnamefont
  {Vargas}},\ }\bibfield  {title} {\bibinfo {title} {{Lorentz and CPT tests
  with hydrogen, antihydrogen, and related systems}},\ }\href
  {https://doi.org/10.1103/PhysRevD.92.056002} {\bibfield  {journal} {\bibinfo
  {journal} {Phys. Rev. D}\ }\textbf {\bibinfo {volume} {92}},\ \bibinfo
  {pages} {056002} (\bibinfo {year} {2015})},\ \Eprint
  {https://arxiv.org/abs/1506.01706} {arXiv:1506.01706 [hep-ph]} \BibitemShut
  {NoStop}%
\bibitem [{\citenamefont {Mariz}\ \emph {et~al.}(2018)\citenamefont {Mariz},
  \citenamefont {Maluf}, \citenamefont {Nascimento},\ and\ \citenamefont
  {Petrov}}]{Mariz:2016ooa}%
  \BibitemOpen
  \bibfield  {author} {\bibinfo {author} {\bibfnamefont {T.}~\bibnamefont
  {Mariz}}, \bibinfo {author} {\bibfnamefont {R.~V.}\ \bibnamefont {Maluf}},
  \bibinfo {author} {\bibfnamefont {J.~R.}\ \bibnamefont {Nascimento}},\ and\
  \bibinfo {author} {\bibfnamefont {A.~Y.}\ \bibnamefont {Petrov}},\ }\bibfield
   {title} {\bibinfo {title} {{On one-loop corrections to the CPT-even
  Lorentz-breaking extension of QED}},\ }\href
  {https://doi.org/10.1142/S0217751X18500185} {\bibfield  {journal} {\bibinfo
  {journal} {Int. J. Mod. Phys. A}\ }\textbf {\bibinfo {volume} {33}},\
  \bibinfo {pages} {1850018} (\bibinfo {year} {2018})},\ \Eprint
  {https://arxiv.org/abs/1604.06647} {arXiv:1604.06647 [hep-th]} \BibitemShut
  {NoStop}%
\bibitem [{\citenamefont {Ding}\ and\ \citenamefont
  {Kosteleck\'y}(2016)}]{Ding:2016lwt}%
  \BibitemOpen
  \bibfield  {author} {\bibinfo {author} {\bibfnamefont {Y.}~\bibnamefont
  {Ding}}\ and\ \bibinfo {author} {\bibfnamefont {V.~A.}\ \bibnamefont
  {Kosteleck\'y}},\ }\bibfield  {title} {\bibinfo {title} {{Lorentz-violating
  spinor electrodynamics and Penning traps}},\ }\href
  {https://doi.org/10.1103/PhysRevD.94.056008} {\bibfield  {journal} {\bibinfo
  {journal} {Phys. Rev. D}\ }\textbf {\bibinfo {volume} {94}},\ \bibinfo
  {pages} {056008} (\bibinfo {year} {2016})},\ \Eprint
  {https://arxiv.org/abs/1608.07868} {arXiv:1608.07868 [hep-ph]} \BibitemShut
  {NoStop}%
\bibitem [{\citenamefont {Reis}\ and\ \citenamefont
  {Schreck}(2017)}]{Reis:2016hzu}%
  \BibitemOpen
  \bibfield  {author} {\bibinfo {author} {\bibfnamefont {J.~A. A.~S.}\
  \bibnamefont {Reis}}\ and\ \bibinfo {author} {\bibfnamefont {M.}~\bibnamefont
  {Schreck}},\ }\bibfield  {title} {\bibinfo {title} {{Lorentz-violating
  modification of Dirac theory based on spin-nondegenerate operators}},\ }\href
  {https://doi.org/10.1103/PhysRevD.95.075016} {\bibfield  {journal} {\bibinfo
  {journal} {Phys. Rev. D}\ }\textbf {\bibinfo {volume} {95}},\ \bibinfo
  {pages} {075016} (\bibinfo {year} {2017})},\ \Eprint
  {https://arxiv.org/abs/1612.06221} {arXiv:1612.06221 [hep-th]} \BibitemShut
  {NoStop}%
\bibitem [{\citenamefont {Kosteleck\'y}\ and\ \citenamefont
  {Vargas}(2018)}]{Kostelecky:2018fmc}%
  \BibitemOpen
  \bibfield  {author} {\bibinfo {author} {\bibfnamefont {V.~A.}\ \bibnamefont
  {Kosteleck\'y}}\ and\ \bibinfo {author} {\bibfnamefont {A.~J.}\ \bibnamefont
  {Vargas}},\ }\bibfield  {title} {\bibinfo {title} {{Lorentz and CPT tests
  with clock-comparison experiments}},\ }\href
  {https://doi.org/10.1103/PhysRevD.98.036003} {\bibfield  {journal} {\bibinfo
  {journal} {Phys. Rev. D}\ }\textbf {\bibinfo {volume} {98}},\ \bibinfo
  {pages} {036003} (\bibinfo {year} {2018})},\ \Eprint
  {https://arxiv.org/abs/1805.04499} {arXiv:1805.04499 [hep-ph]} \BibitemShut
  {NoStop}%
\bibitem [{\citenamefont {Ba\^{e}ta~Scarpelli}\ \emph
  {et~al.}(2018)\citenamefont {Ba\^{e}ta~Scarpelli}, \citenamefont {Brito},
  \citenamefont {Felipe}, \citenamefont {Nascimento},\ and\ \citenamefont
  {Petrov}}]{BaetaScarpelli:2018vsm}%
  \BibitemOpen
  \bibfield  {author} {\bibinfo {author} {\bibfnamefont {A.~P.}\ \bibnamefont
  {Ba\^{e}ta~Scarpelli}}, \bibinfo {author} {\bibfnamefont {L.~C.~T.}\
  \bibnamefont {Brito}}, \bibinfo {author} {\bibfnamefont {J.~C.~C.}\
  \bibnamefont {Felipe}}, \bibinfo {author} {\bibfnamefont {J.~R.}\
  \bibnamefont {Nascimento}},\ and\ \bibinfo {author} {\bibfnamefont {A.~Y.}\
  \bibnamefont {Petrov}},\ }\bibfield  {title} {\bibinfo {title} {{Higher-order
  one-loop contributions in Lorentz-breaking QED}},\ }\href
  {https://doi.org/10.1209/0295-5075/123/21001} {\bibfield  {journal} {\bibinfo
   {journal} {EPL}\ }\textbf {\bibinfo {volume} {123}},\ \bibinfo {pages}
  {21001} (\bibinfo {year} {2018})},\ \Eprint
  {https://arxiv.org/abs/1805.06256} {arXiv:1805.06256 [hep-ph]} \BibitemShut
  {NoStop}%
\bibitem [{\citenamefont {Ferrari}\ \emph {et~al.}(2020)\citenamefont
  {Ferrari}, \citenamefont {Nascimento},\ and\ \citenamefont
  {Petrov}}]{Ferrari:2018tps}%
  \BibitemOpen
  \bibfield  {author} {\bibinfo {author} {\bibfnamefont {A.~F.}\ \bibnamefont
  {Ferrari}}, \bibinfo {author} {\bibfnamefont {J.~R.}\ \bibnamefont
  {Nascimento}},\ and\ \bibinfo {author} {\bibfnamefont {A.~Y.}\ \bibnamefont
  {Petrov}},\ }\bibfield  {title} {\bibinfo {title} {{Radiative corrections and
  Lorentz violation}},\ }\href {https://doi.org/10.1140/epjc/s10052-020-8000-0}
  {\bibfield  {journal} {\bibinfo  {journal} {Eur. Phys. J. C}\ }\textbf
  {\bibinfo {volume} {80}},\ \bibinfo {pages} {459} (\bibinfo {year} {2020})},\
  \Eprint {https://arxiv.org/abs/1812.01702} {arXiv:1812.01702 [hep-th]}
  \BibitemShut {NoStop}%
\bibitem [{\citenamefont {Reis}\ and\ \citenamefont
  {Schreck}(2019)}]{Reis:2019jmm}%
  \BibitemOpen
  \bibfield  {author} {\bibinfo {author} {\bibfnamefont {J.~A. A.~S.}\
  \bibnamefont {Reis}}\ and\ \bibinfo {author} {\bibfnamefont {M.}~\bibnamefont
  {Schreck}},\ }\bibfield  {title} {\bibinfo {title} {{Formal developments for
  Lorentz-violating Dirac fermions and neutrinos}},\ }\href
  {https://doi.org/10.3390/sym11101197} {\bibfield  {journal} {\bibinfo
  {journal} {Symmetry}\ }\textbf {\bibinfo {volume} {11}},\ \bibinfo {pages}
  {1197} (\bibinfo {year} {2019})},\ \Eprint {https://arxiv.org/abs/1909.11061}
  {arXiv:1909.11061 [hep-th]} \BibitemShut {NoStop}%
\bibitem [{\citenamefont {Brito}\ \emph {et~al.}(2020)\citenamefont {Brito},
  \citenamefont {Felipe}, \citenamefont {Nascimento}, \citenamefont {Petrov},\
  and\ \citenamefont {Ba\^eta~Scarpelli}}]{Brito:2020eiy}%
  \BibitemOpen
  \bibfield  {author} {\bibinfo {author} {\bibfnamefont {L.~C.~T.}\
  \bibnamefont {Brito}}, \bibinfo {author} {\bibfnamefont {J.~C.~C.}\
  \bibnamefont {Felipe}}, \bibinfo {author} {\bibfnamefont {J.~R.}\
  \bibnamefont {Nascimento}}, \bibinfo {author} {\bibfnamefont {A.~Y.}\
  \bibnamefont {Petrov}},\ and\ \bibinfo {author} {\bibfnamefont {A.~P.}\
  \bibnamefont {Ba\^eta~Scarpelli}},\ }\bibfield  {title} {\bibinfo {title}
  {{Higher-order one-loop renormalization in the spinor sector of minimal
  Lorentz-violating extended QED}},\ }\href
  {https://doi.org/10.1103/PhysRevD.102.075017} {\bibfield  {journal} {\bibinfo
   {journal} {Phys. Rev. D}\ }\textbf {\bibinfo {volume} {102}},\ \bibinfo
  {pages} {075017} (\bibinfo {year} {2020})},\ \Eprint
  {https://arxiv.org/abs/2007.11538} {arXiv:2007.11538 [hep-th]} \BibitemShut
  {NoStop}%
\bibitem [{\citenamefont {Ding}\ and\ \citenamefont
  {Rawnak}(2020)}]{Ding:2020aew}%
  \BibitemOpen
  \bibfield  {author} {\bibinfo {author} {\bibfnamefont {Y.}~\bibnamefont
  {Ding}}\ and\ \bibinfo {author} {\bibfnamefont {M.~F.}\ \bibnamefont
  {Rawnak}},\ }\bibfield  {title} {\bibinfo {title} {{Lorentz and CPT tests
  with charge-to-mass ratio comparisons in Penning traps}},\ }\href
  {https://doi.org/10.1103/PhysRevD.102.056009} {\bibfield  {journal} {\bibinfo
   {journal} {Phys. Rev. D}\ }\textbf {\bibinfo {volume} {102}},\ \bibinfo
  {pages} {056009} (\bibinfo {year} {2020})},\ \Eprint
  {https://arxiv.org/abs/2008.08484} {arXiv:2008.08484 [hep-ph]} \BibitemShut
  {NoStop}%
\bibitem [{\citenamefont {Ferrari}\ \emph {et~al.}(2021)\citenamefont
  {Ferrari}, \citenamefont {Furtado}, \citenamefont {Assun\c{c}\~ao},
  \citenamefont {Mariz},\ and\ \citenamefont {Petrov}}]{Ferrari:2021eam}%
  \BibitemOpen
  \bibfield  {author} {\bibinfo {author} {\bibfnamefont {A.~F.}\ \bibnamefont
  {Ferrari}}, \bibinfo {author} {\bibfnamefont {J.}~\bibnamefont {Furtado}},
  \bibinfo {author} {\bibfnamefont {J.~F.}\ \bibnamefont {Assun\c{c}\~ao}},
  \bibinfo {author} {\bibfnamefont {T.}~\bibnamefont {Mariz}},\ and\ \bibinfo
  {author} {\bibfnamefont {A.~Y.}\ \bibnamefont {Petrov}},\ }\bibfield  {title}
  {\bibinfo {title} {{One-loop calculations in Lorentz-breaking theories and
  proper-time method}},\ }\Eprint {https://arxiv.org/abs/2109.11901}
  {arXiv:2109.11901 [hep-th]}  (\bibinfo {year} {2021})\BibitemShut {NoStop}%
\bibitem [{\citenamefont {Kosteleck\'{y}}\ and\ \citenamefont
  {Lane}(1999)}]{Kostelecky:1999zh}%
  \BibitemOpen
  \bibfield  {author} {\bibinfo {author} {\bibfnamefont {V.~A.}\ \bibnamefont
  {Kosteleck\'{y}}}\ and\ \bibinfo {author} {\bibfnamefont {C.~D.}\
  \bibnamefont {Lane}},\ }\bibfield  {title} {\bibinfo {title}
  {{Nonrelativistic quantum Hamiltonian for Lorentz violation}},\ }\href
  {https://doi.org/10.1063/1.533090} {\bibfield  {journal} {\bibinfo  {journal}
  {J. Math. Phys.}\ }\textbf {\bibinfo {volume} {40}},\ \bibinfo {pages} {6245}
  (\bibinfo {year} {1999})},\ \Eprint {https://arxiv.org/abs/hep-ph/9909542}
  {arXiv:hep-ph/9909542} \BibitemShut {NoStop}%
\bibitem [{\citenamefont {Ahari}\ \emph {et~al.}(2016)\citenamefont {Ahari},
  \citenamefont {Ortiz},\ and\ \citenamefont {Seradjeh}}]{Ahari_2016}%
  \BibitemOpen
  \bibfield  {author} {\bibinfo {author} {\bibfnamefont {M.~T.}\ \bibnamefont
  {Ahari}}, \bibinfo {author} {\bibfnamefont {G.}~\bibnamefont {Ortiz}},\ and\
  \bibinfo {author} {\bibfnamefont {B.}~\bibnamefont {Seradjeh}},\ }\bibfield
  {title} {\bibinfo {title} {On the role of self-adjointness in the continuum
  formulation of topological quantum phases},\ }\href
  {https://doi.org/10.1119/1.4961500} {\bibfield  {journal} {\bibinfo
  {journal} {American Journal of Physics}\ }\textbf {\bibinfo {volume} {84}},\
  \bibinfo {pages} {858} (\bibinfo {year} {2016})}\BibitemShut {NoStop}%
\bibitem [{\citenamefont {Seradjeh}\ and\ \citenamefont
  {Vennettilli}(2018)}]{Seradjeh_2018}%
  \BibitemOpen
  \bibfield  {author} {\bibinfo {author} {\bibfnamefont {B.}~\bibnamefont
  {Seradjeh}}\ and\ \bibinfo {author} {\bibfnamefont {M.}~\bibnamefont
  {Vennettilli}},\ }\bibfield  {title} {\bibinfo {title} {{Surface spectra of
  Weyl semimetals through self-adjoint extensions}},\ }\href
  {https://doi.org/10.1103/physrevb.97.075132} {\bibfield  {journal} {\bibinfo
  {journal} {Phys. Rev. B}\ }\textbf {\bibinfo {volume} {97}},\ \bibinfo
  {pages} {075132} (\bibinfo {year} {2018})},\ \Eprint
  {https://arxiv.org/abs/1712.04355} {arXiv:1712.04355 [cond-mat.mes-hall]}
  \BibitemShut {NoStop}%
\bibitem [{\citenamefont {Peskin}\ and\ \citenamefont
  {Schroeder}(1995)}]{Peskin:1995ev}%
  \BibitemOpen
  \bibfield  {author} {\bibinfo {author} {\bibfnamefont {M.~E.}\ \bibnamefont
  {Peskin}}\ and\ \bibinfo {author} {\bibfnamefont {D.~V.}\ \bibnamefont
  {Schroeder}},\ }\href@noop {} {\emph {\bibinfo {title} {An Introduction to
  Quantum Field Theory}}}\ (\bibinfo  {publisher} {Addison-Wesley Publishing
  Company},\ \bibinfo {address} {Reading; Massachusetts},\ \bibinfo {year}
  {1995})\BibitemShut {NoStop}%
\bibitem [{\citenamefont {Grushin}(2019)}]{Grushin:2019uuu}%
  \BibitemOpen
  \bibfield  {author} {\bibinfo {author} {\bibfnamefont {A.~G.}\ \bibnamefont
  {Grushin}},\ }\bibfield  {title} {\bibinfo {title} {{Common and Not-So-Common
  High-Energy Theory Methods for Condensed Matter Physics}},\ }in\ \href
  {https://doi.org/10.1007/978-3-319-76388-0_6} {\emph {\bibinfo {booktitle}
  {{Topological Matter {\textemdash} Lectures from the Topological Matter
  School 2017}}}},\ \bibinfo {editor} {edited by\ \bibinfo {editor}
  {\bibfnamefont {D.}~\bibnamefont {Bercioux}}, \bibinfo {editor}
  {\bibfnamefont {J.}~\bibnamefont {Cayssol}}, \bibinfo {editor} {\bibfnamefont
  {M.~G.}\ \bibnamefont {Vergniory}},\ and\ \bibinfo {editor} {\bibfnamefont
  {M.~R.}\ \bibnamefont {Calvo}}}\ (\bibinfo  {publisher} {Springer},\ \bibinfo
  {address} {Cham, Switzerland},\ \bibinfo {year} {2019})\ Chap.~\bibinfo
  {chapter} {6}, pp.\ \bibinfo {pages} {149--176},\ \Eprint
  {https://arxiv.org/abs/1909.02983} {arXiv:1909.02983 [cond-mat.mes-hall]}
  \BibitemShut {NoStop}%
\bibitem [{\citenamefont {{N. McGinnis}}(2020)}]{McGinnis:2020tyj}%
  \BibitemOpen
  \bibfield  {author} {\bibinfo {author} {\bibnamefont {{N. McGinnis}}},\
  }\bibfield  {title} {\bibinfo {title} {{The Standard-Model Extension as an
  Effective Field Theory for Weyl Semimetals}},\ }in\ \href
  {https://doi.org/10.1142/9789811213984_0061} {\emph {\bibinfo {booktitle}
  {{Proceedings of the Eighth Meeting on CPT and Lorentz Symmetry}}}},\
  \bibinfo {editor} {edited by\ \bibinfo {editor} {\bibnamefont {{R.
  Lehnert}}}}\ (\bibinfo  {publisher} {{World Scientific Publishing Co Pte
  Ltd}},\ \bibinfo {address} {Singapore},\ \bibinfo {year} {2020})\ pp.\
  \bibinfo {pages} {231--233}\BibitemShut {NoStop}%
\bibitem [{\citenamefont {Kosteleck\'y}\ \emph {et~al.}(2012)\citenamefont
  {Kosteleck\'y}, \citenamefont {Russell},\ and\ \citenamefont
  {Tso}}]{AlanKostelecky:2012yjr}%
  \BibitemOpen
  \bibfield  {author} {\bibinfo {author} {\bibfnamefont {V.~A.}\ \bibnamefont
  {Kosteleck\'y}}, \bibinfo {author} {\bibfnamefont {N.}~\bibnamefont
  {Russell}},\ and\ \bibinfo {author} {\bibfnamefont {R.}~\bibnamefont {Tso}},\
  }\bibfield  {title} {\bibinfo {title} {{Bipartite Riemann-Finsler geometry
  and Lorentz violation}},\ }\href
  {https://doi.org/10.1016/j.physletb.2012.09.002} {\bibfield  {journal}
  {\bibinfo  {journal} {Phys. Lett. B}\ }\textbf {\bibinfo {volume} {716}},\
  \bibinfo {pages} {470} (\bibinfo {year} {2012})},\ \Eprint
  {https://arxiv.org/abs/1209.0750} {arXiv:1209.0750 [hep-th]} \BibitemShut
  {NoStop}%
\bibitem [{\citenamefont {Russell}(2015)}]{Russell:2015gwa}%
  \BibitemOpen
  \bibfield  {author} {\bibinfo {author} {\bibfnamefont {N.}~\bibnamefont
  {Russell}},\ }\bibfield  {title} {\bibinfo {title} {{Finsler-like structures
  from Lorentz-breaking classical particles}},\ }\href
  {https://doi.org/10.1103/PhysRevD.91.045008} {\bibfield  {journal} {\bibinfo
  {journal} {Phys. Rev. D}\ }\textbf {\bibinfo {volume} {91}},\ \bibinfo
  {pages} {045008} (\bibinfo {year} {2015})},\ \Eprint
  {https://arxiv.org/abs/1501.02490} {arXiv:1501.02490 [hep-th]} \BibitemShut
  {NoStop}%
\bibitem [{\citenamefont {Foster}\ and\ \citenamefont
  {Lehnert}(2015)}]{Foster:2015yta}%
  \BibitemOpen
  \bibfield  {author} {\bibinfo {author} {\bibfnamefont {J.}~\bibnamefont
  {Foster}}\ and\ \bibinfo {author} {\bibfnamefont {R.}~\bibnamefont
  {Lehnert}},\ }\bibfield  {title} {\bibinfo {title} {{Classical-physics
  applications for Finsler $b$ space}},\ }\href
  {https://doi.org/10.1016/j.physletb.2015.04.047} {\bibfield  {journal}
  {\bibinfo  {journal} {Phys. Lett. B}\ }\textbf {\bibinfo {volume} {746}},\
  \bibinfo {pages} {164} (\bibinfo {year} {2015})},\ \Eprint
  {https://arxiv.org/abs/1504.07935} {arXiv:1504.07935 [physics.class-ph]}
  \BibitemShut {NoStop}%
\bibitem [{\citenamefont {Schreck}(2016)}]{Schreck:2015seb}%
  \BibitemOpen
  \bibfield  {author} {\bibinfo {author} {\bibfnamefont {M.}~\bibnamefont
  {Schreck}},\ }\bibfield  {title} {\bibinfo {title} {{Classical Lagrangians
  and Finsler structures for the nonminimal fermion sector of the
  Standard-Model Extension}},\ }\href
  {https://doi.org/10.1103/PhysRevD.93.105017} {\bibfield  {journal} {\bibinfo
  {journal} {Phys. Rev. D}\ }\textbf {\bibinfo {volume} {93}},\ \bibinfo
  {pages} {105017} (\bibinfo {year} {2016})},\ \Eprint
  {https://arxiv.org/abs/1512.04299} {arXiv:1512.04299 [hep-th]} \BibitemShut
  {NoStop}%
\bibitem [{\citenamefont {Reis}\ and\ \citenamefont
  {Schreck}(2018)}]{Reis:2017ayl}%
  \BibitemOpen
  \bibfield  {author} {\bibinfo {author} {\bibfnamefont {J.~A. A.~S.}\
  \bibnamefont {Reis}}\ and\ \bibinfo {author} {\bibfnamefont {M.}~\bibnamefont
  {Schreck}},\ }\bibfield  {title} {\bibinfo {title} {{Leading-order classical
  Lagrangians for the nonminimal standard-model extension}},\ }\href
  {https://doi.org/10.1103/PhysRevD.97.065019} {\bibfield  {journal} {\bibinfo
  {journal} {Phys. Rev. D}\ }\textbf {\bibinfo {volume} {97}},\ \bibinfo
  {pages} {065019} (\bibinfo {year} {2018})},\ \Eprint
  {https://arxiv.org/abs/1711.11169} {arXiv:1711.11169 [hep-th]} \BibitemShut
  {NoStop}%
\bibitem [{\citenamefont {Colladay}(2017)}]{Colladay:2017bon}%
  \BibitemOpen
  \bibfield  {author} {\bibinfo {author} {\bibfnamefont {D.}~\bibnamefont
  {Colladay}},\ }\bibfield  {title} {\bibinfo {title} {{Extended hamiltonian
  formalism and Lorentz-violating lagrangians}},\ }\href
  {https://doi.org/10.1016/j.physletb.2017.07.027} {\bibfield  {journal}
  {\bibinfo  {journal} {Phys. Lett. B}\ }\textbf {\bibinfo {volume} {772}},\
  \bibinfo {pages} {694} (\bibinfo {year} {2017})},\ \Eprint
  {https://arxiv.org/abs/1706.06637} {arXiv:1706.06637 [hep-th]} \BibitemShut
  {NoStop}%
\bibitem [{\citenamefont {Schreck}(2019)}]{Schreck:2019mmr}%
  \BibitemOpen
  \bibfield  {author} {\bibinfo {author} {\bibfnamefont {M.}~\bibnamefont
  {Schreck}},\ }\bibfield  {title} {\bibinfo {title} {{Classical Lagrangians
  for the nonminimal Standard-Model Extension at higher orders in Lorentz
  violation}},\ }\href {https://doi.org/10.1016/j.physletb.2019.04.021}
  {\bibfield  {journal} {\bibinfo  {journal} {Phys. Lett. B}\ }\textbf
  {\bibinfo {volume} {793}},\ \bibinfo {pages} {70} (\bibinfo {year} {2019})},\
  \Eprint {https://arxiv.org/abs/1903.05064} {arXiv:1903.05064 [hep-th]}
  \BibitemShut {NoStop}%
\bibitem [{\citenamefont {Reis}\ and\ \citenamefont
  {Schreck}(2021)}]{Reis:2021ban}%
  \BibitemOpen
  \bibfield  {author} {\bibinfo {author} {\bibfnamefont {J.~A. A.~S.}\
  \bibnamefont {Reis}}\ and\ \bibinfo {author} {\bibfnamefont {M.}~\bibnamefont
  {Schreck}},\ }\bibfield  {title} {\bibinfo {title} {{Classical Lagrangians
  for the nonminimal spin-nondegenerate Standard-Model extension at higher
  orders in Lorentz violation}},\ }\href
  {https://doi.org/10.1103/PhysRevD.103.095029} {\bibfield  {journal} {\bibinfo
   {journal} {Phys. Rev. D}\ }\textbf {\bibinfo {volume} {103}},\ \bibinfo
  {pages} {095029} (\bibinfo {year} {2021})},\ \Eprint
  {https://arxiv.org/abs/2102.10164} {arXiv:2102.10164 [hep-th]} \BibitemShut
  {NoStop}%
\end{thebibliography}%

\end{document}